\documentclass[review,nonatbib]{elsarticle}

\usepackage{graphicx}
\usepackage{amsmath, amssymb, bm}
\usepackage{siunitx}
\usepackage{physics}
\usepackage{booktabs, longtable}
\usepackage{array, tabularx}
\usepackage{subcaption}
\usepackage{microtype}
\usepackage{xcolor}
\usepackage[hidelinks]{hyperref}

\makeatletter
\let\c@author\relax
\let\c@collab\relax
\makeatother
\usepackage[backend=biber,style=numeric-comp,sorting=none,maxbibnames=99]{biblatex}
\usepackage[nomarkers,figuresonly]{endfloat}

\newcolumntype{Y}{>{\raggedright\arraybackslash}X}


\title{Transition Metal Dichalcogenides Multijunction Solar Cells Toward the Multicolor Limit}

\begin{document}

\begin{frontmatter}

\author{Seungwoo Lee}
\address{Department of Integrated Energy Engineering (College of Engineering), KU-KIST Graduate School of Converging Science and Technology, and Department of Biomicrosystem Technology, Korea University, Seoul 02841, Republic of Korea}
\ead{seungwoo@korea.ac.kr}

\begin{abstract}
Transition metal dichalcogenides (TMDs) and other van der Waals (vdW) semiconductors enable transfer-printed, lattice--mismatch-free stacking of many photovoltaic junctions, motivating a re-examination of multijunction detailed-balance limits under realistic material and optical constraints. Here, we develop a unified thermodynamic framework for a multijunction photovoltaic device, which can define a set of device bandgap-window constraints, optical boundary conditions, and luminescence/entropy penalties and therefore define how closely any realistic multijunction photovoltaic device can approach multicolor limit. By applying it to a conservative TMD bandgap window (1.0--2.1~eV), we show that the accessible bandgap window imposes a large-junction number ($N$) efficiency limit: under full concentration, unconstrained ladders approach 84.5\% at $N=50$, whereas the TMD window plateaus near 63.4\%. This efficiency plateau is set by photons outside the bandgaps, so radiative quality and optics dominate beyond $N=5$ junctions for realistic transfer-printed device stacks. We distinguish this thermodynamic limit from implementation-level derating by interface recombination, parasitic absorption, exciton collection, contact/series resistance, and external power-management losses. We also identify an experimentally achievable $N=5$ ladder $E_g\approx(2.10,1.78,1.50,1.24, 1.00)$~eV and map each rung to candidate vdW/TMD absorbers. Using reciprocity and luminescence thermodynamics, we quantify penalties from finite external radiative efficiency, two-sided emission, and luminescent coupling, and introduce the upward-emitted luminescence power as an indicator of entropy-loss proxy. Incorporating excitonic absorptance and nanophotonic thickness bounds yields practical thickness and light-management targets for transfer-printed stacks. Finally, inserting an idealized nonreciprocal multijunction model into the reciprocity-optimized ladders provides conservative efficiency advantage estimates, which are consistent with negligible benefit for single junctions but measurable efficiency gains for multijunction TMD devices.
\end{abstract}

\begin{keyword}
multijunction photovoltaics \sep transition metal dichalcogenides \sep detailed balance \sep external radiative efficiency \sep luminescent coupling
\end{keyword}

\end{frontmatter}

\section{Introduction}
The Shockley--Queisser (SQ) detailed-balance limit for a single-junction solar cell formalizes the thermodynamic tradeoffs between (i) transmission loss from sub-bandgap photons, (ii) thermalization loss from above-bandgap photons, and (iii) radiative emission that reduces the achievable open-circuit voltage $V_\mathrm{oc}$.\cite{ShockleyQueisser1961}
Multijunction photovoltaics mitigates these losses by partitioning the solar spectrum across multiple bandgaps.\cite{DeVos1980,Green2006,KimNanophotonics2025} 
In the radiative and ideal-optics limit, the efficiency of multijunction devices approaches the reciprocal multicolor limit as the number of junctions ($N$) grows.

Despite decades of progress, pushing beyond triple-junction stacks remains a challenge in conventional epitaxial material systems, because lattice matching, thermal expansion mismatch, and defect formation lead to practical constraints.
Van der Waals (vdWs) semiconductors, especially transition metal dichalcogenides (TMDs), offer a fundamentally different integration pathway: 
their layered nature enables mechanical exfoliation, deterministic transfer printing, and assembly of many junctions without lattice matching.\cite{JariwalaAtwater2017,TMDReview2024,ReichPop2023}
This raises a natural theoretical question: if we could stack an arbitrary number of vdW junctions and could select bandgaps broadly across the solar spectrum, how close could we approach multicolor or even Landsberg limits?

Answering this question requires more than bandgap optimization. 
As emphasized by luminescence thermodynamics, high-efficiency photovoltaics must exhibit strong internal radiative recombination and high external photon extraction.\cite{MillerYablonovitch2012, KimLeeOE2019, KirchartzRau2018, Rau2007, LeeAFM2023LSC} 
Reciprocity further implies that absorption and emission are inseparable: thickness- and nanophotonics-controlled absorptance defines not only the short-circuit current $J_\mathrm{sc}$ but also the radiative dark current and therefore $V_\mathrm{oc}$.\cite{Rau2007,Rau2014} 
In stacked junctions without intermediate mirrors,\cite{KimNanophotonics2025} an upper junction can emit externally both upward (to the environment) and downward (into the stack), which (i) opens additional optical emission channels and thus introduces an entropy and voltage penalty, and (ii) enables luminescent coupling to lower junctions.\cite{DeVos1980,Green2006,KimNanophotonics2025}
These coupling processes can give rise to additional thermalization or entropic losses in the lower junctions, particularly when the bandgaps are closely spaced and the upper junction is highly radiative.
Recent detailed-balance analyses of nanophotonic intermediate mirrors in tandem photovoltaics highlight that the net gain can be strongly device configuration-dependent, highlighting the need for stack-specific photon-entropy accounting.\cite{KimNanophotonics2025}

Motivated by this perspective, we pursue a conservative and practical goal: 
maximize the reciprocal multijunction efficiency achievable with a realistic TMD bandgap window (1.0--2.1~eV), and quantify how both luminescence extraction and thickness constraints define the approach to the multicolor limit.
Nonreciprocal optics is treated as an optional strategy: we benchmark its theoretical efficiency gain by applying the idealized nonreciprocal multijunction model of Fan et al.\cite{FanNonreciprocalMJ2022} to the same bandgap ladders, which are optimized under reciprocity (a conservative ``plug-in'' comparison). Then, we systematically contrast this multijunction efficiency gain with the absence of benefit in single-junction photovoltaics.\cite{FanAPL2022} These developments reinforce the motivation for a unified thermodynamic treatment of emission, coupling, and thickness constraints when designing transfer-printed TMD multijunction stacks.

% =========================================================
\section{The device concept and photon-flow constraints}
\begin{figure}[t]
  \centering
  \includegraphics[width=0.95\linewidth]{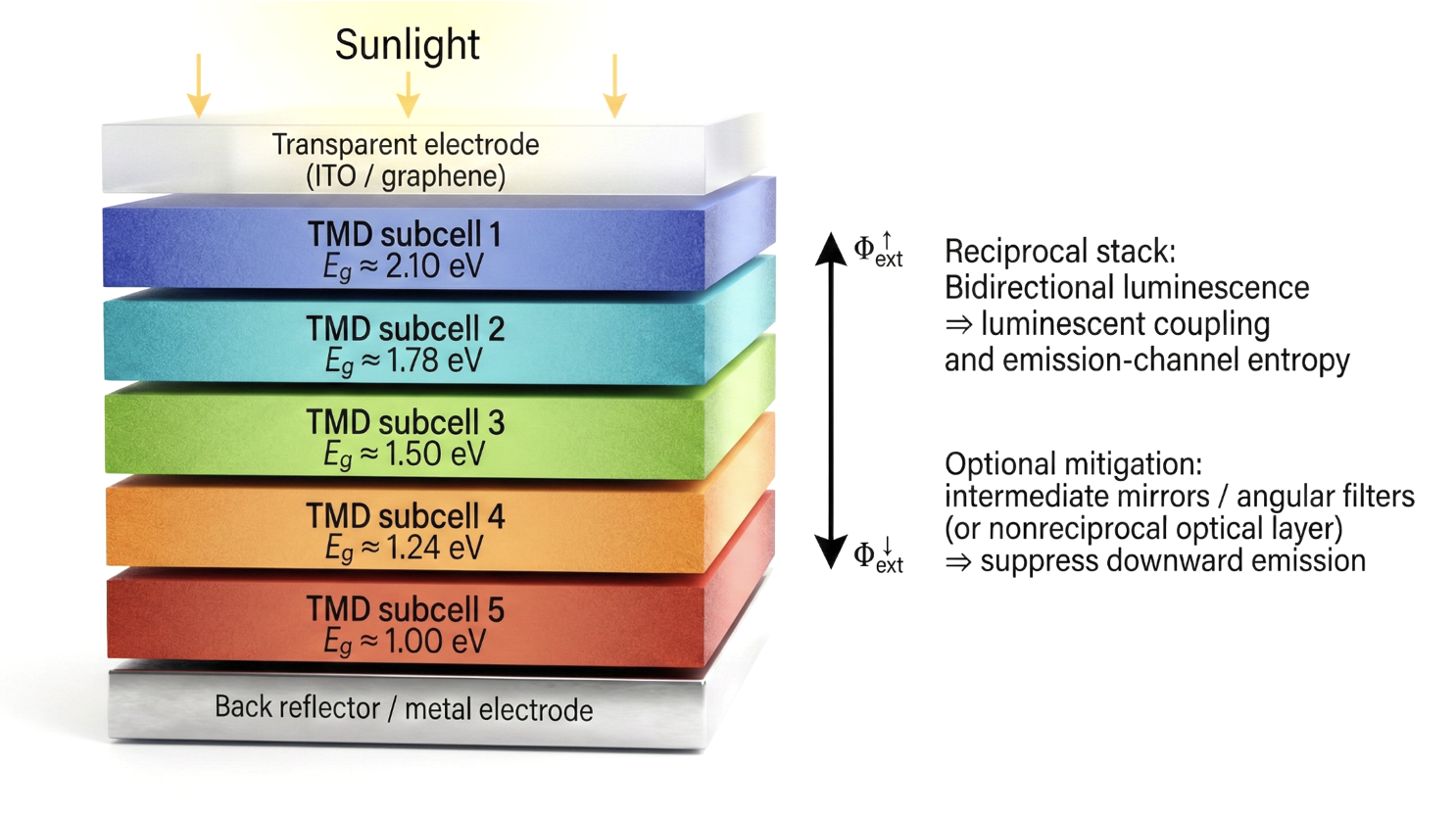}
  \caption{\textbf{Conceptual schematic of a transfer-printed vdW/TMD multijunction solar cell (representative $N=5$ ladder).} 
  In a reciprocal stack without intermediate mirrors, external luminescence from an upper junction can exit upward ($\Phi^\uparrow_\mathrm{ext}$) and propagate downward ($\Phi^\downarrow_\mathrm{ext}$), enabling luminescent coupling to lower junctions and opening additional emission channels (entropy/voltage penalty). 
  Intermediate mirrors or angular filters can suppress unwanted emission channels and recover higher radiative voltage.\cite{MillerYablonovitch2012,Rau2014} 
  An optional nonreciprocal optical layer can, in principle, further reshape emission pathways and access additional thermodynamic efficiency gain in multijunction stacks.\cite{FanNonreciprocalMJ2022}
  }
  \label{fig:fig1}
\end{figure}

Figure~\ref{fig:fig1} summarizes the transfer-printed TMD multijunction architecture, studied in this work: a stack of independently contacted subcells assembled by TMD stacking or transfer printing, with selective contacts/interlayers between subcells, a transparent top electrode, and a reflective back electrode. 
Throughout, we focus on multi-terminal extraction (each TMD subcell operates at its own maximum power point), which isolates the fundamental spectral-partitioning thermodynamics from current-matching constraints.

This multi-terminal, lossless-contact description should be interpreted as an upper-bound diagnostic rather than as a claim that all interfaces, contacts, and external electronics are ideal in a fabricated stack. Herein, the device schematic is therefore tied to explicit implementation components and loss channels (Table~\ref{tab:components}): interface recombination and transfer-induced defects enter through subcell-specific external radiative efficiency (ERE) values, parasitic electrode/interlayer absorption enters through the active absorptance used in $J_{\mathrm{sc}}$, contact and sheet resistance enter through terminal-voltage derating, and independent maximum-power-point tracking (MPPT) enters as a system-level electrical derating. This separation is useful, because it preserves the thermodynamic meaning of the bandgap-ladder limit, while making clear which fabrication issues must be minimized experimentally.

\begin{table}[t]
\centering
\caption{\textbf{Representative implementation components and nonideal loss channels for a transfer-printed vdW/TMD multijunction stack.} The ideal detailed-balance curves in this work are thermodynamic ceilings; the listed terms indicate how practical losses can be included as derating factors without changing the bandgap-ladder optimization framework.}
\label{tab:components}
\small
\setlength{\tabcolsep}{4pt}
\renewcommand{\arraystretch}{1.12}
\begin{tabularx}{\linewidth}{@{}p{2.6cm} p{3.5cm} Y@{}}
\toprule
Stack element / issue & Example components & Model entry and design implication\\
\midrule
Top transparent electrode & Graphene, ITO, ultrathin TCO/metal grid & Parasitic absorption reduces the active absorptance $A_i(E)$; sheet/contact resistance is a series-resistance derating.\\
Transfer-printed TMD absorber and vdW interface & Monolayer/few-layer WS$_2$, MoS$_2$, MoSe$_2$, WSe$_2$, MoTe$_2$ with hBN, Al$_2$O$_3$, or polymer/oxide encapsulation & Transfer-induced defects and interface recombination lower $\mathrm{ERE}_i$ and collection efficiency; passivation and clean transfer raise the attainable voltage.\\
Carrier-selective interlayers / contacts & Graphene, metal oxides, low-work-function metals, vdW tunnel contacts & Optical loss enters as parasitic absorption; contact barriers and lateral transport enter as $R_{s,i}$ and possible crosstalk.\\
Intermediate optical control & Dielectric intermediate mirrors, angular filters, photonic cavities, optional nonreciprocal layer & Changes $\Omega_{\mathrm{emit}}$ and the downward-coupling term; suppresses unwanted emission channels and coupling thermalization.\\
Back reflector / bottom electrode & Ag or Al reflector with passivation and selective contact & Approximates the one-sided emission boundary condition; imperfect reflectivity moves the stack toward the two-sided-emission case.\\
External power electronics & Independent MPPT or submodule-level DC--DC conversion & Multiplies the summed terminal power by an MPPT/electrical efficiency; not included in the thermodynamic ceiling.\\
\bottomrule
\end{tabularx}
\end{table}

Two points are central. First, in the detailed-balance limit, each junction is a light-emitting diode under forward bias: at open circuit and near maximum power, the device emits a substantial photon flux. Consequently, strong external luminescence is not a loss to be suppressed, but a signature of a high-quality device approaching the SQ limit.\cite{MillerYablonovitch2012} Second, in a stacked architecture, the direction and angular distribution of luminescence matter. If an upper junction emits downward into the stack, that emission is not necessarily ``wasted.'' It can be re-absorbed by lower junctions (luminescent coupling). However, it can still introduce (i) a voltage penalty for the emitter by opening emission channels, and (ii) a thermalization penalty in the absorber, if the photon energy significantly exceeds the lower-junction bandgap. As such, a quantitative framework must track both photon flux and power flux of luminescence.

% =========================================================
\section{Theory and Methods}
\subsection{Spectral model, concentration, and incident power}
We model the Sun as a blackbody at temperature $T_s$ and the cell as a blackbody at temperature $T_c$ (unless otherwise stated $T_s=\SI{5778}{K}$ and $T_c=\SI{300}{K}$). 
The spectral photon flux per unit area per unit energy per steradian is
\begin{equation}
\Phi_\mathrm{bb}(E,T)=\frac{2E^2}{h^3c^2}\,\frac{1}{\exp(E/k_BT)-1}
\label{eq:planck}
\end{equation}
where $E$ is photon energy, $h$ is Planck's constant, and $c$ is the speed of light. 
The Sun subtends solid angle $\Omega_s=\pi\sin^2\theta_s$, where $\theta_s\simeq\SI{0.266}{degree}$ is the solar half-angle.
Optical concentration is represented by scaling the incident flux by a factor $C$, so that the spectral photon flux, which is incident on the device, is
\begin{equation}
\Phi_\odot(E)=C\,\Omega_s\,\Phi_\mathrm{bb}(E,T_s)
\end{equation}
The incident radiative power density is
\begin{equation}
P_\mathrm{in}(C)=\int_0^\infty E\,\Phi_\odot(E)\,dE
\end{equation}
In full concentration, we use $C=C_\mathrm{max}=1/\sin^2\theta_s$ (consistent with the SQ convention in which the Sun's étendue is expanded to $\pi$ steradians).\cite{ShockleyQueisser1961,Green2006}
All numerical integrations are performed on an energy grid spanning \SIrange{0.01}{10}{eV}, which captures $\approx 98\%$ of the blackbody solar power for $T_s=\SI{5778}{K}$. Other important physical constants and more detailed information for geometry and spectral model, which are used in this work, are summarized in Supporting Information (Section 1 and Table S1).

\subsection{Single-junction detailed balance}
For an ideal diode, the current--voltage relation is
\begin{equation}
J(V)=J_\mathrm{sc}-J_0\left[\exp\left(\frac{qV}{k_BT_c}\right)-1\right]
\label{eq:JV}
\end{equation}
where $J_\mathrm{sc}$ and $J_0$ are the short-circuit current density and the dark-current prefactor, respectively. For an absorber with absorptance $A(E)$ (not necessarily unity), the photogenerated current density is
\begin{equation}
J_\mathrm{sc}=q\int_0^\infty A(E)\,\Phi_\odot(E)\,dE
\label{eq:Jsc_general}
\end{equation}
Radiative recombination produces an emitted photon flux proportional to $A(E)\,\Phi_\mathrm{bb}(E,T_c)$ by both Kirchhoff's law and reciprocity.\cite{Rau2007}
Thus, the radiative dark current is
\begin{equation}
J_{0,\mathrm{rad}}=q\int_0^\infty A(E)\,\Phi_\mathrm{bb}(E,T_c)\,\Omega_\mathrm{emit}\,dE
\label{eq:J0rad_general}
\end{equation}
where $\Omega_\mathrm{emit}$ is the emission solid angle. For one-sided emission (perfect back reflector), $\Omega_\mathrm{emit}=2\pi$, whereas two-sided emission corresponds to $\Omega_\mathrm{emit}=4\pi$.

Nonradiative recombination and parasitic optical loss reduce the fraction of recombination events, which are responsible for externally emitted photons. 
A convenient device-quality metric is the $\mathrm{ERE}$, which can be introduced via
\begin{equation}
J_0=\frac{J_{0,\mathrm{rad}}}{\mathrm{ERE}}
\label{eq:ERE}
\end{equation}
At fixed $J_\mathrm{sc}$, Eq.~\eqref{eq:ERE} produces a voltage penalty
\begin{equation}
\Delta V_\mathrm{ERE}=\frac{k_BT_c}{q}\ln(\mathrm{ERE})
\end{equation}
which is identical in form to the $\ln Q^\mathrm{LED}_e$ term that appears in the thermodynamic voltage decomposition of Rau and co-workers.\cite{Rau2014,Rau2017,KirchartzRau2018} More detailed derivations and validation for single-junction detailed balance are summarized in Supporting Information (Section 2).

\subsection{Practical loss channels beyond the thermodynamic limit}
The ideal curves reported in Figures~\ref{fig:fig2}--\ref{fig:fig10} intentionally separate the thermodynamic limit from device-specific derating. For a fabricated TMD stack, the absorptance that contributes to current should be the collected active absorptance,
\begin{equation}
A^{\mathrm{col}}_i(E;t)=\eta_{\mathrm{diss},i}(E)\,\eta_{\mathrm{tr},i}(E,t)\,A^{\mathrm{act}}_i(E;t),
\label{eq:Acol_main}
\end{equation}
where $\eta_{\mathrm{diss},i}$ represents exciton/free-carrier dissociation at the selective contact or vdW heterointerface, $\eta_{\mathrm{tr},i}$ represents transport and collection, and $A^{\mathrm{act}}_i$ excludes parasitic absorption in electrodes and interlayers. Nonradiative recombination at transfer-induced defects or vdW interfaces is captured by $\mathrm{ERE}_i<1$. Series and contact resistance can be introduced by the terminal-voltage relation $V_{\mathrm{term},i}=V_i-J_iR_{s,i}$, so that $P_i=V_{\mathrm{term},i}J_i$ rather than $V_iJ_i$ (see more details in Section 2.3 of Supporting Information). Therefore, these practical nonidealities lower the achievable efficiency below the limit curves, but they do not change the purpose of the present model: to identify the best bandgap window, emission boundary conditions, and optical targets before applying material- and process-specific derating factors.

The maximum power point (MPP) can be obtained analytically using the Lambert-$W$ function. 
Starting from Eq.~\eqref{eq:JV}, the condition $d(VJ)/dV=0$ leads to $(1+v)\exp(v)=1+J_\mathrm{sc}/J_0$ where $v=qV/(k_BT_c)$. This gives
\begin{equation}
v_\mathrm{mpp}=W\!\left(e\left[1+J_\mathrm{sc}/J_0\right]\right)-1
\label{eq:vmpp}
\end{equation}
with $V_\mathrm{mpp}=(k_BT_c/q)\,v_\mathrm{mpp}$ and $P_\mathrm{mpp}=V_\mathrm{mpp}J(V_\mathrm{mpp})$.
A full derivation is provided in the Supporting Information (Section 2.1--2.5).

\subsection{Unlimited-junction split-spectrum multijunction model}
We consider a $N$-junction multijunction device with bandgaps $E_{g,1}>E_{g,2}>\cdots>E_{g,N}$. 
In the split-spectrum and multi-terminal limit, junction $i$ interacts with photon energies in a window $E\in[E_{g,i},E_{g,i-1}]$, where $E_{g,0}\equiv E_\mathrm{max}$ is a numerical high-energy cutoff. For step absorbers, $A_i(E)=1$ in this window and $A_i(E)=0$ outside it.
The total maximum power density is
\begin{equation}
P^\ast_\mathrm{tot}=\sum_{i=1}^N P_i^\ast(E_{g,i};E_{g,i-1})
\label{eq:Ptot}
\end{equation}
where $P_i^\ast$ is obtained from Eqs.~\eqref{eq:Jsc_general}--\eqref{eq:vmpp}, which are restricted to the window. We compute optimal bandgap ladders by maximizing Eq.~\eqref{eq:Ptot} with dynamic programming (DP) on a discretized bandgap grid (see more details in Section 3, Supporting Information). We study two scenarios:
(i) unconstrained $E_g$ (theoretical convergence) and 
(ii) a conservative, experimentally available TMD bandgap window $E_g\in[1.0,2.1]$~eV.

\subsection{Reciprocity, emission channels, and luminescent coupling in reciprocal stacks}
The reciprocity relation between photovoltaic quantum efficiency and electroluminescent emission connects absorption and emission spectra.\cite{Rau2007}
A compact form for the emitted photon flux spectrum under bias $V$ is
\begin{equation}
\phi_\mathrm{em}(E,V)=A(E)\,\Phi_\mathrm{bb}(E,T_c)\,\Omega_\mathrm{emit}\left[\exp\left(\frac{qV}{k_BT_c}\right)-1\right]
\label{eq:emission_spectrum}
\end{equation}
where $A(E)$ is the device absorptance (or external quantum efficiency under ideal collection).
Equation~\eqref{eq:emission_spectrum} implies that any thickness- or nanophotonics-induced change in $A(E)$ affects both $J_\mathrm{sc}$ and $J_{0,\mathrm{rad}}$ (also, see Section 4, Supporting Information).

In a reciprocal multijunction stack lacking intermediate mirrors, each junction can emit both upward and downward (two hemispheres). Allowing both forward and downward hemispheres doubles the radiative dark current prefactor and produces a radiative-voltage reduction $\Delta V=(k_BT_c/q)\ln 2$ at fixed generation.
More importantly, downward emission from junction $i$ can be absorbed by junction $i+1$, if $E_{g,i}>E_{g,i+1}$, producing a luminescent-coupling current $J_\mathrm{LC}$.
We model this effect using a minimal coupled detailed-balance chain (also, called luminescent coupling chain model, as further detailed in Section 5, Supporting Information), in which
\begin{align}
J_i(V_i) &= J_{\odot,i} + J^\downarrow_{\mathrm{rad},i-1}(V_{i-1}) - \frac{J_{0,i}^\uparrow+J_{0,i}^\downarrow}{\mathrm{ERE}_i}\left[\exp\left(\frac{qV_i}{k_BT_c}\right)-1\right] \\
J^\downarrow_{\mathrm{rad},i}(V_i) &= J_{0,i}^\downarrow\left[\exp\left(\frac{qV_i}{k_BT_c}\right)-1\right]
\end{align}
with $J^\downarrow_{\mathrm{rad},0}\equiv 0$. 
This model captures (i) the emission-channel entropy penalty (i.e., two-sided emission increases $J_0$), and (ii) the photon recycling and coupling that can boost the lower-junction current.
We solve for the system MPP by coordinate-ascent optimization over the set of subcell voltages $\{V_i\}$.

\subsection{Entropy loss as upward-emitted luminescence power and coupling thermalization}
A recurring conceptual challenge is to interpret luminescence in multijunction stacks: luminescence is both a necessity for high $V_\mathrm{oc}$ or a potential loss channel (if it can be emitted into parasitic modes).
To make this quantitative, we compute the upward-emitted luminescence power
\begin{equation}
P^\uparrow_\mathrm{lum}=\sum_i \int E\,\phi^\uparrow_{\mathrm{em},i}(E,V_i)\,dE
\end{equation}
and report $P^\uparrow_\mathrm{lum}/P_\mathrm{in}$ as an entropy-loss proxy for emission-channel losses. 
In reciprocal stacks without intermediate mirrors, $P^\uparrow_\mathrm{lum}$ is nonzero, because the stack must emit externally. 
In the idealized nonreciprocal model, $P^\uparrow_\mathrm{lum}$ can be suppressed in multijunction architectures, enabling additional efficiency gain.\cite{FanNonreciprocalMJ2022}

Downward luminescence, in contrast, is not necessarily lost: it can be absorbed by the next junction and partially converted to electrical work. 
However, the portion of downward luminescent power that exceeds the electrical work extracted by the lower junction becomes heat (i.e., thermalization loss).
We compute a coupling-thermalization proxy:
\begin{equation}
P_\mathrm{therm}^{\mathrm{LC}}=\sum_{i=1}^{N-1}\left(P^\downarrow_{\mathrm{lum},i}-V_{i+1}J_{\mathrm{LC},i}\right)
\end{equation}
where $P^\downarrow_{\mathrm{lum},i}$ is the downward luminescent power emitted by junction $i$, while $J_{\mathrm{LC},i}$ is the corresponding coupling current absorbed by junction $i+1$.
We use these definitions to separate upward entropy loss from coupling-induced thermalization for a representative $N=5$ ladder.

\subsection{Finite absorptance, excitonic TMD absorption, and thickness constraints}
Although the SQ limit assumes unity absorptance above $E_g$, real TMD absorbers are ultrathin (less than 1 nm for a monolayer); consequently, their absorption cannot be unity. Also, it is dominated by strong excitonic resonances and large absorption coefficients in the visible; nevertheless, tens-of-nanometers thickness is typically required to approach unity absorptance over broad spectral windows without strong light trapping.\cite{ReichPop2023,JariwalaAtwater2017}
We therefore supplement the step-absorber model with a thickness-dependent absorptance model:
\begin{equation}
A(E;t)=1-\exp[-\alpha(E)L_\mathrm{eff}(E)], \quad L_\mathrm{eff}(E)=F(E)\,t
\label{eq:beerlambert}
\end{equation}
where $\alpha(E)$ and $F(E)$ are the absorption coefficient and an optional path-length enhancement factor, respectively.
To capture TMD-like spectral shape, we adopt a minimal excitonic absorption coefficient model consisting of an Urbach tail, a continuum onset, and A/B exciton peaks (Section 6, Supporting Information). 
Because reciprocity requires the same $A(E;t)$ in Eq.~\eqref{eq:J0rad_general}, thickness directly affects both $J_\mathrm{sc}$ and $V_\mathrm{oc}$.

The excitonic peaks in this model describe optical absorption, not an assumption that every exciton is collected as useful current. Strong exciton binding in TMDs, finite exciton diffusion length, and field- or interface-assisted dissociation can be included by replacing $A(E;t)$ with $A^{\mathrm{col}}(E;t)$ in Eq.~\eqref{eq:Acol_main}. In this representation, exciton recombination and transfer-induced traps reduce $\eta_{\mathrm{diss}}\eta_{\mathrm{tr}}$ and/or $\mathrm{ERE}$; consequently, type-II vdW junctions, built-in fields, low-barrier contacts, and encapsulated clean interfaces are required for an experimental device to approach the thermodynamic limit.

In the ultrathin regime, geometric-optics light-trapping factors (e.g., $4n^2$, where $n$ is the refractive index) may be overly optimistic.
The path-length enhancement factor is simply $F(E)=L_{\mathrm{eff}}(E)/t$, i.e., the ratio between the average optical distance traveled by photons inside the absorber and the physical thickness $t$:
$F(E)=1$ describes a single-pass planar film, while $F(E)>1$ reflects multi-pass propagation and light trapping (back reflectors, resonant cavities, and scattering/angle randomization).
In thick absorbers with randomized ray directions, the broadband average is bounded by the Yablonovitch limit $F\le 4n^2$.\cite{YuRamanFan2010}

When the thickness becomes comparable to the wavelength, however, light is better viewed as coupling through a finite set of radiative and guided modes, and the number of usable optical channels grows with thickness.
Yu, Raman, and Fan derived wave-optics bounds on the maximum achievable enhancement in nanophotonic textures as a function of thickness and in-plane periodicity,\cite{YuRamanFan2010}
and Miller emphasized the more general thickness--functionality connection for broadband optics.\cite{MillerThickness2023}
To capture this thickness-limited behavior without committing to a specific nanophotonic design, we introduce a normalized broadband enhancement function $g(t)\in[0,1]$ and define a thickness-dependent path-length enhancement factor: 
\begin{equation}
F_{\mathrm{proxy}}(t)=1+\left(4n^2-1\right)g(t)
\label{eq:Fproxy}
\end{equation}
so that $F_{\mathrm{proxy}}=1$ in the no-light-trapping limit and $F_{\mathrm{proxy}}\to 4n^2$ only when sufficient thickness (and optical degrees of freedom) is available.
In this work, we use $n=4.5$ as a representative TMD $n$ value (in general, $n=4.5$ of representative TMDs are ranged from 4.17 to 4.70, as summarized in Table S2, Supporting Information) and take a simple saturating proxy $g(t)=1-\exp(-t/t_0)$ with $t_0=200~\mathrm{nm}$ (see Section 7, Supporting Information).
When using this proxy in Eq.~(\ref{eq:beerlambert}), we approximate $F(E)\!\approx\!F_{\mathrm{proxy}}(t)$, and also discuss how wavelength-resolved bounds can be incorporated, when a specific device architecture is known.

% =========================================================
\section{Results}
\begin{figure}[t]
  \centering
  \includegraphics[width=0.9\linewidth]{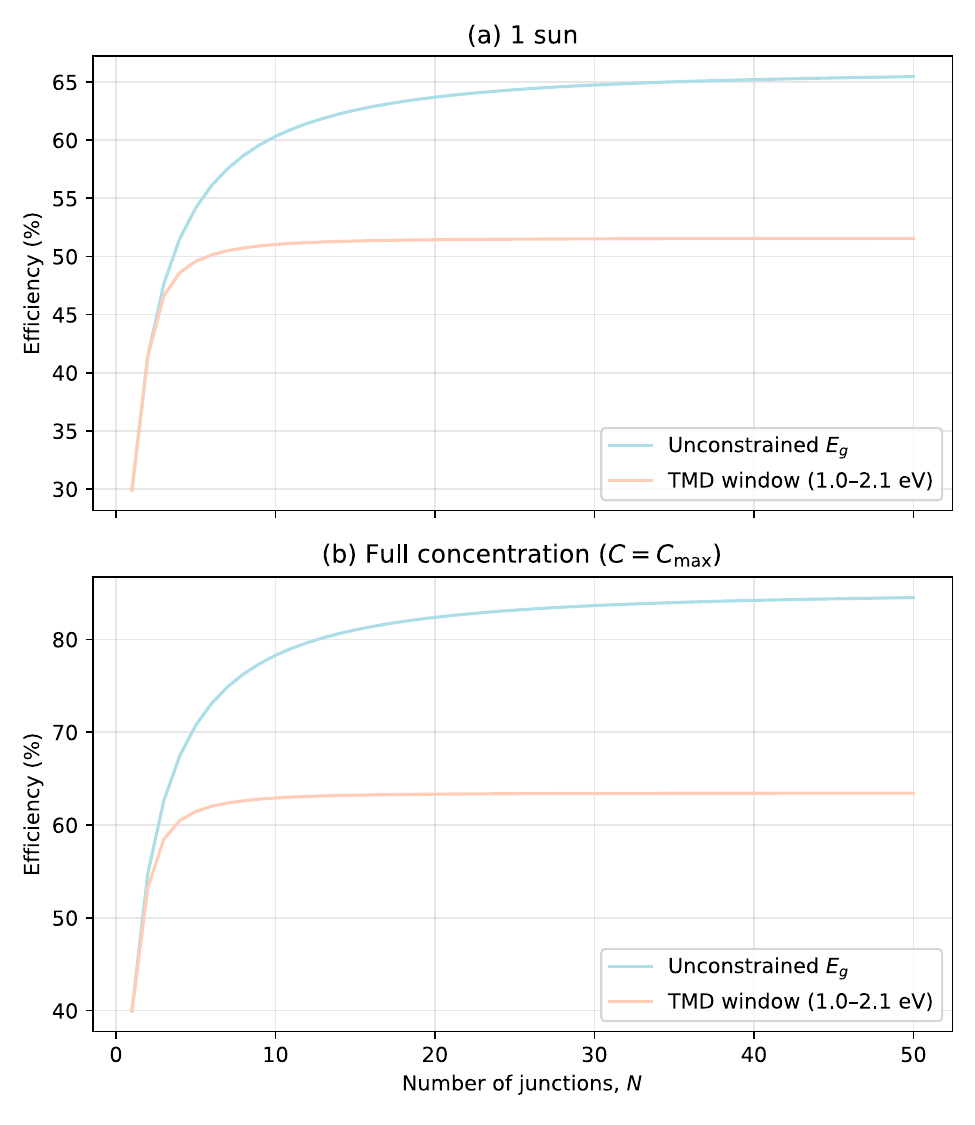}
  \caption{\textbf{Detailed-balance efficiency versus junction number.} 
  (a) 1 sun and (b) full concentration for unconstrained optimal bandgaps and a conservative TMD bandgap window (1.0--2.1~eV), computed for split-spectrum (multi-terminal) multijunctions with one-sided emission and $\mathrm{ERE}=1$. 
  Numerical tables and bandgap ladders are provided in the SI and the accompanying code package.}
  \label{fig:fig2}
\end{figure}

\noindent\textbf{Efficiency versus junction $N$ and the role of the bandgap window.}
Figure~\ref{fig:fig2} summarizes the detailed-balance efficiency versus junction number $N$ for (i) unconstrained bandgaps and (ii) the conservative TMD bandgap window (1.0--2.1~eV), under both 1 sun and full concentration: more detailed information for these calculations is summarized in Tables~S3--S6, Supporting Information such as candidate vdW/TMD absorber library, prioritized mapping of realistic vdW/TMD candidates, generic parameter values for the excitonic absorption coefficient model, and illustrative material-specific excitonic parameters for sensitivity analysis. Under full concentration, the unconstrained efficiency increases from $\sim 40\%$ (single junction) to $\sim 70.7\%$ for $N=5$ and $\sim 84.5\%$ by $N=50$ (approaching the reciprocal multicolor limit). 
Imposing the TMD window yields a pronounced large-$N$ plateau: $\sim 61.5\%$ for $N=5$ and $\sim 63.4\%$ by $N=50$.
This plateau reflects an intrinsic materials constraint: once the accessible bandgap set no longer spans the entire solar spectrum, additional junctions provide diminishing returns.

In principle, the reciprocal multicolor limit is reached, when the spectrum assigned to each junction becomes narrower as $N$ increases. This multijunction approach suppresses thermalization, while maintaining a high radiative voltage under concentration. In the conservative TMD window, photons below 1.0 eV (near-IR) and above 2.1 eV (blue/UV) are irreversibly lost to transmission, so increasing $N$ primarily repartitions the same limited spectral band. The rapid saturation by $N$ on the order of five, therefore, indicates that, within TMD-only bandgaps, further progress hinges more on luminescence quality (ERE), emission-channel control, and ultrathin optics than on pushing junction count. For example, under full concentration, the TMD-window ladder rises from $\eta\approx 53.3\%$ at $N=2$ to $\eta\approx 61.5\%$ at $N=5$, whereas increasing the junction number to $N=10$ gives only $\eta\approx 62.9\%$ and the large-$N$ value at $N=50$ is $\eta\approx 63.4\%$ (bandgap ladders used herein are summarized in Tables~S7--S10 of Section 8, Supporting Information).

This plateau statement refers to the ideal split-spectrum, one-sided-emission calculation in Figure~\ref{fig:fig2}. It is not inconsistent with the later luminescent-coupling analysis. In a reciprocal stack without intermediate mirrors, increasing $N$ also decreases some adjacent bandgap spacings, which increases downward luminescent coupling and coupling-induced thermalization. Therefore, once the spectral-partitioning gain has saturated, the additional coupling/optical complexity provides a practical reason to stop near $N\sim5$ rather than to keep increasing $N$ (also see more detailed note in Section 5.5, Supporting Information).

\begin{figure}[t]
  \centering
  \includegraphics[width=0.92\linewidth]{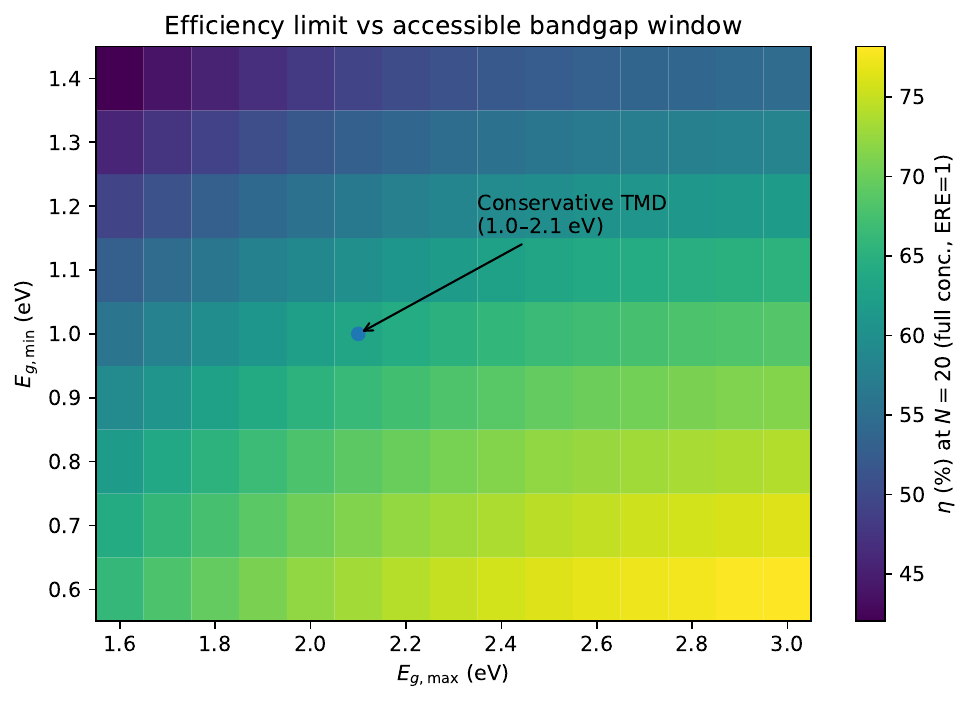}
  \caption{\textbf{Efficiency limit versus accessible bandgap window.}
  Heatmap of the maximum achievable efficiency for a representative high-$N$ case ($N=20$, multi-terminal, $\mathrm{ERE}=1$, full concentration), as a function of the allowed bandgap window $(E_{g,\min},E_{g,\max})$. 
  The conservative TMD window (1.0--2.1~eV) is highlighted.}
  \label{fig:fig3}
\end{figure}

\noindent\textbf{Efficiency limit versus accessible bandgap window.}
To quantify how strongly the accessible bandgap window constrains large-$N$ performance, Figure~\ref{fig:fig3} maps the efficiency limit (herein, shown for $N=20$ at full concentration) versus $(E_{g,\min},E_{g,\max})$.
The conservative TMD bandgap point (1.0--2.1~eV) are located far from an unconstrained optimum, explaining the plateau in Figure~\ref{fig:fig2}.
This map provides a compact way to evaluate how expanding the accessible bandgap range (e.g., by including other vdW semiconductors beyond conventional TMDs, or by alloying/strain) could raise the efficiency plateau.

The map also highlights an asymmetry between extending the lowest- and highest-bandgap bounds. Lowering $E_{g,\min}$ generally leads to a larger efficiency gain, because the solar photon flux is relatively high in the near-infrared (NIR), whereas raising $E_{g,\max}$ mainly improves voltage and reduces high-energy thermalization. This suggests that identifying a stable narrow-gap bottom absorber in the 0.8--1.0 eV range can provide disproportionate leverage, even if the top bandgap remains near 2 eV. In principle, the lower cutoff $E_{g,\min}$ directly sets the fraction of NIR photons that are transmitted without conversion; within a fixed-width window, extending $E_{g,\min}$ downward, therefore, provides a larger efficiency gain potential than modestly increasing $E_{g,\max}$.

\begin{figure}[t]
  \centering
  \includegraphics[width=0.85\linewidth]{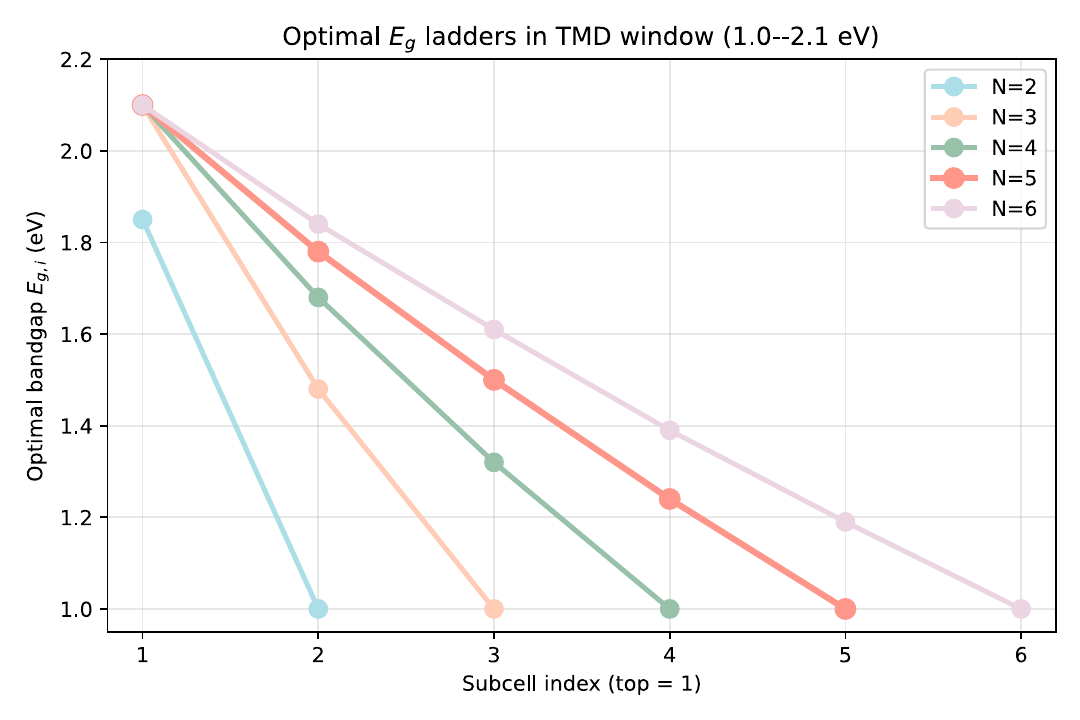}
  \caption{\textbf{Optimal bandgap ladders in the conservative TMD window.}
  $E_{g,i}$ vs subcell index for selected $N$ values (top cell is index 1). 
  The $N=5$ ladder in Eq.~\eqref{eq:ladder5} is highlighted as a representative design.}
  \label{fig:fig4}
\end{figure}

\noindent\textbf{Optimal bandgap ladders and a representative $N=5$ TMD design.}
Figure~\ref{fig:fig4} shows optimal bandgap ladders for selected $N$ within the conservative TMD window. 
As $N$ increases, the ladder becomes denser, with the top junction pushed toward the maximum available gap ($\sim 2.1$~eV) and the bottom junction pushed toward $\sim 1$~eV.

As mentioned above, we adopt $N=5$ as a representative experimentally achievable design. 
For $N=5$ in the TMD window, the DP optimizer yields the ladder
\begin{equation}
E_g^{(N=5)}\approx(2.10,\,1.78,\,1.50,\,1.24,\, and \ 1.00)\ \mathrm{eV}\quad \text{(from top to bottom)}
\label{eq:ladder5}
\end{equation}
This ladder is used throughout subsequent analyses of luminescence, coupling, and thickness. Layer-resolved operating metrics for this representative $N=5$ stack (including ERE and finite-absorptance effects) are provided in Figures~S1 and S2, Supporting Information.

Herein, the rung spacing is not uniform, because the optimizer concentrates rungs, where the solar spectrum has relatively high photon flux, while keeping a comparatively wide spacing between the lowest two bandgaps. This nonuniformity matters beyond spectral splitting: in reciprocal stacks without intermediate mirrors, smaller bandgap spacing increases downward luminescent coupling and the associated thermalization burden in the lower cell. Thus, optical design choices that control two-sided emission can be as important as the bandgap selection itself.

\begin{figure}[t]
  \centering
  \includegraphics[width=0.92\linewidth]{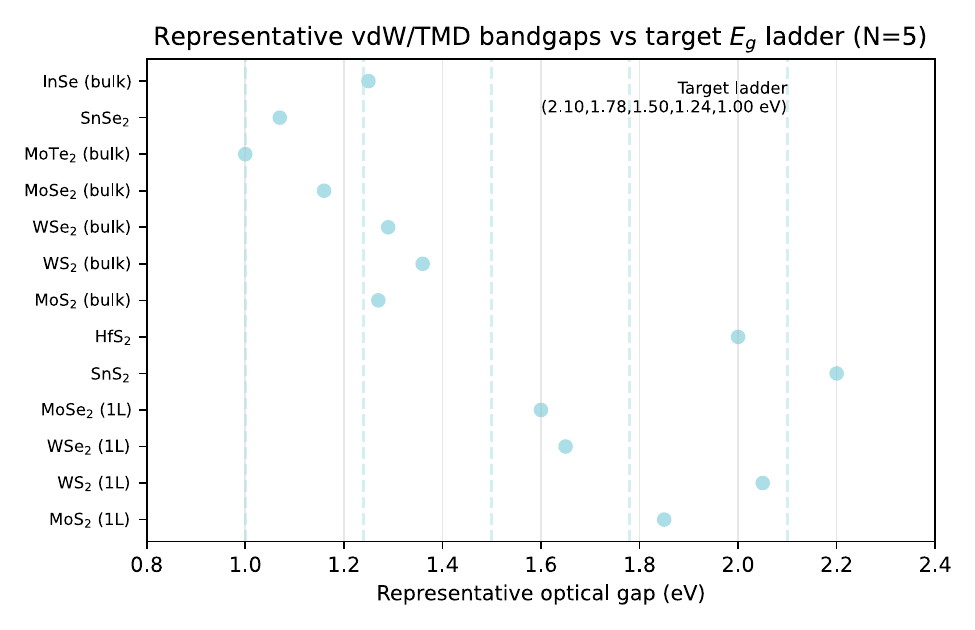}
  \caption{\textbf{Illustrative mapping between target bandgaps and vdW/TMD candidates.}
  Symbols denote representative optical gaps drawn from the vdW/TMD literature (see Supplement Information, Table~S3 for notes and sources), and dashed vertical lines denote the representative $N=5$ target ladder (Eq.~\eqref{eq:ladder5}).}
  \label{fig:fig5}
\end{figure}

\noindent\textbf{Material mapping: candidate vdW and TMD absorbers for the $N=5$ ladder.}
To connect target bandgaps to realizable absorbers, Figure~\ref{fig:fig5} overlays representative optical bandgaps, which were reported for a selection of vdW semiconductors and TMD thickness variants.\cite{JariwalaAtwater2017,ReichPop2023,TMDReview2024,DuMser2024}
This mapping is necessarily approximate, because optical bandgaps depend on thickness, strain, dielectric environment, and exciton binding energy.
Nevertheless, this comparison still provides a practical starting point for selecting TMD subcell candidates for stacks.

Two practical subtleties are noteworthy, when translating target optical bandgaps into device layers. First, monolayer TMD optical bandgaps are excitonic and depend on dielectric environment and strain, so the same nominal absorber can shift in both absorptance and radiative efficiency, depending on encapsulation and adjacent layers. Second, many multilayer TMDs tend to retain indirect-bandgap as thickness increases, which can reduce luminescence and therefore reduce voltage, unless light management and passivation mitigate nonradiative recombination. Table~\ref{tab:tab1} provides a compact rung-to-material mapping for the representative $N=5$ ladder, emphasizing TMD-first choices within the conservative 1.0--2.1~eV window. The broader candidate database, used in the thickness-dependent calculations, such as thickness, strain, alloy tunability ranges, and optical-constants, is provided in the Supplement Information (Tables~S3--S6).

Among the five rungs, the bottom cell target near 1.0 eV is the most challenging within conventional TMDs, as MoTe$_2$ and related narrow-gap vdW absorbers require extremely precise phase control, inert encapsulation, low-barrier contacts, and high ERE;\cite{JariwalaAtwater2017,ReichPop2023,TMDReview2024} but also the most consequential for the efficiency limit, because it controls NIR photon harvesting. For example, as shown in a bottom-gap sensitivity analysis in the Supporting Information (Section 8): raising the minimum accessible gap from 1.0 to 1.1~eV lowers the full-concentration $N=5$ efficiency from 61.45\% to 58.63\% and the $N=50$ plateau from 63.43\% to 60.14\%; raising it to 1.2~eV gives 55.54\% ($N=5$) and 56.66\% ($N=50$).

{\footnotesize
\setlength{\tabcolsep}{3pt}
\renewcommand{\arraystretch}{1.06}
\begin{longtable}{@{}p{0.09\linewidth} p{0.08\linewidth} p{0.17\linewidth} p{0.28\linewidth} p{0.31\linewidth}@{}}
\caption{\textbf{Representative rung-to-material mapping for the $N=5$ target ladder within the conservative 1.0--2.1~eV window.}
Primary choices emphasize widely studied group-VI TMDs compatible with transfer printing; alternates illustrate thickness/strain/alloy routes and broader vdW options.
See the Supplement Information, Tables~S3--S6 for a more comprehensive candidate library and parameter notes.}
\label{tab:tab1}\\
\toprule
Target $E_g$ (eV) & Rung & Primary candidate & Alternates & Practical notes (tunability / caveats)\\
\midrule
\endfirsthead
\toprule
Target $E_g$ (eV) & Rung & Primary candidate & Alternates & Practical notes (tunability / caveats)\\
\midrule
\endhead
\midrule
\multicolumn{5}{r}{Continued on next page}\\
\endfoot
\bottomrule
\endlastfoot
2.10 & Top & WS$_2$ (1L) & MoS$_2$ (1L, strain/dielectric); HfS$_2$ (few-L) & Wide-gap top cell; strong visible excitons; encapsulation and work-function control mitigate contact loss.\\
1.78 & 2 & MoS$_2$ (1L) & WS$_2$ (2L--few-L); MoS$_{2(1-x)}$Se$_{2x}$ (alloy) & Gap tunable by thickness and dielectric screening; prioritize high ERE and low series resistance.\\
1.50 & 3 & MoSe$_2$ (1L) & ReS$_2$; WSe$_2$ (1L, strain) & Mid-gap cell with strong excitonic absorption; minimize parasitic absorption in interlayers.\\
1.24 & 4 & WSe$_2$ (few-L/bulk) & MoS$_2$ (bulk); InSe (few-L) & Near-IR edge; multilayer TMDs may become indirect; thickness increase can compensate finite absorptance.\\
1.00 & Bottom & MoTe$_2$ (few-L/bulk) & MoSe$_2$ (bulk, $\sim$1.1); BP (encapsulated, thickness-tuned) & Narrow-gap bottom cell; stability/oxidation and contact barriers require encapsulation and barrier-free contacts. If this rung cannot reach high ERE, the stack should be re-optimized with a higher $E_{g,\min}$.\\
\end{longtable}
}

\begin{figure}[t]
  \centering
  \includegraphics[width=0.92\linewidth]{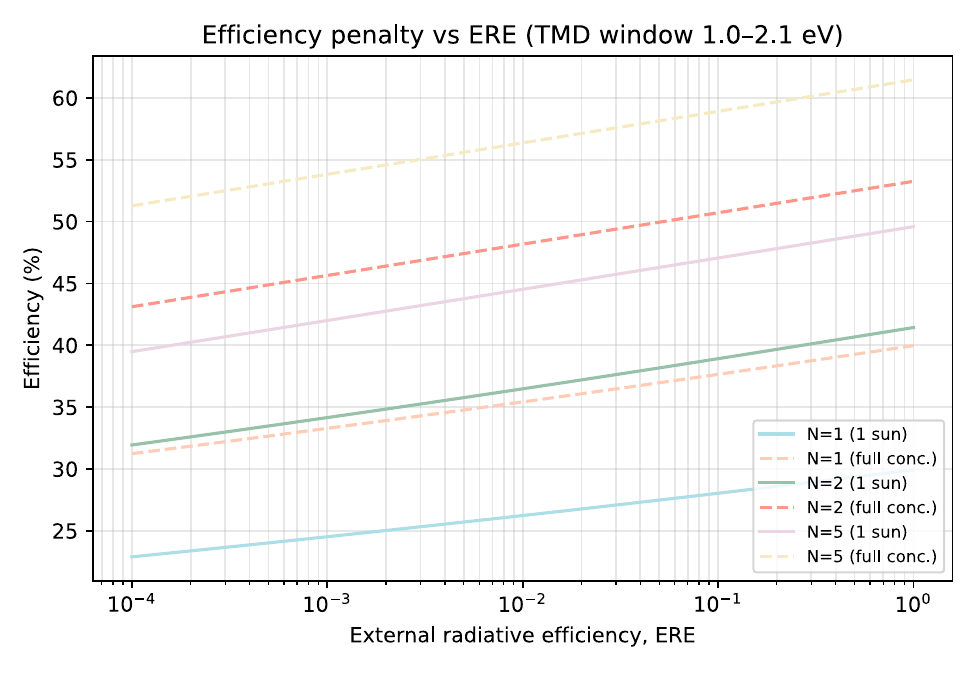}
  \caption{\textbf{Efficiency versus external radiative efficiency (ERE) in the TMD window.}
  Efficiency computed for $N=1,2,5$ under 1 sun (solid) and full concentration (dashed), illustrating the SQ-triangle voltage penalty and its correspondence to the thermodynamic $Q^\mathrm{LED}_e$ term.\cite{Rau2014,Rau2017}}
  \label{fig:fig6}
\end{figure}

\noindent\textbf{System-level penalty from finite ERE.}
Figure~\ref{fig:fig6} shows how efficiency degrades as $\mathrm{ERE}$ decreases below unity, for selected $N$ values in the conservative TMD window. 
Because the voltage penalty is logarithmic, efficiencies remain relatively robust for $\mathrm{ERE}\gtrsim 10^{-2}$, but decline rapidly for lower $\mathrm{ERE}$. 
Importantly, in multijunction systems, the penalty compounds across junctions: one low-ERE subcell can dominate the system loss.\cite{MillerYablonovitch2012,Rau2017,KirchartzRau2018}
This motivates aggressive light-management and contact designs (e.g., low parasitic absorption, high-reflectivity mirrors, and low-loss transparent electrodes) that preserve external photon extraction in every TMD subcell.

A simple quantitative rule-of-thumb illustrates the compounding nature of the loss: at room temperature, an ERE of $10^{-2}$ corresponds to an open-circuit voltage penalty on the order of one tenth of a volt per junction, so a five-junction stack can lose several tenths of a volt in total, even if current matching is ideal. This is why maintaining high ERE in every TMD subcell (not only the top) is a prerequisite for approaching multicolor efficiencies. For the representative $N=5$ ladder, reducing ERE from 1 to $10^{-2}$ lowers the full-concentration stack efficiency by about six absolute percentage points in the thick-limit regime (Figure~\ref{fig:fig6}).

\begin{figure}[t]
  \centering
  \includegraphics[width=0.92\linewidth]{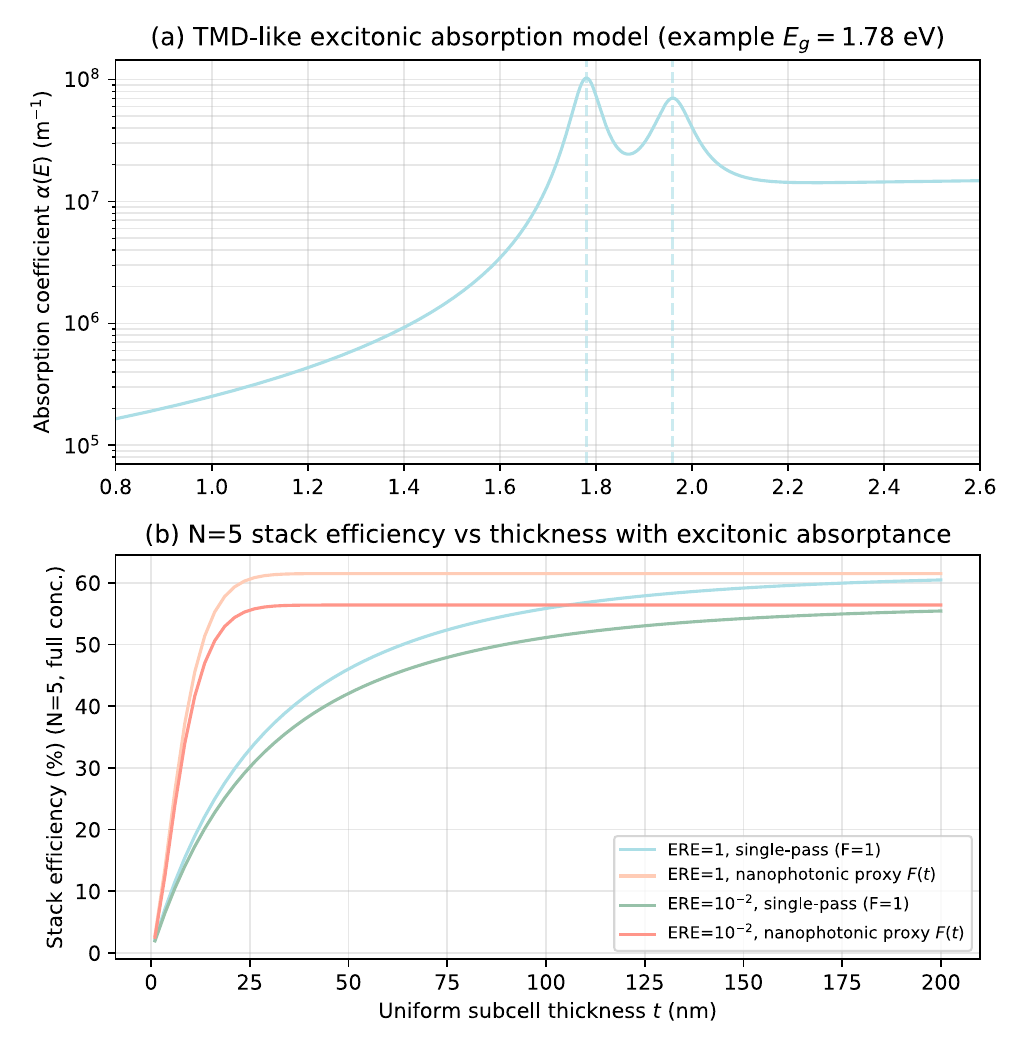}
  \caption{\textbf{Thickness dependence for an $N=5$ TMD multijunction stack with excitonic absorptance.}
  (a) Example absorption coefficient model $\alpha(E)$ for a representative bandgap ($E_g=1.78$~eV) including A/B exciton peaks and a continuum onset. 
  (b) Total $N=5$ stack efficiency under full concentration versus uniform subcell thickness $t$, comparing single-pass absorption ($F=1$) to the thickness-dependent broadband light-trapping proxy $F_{\mathrm{proxy}}(t)$ of Eq.~(\ref{eq:Fproxy}). 
  Reciprocity is enforced by using the same $A(E;t)$ in both $J_\mathrm{sc}$ and the radiative dark current.}
  \label{fig:fig7}
\end{figure}

\noindent\textbf{Thickness dependence with excitonic TMD absorptance.}
Figure~\ref{fig:fig7} moves beyond step absorbers to incorporate finite absorptance in ultrathin TMD layers. Figure 7(a) shows the excitonic absorption coefficient model used herein, including A/B exciton peaks and a continuum onset; the model is intentionally minimal but captures the qualitative spectral features of many group-VI TMDs.\cite{JariwalaAtwater2017,ReichPop2023} The underlying optical-constant conventions ($n$, $k$, and $\alpha$) and representative material parameters used in this work are summarized in the Supporting Information (Tables~S2--S6).
Figure 7(b) shows the resulting $N=5$ stack efficiency versus a uniform TMD subcell thickness $t$ for $\mathrm{ERE}=1$ and $\mathrm{ERE}=10^{-2}$. 
Even though monolayer and few-layer TMDs exhibit pronounced excitonic absorption near the band edge, the usable solar spectrum spans a broad range of above-bandgap energies and the continuum absorption remains finite; consequently, a very thin film does not absorb all incident photons in a single pass. In the Beer--Lambert regime ($\alpha t\!\ll\!1$), the absorptance scales approximately as $A\approx \alpha t$, so the short-circuit current and efficiency rise only gradually with thickness.

\noindent\textbf{Nanophotonic bounds and the role of thickness.}
The need for a thickness-dependent broadband proxy follows from the distinction between geometric optics and wavelength-scale optics, as mentioned above.
In thick absorbers with randomized ray directions, the average internal path length can be increased up to the Yablonovitch bound $F\le 4n^2$. When $t\lesssim \lambda$, however, the optical response is mediated by a finite set of radiative and guided modes; the number of accessible channels (and therefore the achievable broadband response) grows with thickness.
As a result, nanophotonic textures may yield strong enhancement at selected wavelengths or angles, but sustaining large enhancement over the broad solar spectrum is fundamentally thickness-limited.

\begin{figure}[p]
  \centering
  % Enlarged rendering for readability; caption style follows the global ACS template.
  \includegraphics[height=0.94\textheight,keepaspectratio]{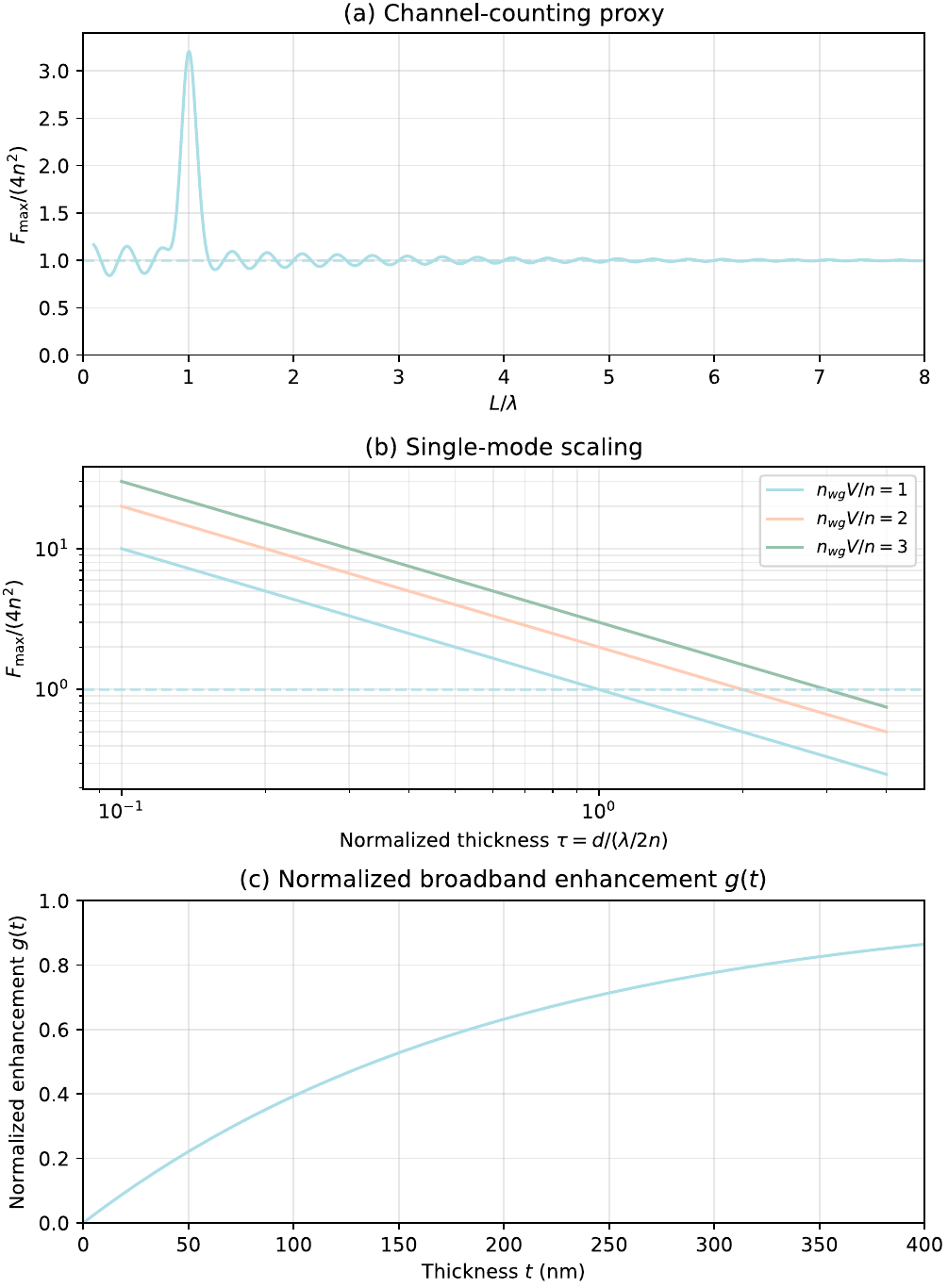}
  \caption{\textbf{Nanophotonic bounds and a broadband path-length enhancement proxy.}
  (a) Channel-counting proxy illustrating deviations from the geometric-optics $4n^2$ limit in the wavelength-scale regime.
  (b) Single-mode scaling proxy for $F_{\max}/(4n^2)$ versus normalized thickness $\tau=d/(\lambda/2n)$.
  (c) Normalized broadband enhancement $g(t)=(F_{\mathrm{proxy}}(t)-1)/(4n^2-1)$ used to construct the thickness-dependent path-length enhancement factor $F_{\mathrm{proxy}}(t)=1+(4n^2-1)g(t)$ in Figure~\ref{fig:fig7}.\cite{YuRamanFan2010,MillerThickness2023,KimLeeOE2019}}
  \label{fig:fig8}
\end{figure}

Figure~\ref{fig:fig8} summarizes representative results from nanophotonic light-trapping theory and motivates the normalized enhancement function $g(t)$ used throughout this work.
Figure 8(a) illustrates the Yu--Raman--Fan wave-optics enhancement oscillations around the $4n^2$ limit for periodic textures, highlighting the discrete-channel nature of wavelength-scale light trapping.\cite{YuRamanFan2010} Figure 8(b) illustrates single-mode scaling trends emphasized in nanophotonic bounds.\cite{YuRamanFan2010} Figure 8(c) plots the simple saturating function $g(t)$ adopted here, which can be interpreted as the fraction of the $4n^2$ enhancement that is experimentally achievable as a broadband average at thickness $t$.
We then construct the effective path-length enhancement factor $F_{\mathrm{proxy}}(t)=1+(4n^2-1)g(t)$ [Eq.~(\ref{eq:Fproxy})], which is used in the absorptance model of Eq.~(\ref{eq:beerlambert}) and in the thickness sweeps of Figure~\ref{fig:fig7}.\cite{MillerThickness2023,KimLeeOE2019} For multijunction stacks, this thickness constraint is wavelength selective: the longest-wavelength TMD subcells are the hardest to enhance in the ultrathin regime, so the bottom junction often dictates the minimum practical thickness or the required strength of light trapping. As such, optical design should be co-optimized with the bandgap ladder, rather than treated as a post-processing step.\cite{MillerThickness2023}

\begin{figure}[p]
  \centering
  % Enlarged rendering for readability; caption style follows the global ACS template.
  \includegraphics[height=0.88\textheight,keepaspectratio]{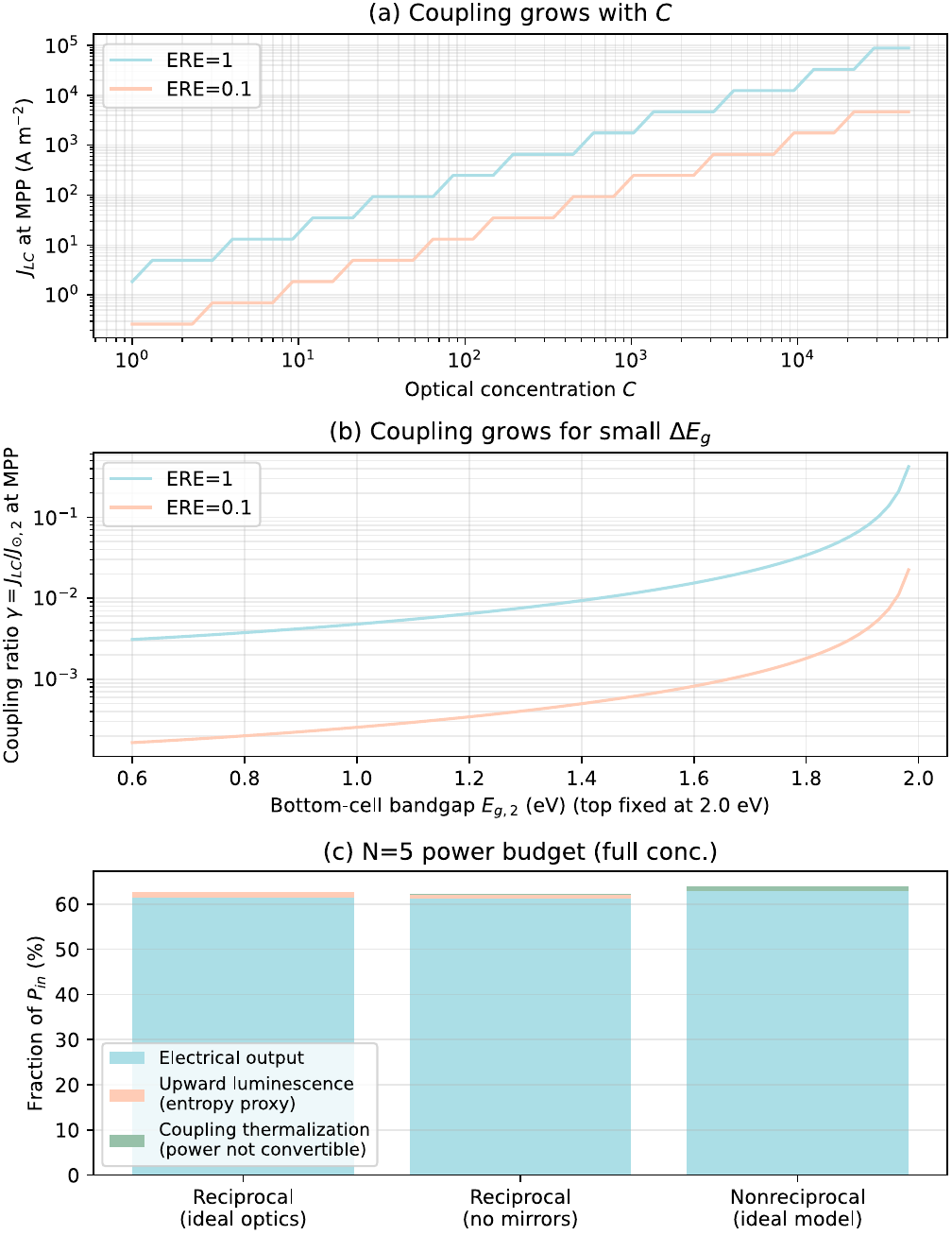}
  \caption{\textbf{Luminescent coupling conditions and entropy/thermalization power budget.}
  (a) Downward luminescent coupling current at MPP increases strongly with optical concentration and is larger for higher ERE.
  (b) Coupling ratio $\gamma=J_\mathrm{LC}/J_{\odot,2}$ increases rapidly as the bandgap spacing decreases.
  (c) Power-budget decomposition for the representative $N=5$ ladder under full concentration, comparing reciprocal ideal optics (no coupling), reciprocal stacks without intermediate mirrors (two-sided emission + coupling), and an idealized nonreciprocal model. 
  Upward-emitted luminescence is shown as an entropy-loss proxy; coupling thermalization quantifies luminescent power not convertible to electrical work in the lower junctions.}
  \label{fig:fig9}
\end{figure}

\noindent\textbf{Luminescent coupling and a power-budget view of entropy vs thermalization.}
Figure~\ref{fig:fig9} quantifies when luminescent coupling becomes appreciable in reciprocal stacks and provides a power-budget decomposition for the representative $N=5$ ladder. Figure 9(a) shows that the downward coupling current $J_\mathrm{LC}$ at the system MPP increases strongly with concentration and is amplified at high ERE due to higher operating voltage. Figure 9(b) shows that the coupling ratio grows rapidly, when adjacent bandgaps become closely spaced (small $\Delta E_g$) (i.e., a condition that is naturally approached in dense high-$N$ ladders). Figure 9(c) provides a compact power budget for the representative $N=5$ ladder: the upward luminescence power is plotted as an entropy-loss proxy, while coupling-induced thermalization is plotted separately. This decomposition clarifies the key design point: suppressing upward emission can reduce entropy loss (and increase voltage), but downward emission into lower junctions can still lead to thermalization loss, if energy is not efficiently converted.

Importantly, coupling is not universally detrimental. When the lower cell converts the coupled photons efficiently and the bandgap spacing is sufficiently large, downward luminescence can partially compensate current deficits and relax the need for perfect spectral splitting. However, in dense ladders with small bandgap spacing, coupled photons arrive with excess energy relative to the lower bandgap. As a result, these coupled photons are more likely to be dissipated as heat, effectively reintroducing a thermalization channel that grows with concentration and ERE. For the representative $N=5$ ladder at full concentration, the reciprocal ideal-optics stack loses $\sim 1.24\%$ of the incident solar power as upward-emitted luminescence, whereas in the idealized nonreciprocal model, this upward channel is eliminated and the electrical output increases by $\sim 1.5\%$ absolute, with part of the redirected luminescence, which appear as coupling-induced thermalization (Figure~\ref{fig:fig9}c).
For higher-$N$ ladders, the spectral gain from adding another junction is already small within the 1.0--2.1~eV window, while the risk of small-$\Delta E_g$ coupling and additional interlayer loss increases. This interpretation is therefore that the plateau and coupling analyses point in the same direction: experimentally useful stacks should first optimize a small number of high-quality, optically isolated junctions before pursuing larger $N$.

\noindent\textbf{Reciprocal vs nonreciprocal efficiency gain.}
Nonreciprocal multijunction photovoltaics has been proposed as a route to approach the Landsberg limit by suppressing entropy losses, which are associated with reciprocal emission pathways.\cite{FanNonreciprocalMJ2022}
However, for single-junction solar cells, nonreciprocity does not increase the SQ efficiency limit under detailed balance.\cite{FanAPL2022} To place these insights in the present context, we benchmark the potential efficiency gain by applying the idealized nonreciprocal multijunction model to the same bandgap ladders, that are optimized under reciprocity (i.e., conservative comparison; also see further detailed information in Section 9, Supporting Information).

\begin{figure}[t]
  \centering
  \includegraphics[width=0.96\linewidth]{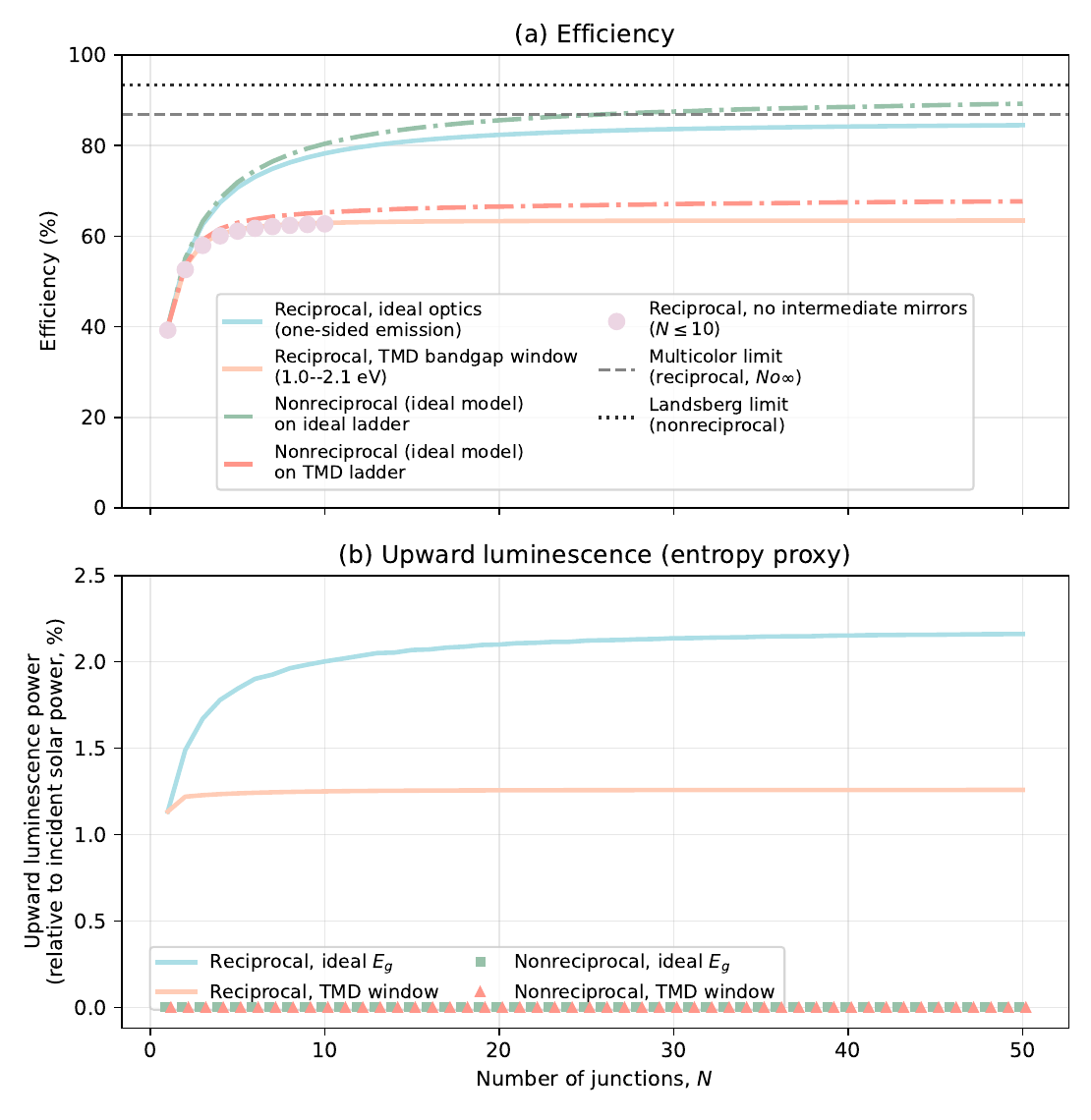}
  \caption{\textbf{Reciprocal vs nonreciprocal efficiency gain under full concentration.}
  (a) Efficiency versus junction number for reciprocal split-spectrum multijunctions (one-sided emission, no coupling), and for the idealized nonreciprocal multijunction model of Fan et al.\cite{FanNonreciprocalMJ2022} applied to the same reciprocity-optimized bandgap ladders.
  Filled markers show a reciprocal-stack estimate without intermediate mirrors (two-sided emission + downward luminescent coupling) for $N\le 10$.
  (b) Upward-emitted luminescence power fraction $P^\uparrow_\mathrm{lum}/P_\mathrm{in}$ as an entropy-loss proxy.
  In the idealized nonreciprocal model used here, the upward-emitted luminescence is identically zero, so the nonreciprocal series in (b) are shown with a slight horizontal offset for visibility.
  Horizontal lines indicate literature multicolor and Landsberg limits for comparison (values depend weakly on $T_s$).\cite{FanNonreciprocalMJ2022}}
  \label{fig:fig10}
\end{figure}

Figure~\ref{fig:fig10} compares full-concentration efficiencies versus $N$ for reciprocal and nonreciprocal cases, for both unconstrained bandgaps and the conservative TMD window.
In the TMD window, the reciprocal efficiency saturates near $\sim 63.4\%$ at large $N$, whereas the nonreciprocal model reaches to $\sim 67.7\%$ at $N=50$ (Figure~\ref{fig:fig10}a): extended bandgap ladder-tables are available in Tables~S7--S11, Supporting Information (Section 8); .
Figure~\ref{fig:fig10}b shows the critical physical signature: the nonreciprocal model suppresses upward-emitted luminescence power (entropy-loss proxy), whereas the reciprocal case must emit upward to maintain detailed balance.
The filled markers in Figure~\ref{fig:fig10}a further show that reciprocal stacks without intermediate mirrors incur additional penalties from two-sided emission and coupling, highlighting the practical importance of intermediate mirrors or angular filters\cite{KimNanophotonics2025} in realistic reciprocal designs. In the conservative comparison, the nonreciprocal model is implemented to the reciprocity-optimized ladder without re-optimizing bandgaps, so the improvement should be interpreted as a lower bound on the possible efficiency gain. In practice, achieving the ideal nonreciprocal behavior is nontrivial, but the comparison clarifies what must be suppressed in reciprocal stacks: upward emission that carries entropy and downward emission that drives coupling thermalization. Intermediate mirrors, angular filters, and spectral-selective contacts\cite{KimNanophotonics2025} can address part of this gap without requiring fully nonreciprocal photonics.

% =========================================================
\section{Discussion: design rules for TMD multijunction cells}
\paragraph{(1) Bandgap-ladder engineering is necessary but not sufficient.}
The DP optimizer provides optimal ladders and reveals the large-$N$ plateau imposed by a constrained bandgap window. 
Within the conservative 1.0--2.1~eV TMD window, $N\approx 5$ already approaches the efficiency plateau, suggesting that the most immediate experimental objective is to realize high-quality $N\le 5$ stacks with excellent optics rather than extremely large $N$ (full ladder list is summarized in Section 10, Supporting Information).

\paragraph{(2) ERE is a system-level requirement.}
Voltage penalties from finite ERE accumulate across subcells (Figure~\ref{fig:fig6}). High-$N$ stacks, therefore, demand high radiative quality in every junction, highlighting the importance of passivation, low-loss contacts, and photon extraction.\cite{MillerYablonovitch2012,KirchartzRau2018}

\paragraph{(3) Control of emission channels and luminescent coupling are essential.}
Without intermediate mirrors, two-sided emission introduces an entropy penalty and enables downward luminescent coupling (Figure~\ref{fig:fig9}a--b). 
Coupling can be beneficial for current in lower cells. However, it also introduces thermalization losses and reduces the voltage. Practical multijunction architectures will likely require selective optical elements (intermediate mirrors, angular filters, or photonic cavities) designed to maximize useful coupling while minimizing entropy penalties.\cite{Rau2014,YuRamanFan2010,KimNanophotonics2025}

\paragraph{(4) Thickness and nanophotonics cannot be ignored in vdW stacks.}
Ultrathin absorbers cannot be treated as step absorbers; finite absorptance reduces both $J_\mathrm{sc}$ and $V_\mathrm{oc}$ through reciprocity (Figure~\ref{fig:fig7}). 
Nanophotonic bounds and thickness constraints restrict the achievable broadband path-length enhancement (Figure~\ref{fig:fig8}). 
These effects motivate moderate absorber thicknesses (tens to hundreds of nm) and co-designed photonic structures rather than extreme monolayer limits for high-efficiency multijunction photovoltaics.

\paragraph{(5) Nonreciprocity is an optional asymptotic strategy with multijunction-specific benefits.}
Nonreciprocity does not improve the SQ limit for single-junction devices.\cite{FanAPL2022} In multijunction architectures, however, rerouting emission pathways can reduce entropy losses and provide efficiency gain toward Landsberg-like limits.\cite{FanNonreciprocalMJ2022}
Our conservative plug-in comparison (Figure~\ref{fig:fig10}) quantifies this efficiency gain for TMD-optimized ladders, while preserving our work's main scope: reciprocal multijunction design and realistic constraints.

\paragraph{(6) The multi-terminal assumption defines an upper-bound diagnostic, not a lossless fabrication prescription.}
Independent extraction is used to remove current-matching constraints and identify the best spectral partition for transfer-printed vdW stacks. A practical realization would require transparent or lateral selective contacts, interlayer insulation/passivation, suppressed optical crosstalk, and independent MPPT or submodule power conversion. These elements introduce parasitic absorption, series resistance, and electronic conversion losses; in the revised framework they are included as $A^{\mathrm{col}}_i(E)$, $\mathrm{ERE}_i$, $R_{s,i}$, and an external MPPT/electrical derating rather than hidden inside the ideal efficiency limit. Two-terminal or monolithically series-connected versions can be analyzed with the same optical model by adding current-matching constraints, but that is a different device-design problem from the thermodynamic ceiling emphasized here.

% =========================================================
\section{Conclusions}
We developed an unlimited-junction detailed-balance framework tailored to transfer-printed TMD multijunction photovoltaics and used it to compute optimal bandgap ladders under unconstrained and conservative TMD bandgap windows. 
A realistic bandgap window produces a large-$N$ efficiency plateau, motivating $N\sim 5$ as a practical near-term target for TMD-based stacks.
We incorporated luminescence thermodynamics and reciprocity to quantify ERE penalties, emission-channel entropy, and luminescent coupling. Also, we introduced a power-budget view, in which upward-emitted luminescence serves as a directly computable entropy-loss proxy, while coupling thermalization is tracked, separately.
Finally, we benchmarked the remaining thermodynamic efficiency gain by applying an idealized nonreciprocal multijunction model to our reciprocally optimized ladders, showing negligible benefit for single junctions but substantial efficiency gain for multijunction stacks.
Our framework also makes explicit how interface defects, exciton collection, parasitic absorption, series resistance, bottom-cell instability, and MPPT losses should be applied as practical derating factors to the thermodynamic limit.
These results provide quantitative design rules for approaching multicolor efficiencies with transfer-printed TMD multijunction photovoltaics.

\section*{Data and code availability}
The numerical results and figures in this work are reproducible from the Python implementation of the detailed-balance, luminescent-coupling, and optimization models described in the main text and Supporting Information. The code is available at https://github.com/NEOlab-code/Multijunction-TMD-solar-cells.git. To keep the manuscript self-contained for submission, the Supporting Information includes the full derivations, parameter tables, and additional figures.

\section*{Supporting Information}
The Supporting Information consists of (i) Section~1 (constants, geometry, and spectral model): solar/blackbody spectral definitions and constants;
(ii) Section~2 (single-junction detailed balance): full detailed-balance derivations, including the Lambert-$W$ maximum-power condition;
(iii) Section~3 (unlimited-junction split-spectrum multijunction model): the dynamic-programming (DP) optimizer used to compute optimal bandgap ladders for arbitrary $N$ and its discretization/convergence checks;
(iv) Section~4 (reciprocity, emission channels, and the role of mirrors): reciprocity-based voltage-loss and entropy-loss decompositions, including the role of ERE and optical outcoupling;
(v) Section~5 (luminescent coupling chain model): the luminescent-coupling chain model, together with the definition and computation of the upward emitted luminescence power and a coupling thermalization/heat proxy;
(vi) Section~6 (TMD optical response and excitonic absorptance model): a TMD material database (representative bandgaps, excitonic absorption spectra, and representative $n(\lambda)$/$k(\lambda)$ dispersion conventions), exciton-dissociation/transport derating, mapping of candidate stacks to the representative $N=5$ target ladder, and additional figures for the representative $N=5$ stack (Figures~S1 and S2));
(vii) Section~7 (nanophotonic bounds and Miller thickness constraints): nanophotonic bounds and thickness constraints relevant to ultrathin absorbers;
(viii) Section~8 (extended bandgap-ladder tables for reproducibility and bottom-cell bandgap sensitivity);
(ix) Section~9 (nonreciprocal multijunction benchmark);
(x) Section~10 (full ladder lists in text form).

\section*{Author Information}
\subsection*{Corresponding Author}
Seungwoo Lee --- E-mail: \href{mailto:seungwoo@korea.ac.kr}{seungwoo@korea.ac.kr}

\subsection*{Author Contributions}
S.L. conceived the original research idea at the initial stage, computed all the results, and wrote the paper.

\section*{Notes}
The author declares no competing financial interest.

\section*{Acknowledgments}
We acknowledge funding from National Research Foundation of Korea (RS-2022-NR068141) and from the KIST Institutional Program (Project No.: 2V09840-23-P023). This research was also supported by a grant of the Korea-US Collaborative Research Fund (KUCRF), funded by the Ministry of Science and ICT and Ministry of Health \& Welfare, Republic of Korea (grant number: RS-2024-00468463) and by Korea University grant.

\printbibliography

@article{ShockleyQueisser1961,
  title={Detailed Balance Limit of Efficiency of p-n Junction Solar Cells},
  author={Shockley, W. and Queisser, H. J.},
  journal={Journal of Applied Physics},
  volume={32},
  pages={510--519},
  year={1961}
}

@article{DeVos1980,
  title={Detailed balance limit of the efficiency of tandem solar cells},
  author={De Vos, A.},
  journal={Journal of Physics D: Applied Physics},
  volume={13},
  pages={839--846},
  year={1980}
}

@article{Green2006,
  title={Efficiency limits for single-junction and tandem solar cells},
  author={Green, M. A.},
  journal={Progress in Photovoltaics: Research and Applications},
  volume={14},
  pages={383--395},
  year={2006}
}

@article{JariwalaAtwater2017,
  title={Van der Waals Materials for Atomically-Thin Photovoltaics: Promise and Outlook},
  author={Jariwala, D. and Davoyan, A. R. and Wong, J. and Atwater, H. A.},
  journal={ACS Photonics},
  volume={4},
  number={12},
  pages={2962--2970},
  year={2017},
  doi={10.1021/acsphotonics.7b01103}
}

@article{ReichPop2023,
  title={Efficiency limit of transition metal dichalcogenide solar cells},
  author={Reich, N. and Yu, Y. and Furchi, M. and others},
  journal={Communications Physics},
  volume={6},
  pages={1--11},
  year={2023}
}

@article{Rau2007,
  title={Reciprocity relation between photovoltaic quantum efficiency and electroluminescent emission of solar cells},
  author={Rau, U.},
  journal={Physical Review B},
  volume={76},
  pages={085303},
  year={2007}
}

@article{Rau2014,
  title={Thermodynamics of light management in photovoltaic devices},
  author={Rau, U. and Paetzold, U. W. and Kirchartz, T.},
  journal={Physical Review B},
  volume={90},
  pages={035211},
  year={2014}
}

@article{Rau2017,
  title={Efficiency potential of photovoltaic materials and devices unveiled by detailed-balance analysis},
  author={Rau, U. and Kirchartz, T.},
  journal={Physical Review B},
  volume={95},
  pages={245201},
  year={2017}
}

@article{KirchartzRau2018,
  title={What makes a good solar cell?},
  author={Kirchartz, T. and Rau, U.},
  journal={Advanced Energy Materials},
  volume={8},
  pages={1703385},
  year={2018}
}

@article{MillerYablonovitch2012,
  title={Strong internal and external luminescence as solar cells approach the Shockley--Queisser limit},
  author={Miller, O. D. and Yablonovitch, E.},
  journal={IEEE Journal of Photovoltaics},
  volume={2},
  pages={303--311},
  year={2012}
}

@article{YuRamanFan2010,
  title={Fundamental limit of nanophotonic light trapping in solar cells},
  author={Yu, Z. and Raman, A. and Fan, S.},
  journal={Proceedings of the National Academy of Sciences},
  volume={107},
  pages={17491--17496},
  year={2010}
}

@article{MillerThickness2023,
  title={Why optics needs thickness},
  author={Miller, D. A. B.},
  journal={Science},
  volume={380},
  pages={742--746},
  year={2023}
}

@article{FanNonreciprocalMJ2022,
  title={Reaching the Ultimate Efficiency of Solar Energy Harvesting with a Nonreciprocal Multijunction Solar Cell},
  author={Park, Y. and Zhao, B. and Fan, S.},
  journal={Nano Letters},
  volume={22},
  pages={448--452},
  year={2022},
  doi={10.1021/acs.nanolett.1c04288}
}

@article{FanAPL2022,
  title={Does Non-Reciprocity Break the Shockley--Queisser Limit in Single-Junction Solar Cells?},
  author={Park, Y. and Fan, S.},
  journal={Applied Physics Letters},
  volume={121},
  pages={111102},
  year={2022},
  doi={10.1063/5.0118129}
}

@article{TMDReview2024,
  title={Emerging Frontiers of 2D Transition Metal Dichalcogenides in Photovoltaics Solar Cell},
  author={Zhou, Z. and Lv, J. and Tan, C. and Yang, L. and Wang, Z.},
  journal={Advanced Functional Materials},
  volume={34},
  pages={2316175},
  year={2024},
  doi={10.1002/adfm.202316175}
}

@article{DuMser2024,
  title={Room-temperature polarization-sensitive photodetectors: Materials, device physics, and applications},
  author={Du, X. and Wu, H. and Peng, Z. and Tan, C. and Yang, L. and Wang, Z.},
  journal={Materials Science and Engineering: R: Reports},
  volume={161},
  pages={100839},
  year={2024},
  doi={10.1016/j.mser.2024.100839}
}

@article{KimNanophotonics2025,
  author  = {Kim, Kwangjin and Lee, Jieun and Lee, Jaewon and Kim, Jin-Young and Lee, Hae-Seok and Lee, Seungwoo},
  title   = {Are nanophotonic intermediate mirrors really effective in enhancing the efficiency of perovskite tandem solar cells?},
  journal = {Nanophotonics},
  year    = {2025},
  volume  = {14},
  number  = {8},
  pages   = {1239--1248},
  doi     = {10.1515/nanoph-2024-0658}
}

@article{KimLeeOE2019,
  author  = {Kim, Kwangjin and Lee, Seungwoo},
  title   = {Detailed balance analysis of plasmonic metamaterial perovskite solar cells},
  journal = {Optics Express},
  year    = {2019},
  volume  = {27},
  number  = {16},
  pages   = {A1241--A1260},
  doi     = {10.1364/OE.27.0A1241}
}

@article{LeeAFM2023LSC,
  author  = {Lee, Shin Hyung and Baek, Dongjae and Cho, Whibeom and Lee, Nohyun and Kim, Kwangjin and Kim, Jae-Hun and Kim, Han-Jun and Kim, Hyeon Ho and Kim, Hyo Jin and Lee, Seungwoo and Lee, Sung-Min},
  title   = {Tailoring Luminescent Solar Concentrators for High-Performance Flexible Double-Junction {III-V} Photovoltaics},
  journal = {Advanced Functional Materials},
  year    = {2023},
  volume  = {33},
  number  = {6},
  pages   = {2210357},
  doi     = {10.1002/adfm.202210357}
}

\end{document}

\maketitle

\newpage

% =========================================================
\section{Constants, geometry, and spectral model}
\subsection{Physical constants (SI units)}
We use the physical constants, which are summarized in Table~\ref{tab:constants}:
\begin{table}[h]
\centering
\caption{Physical constants used throughout the detailed-balance and reciprocity derivations.}
\label{tab:constants}
\begin{tabular}{@{}lll@{}}
\toprule
Symbol & Meaning & Value \\
\midrule
$q$ & elementary charge & $1.602176634\times 10^{-19}$ C \\
$h$ & Planck constant & $6.62607015\times 10^{-34}$ J\,s \\
$c$ & speed of light & $2.99792458\times 10^{8}$ m/s \\
$k_B$ & Boltzmann constant & $1.380649\times 10^{-23}$ J/K \\
\bottomrule
\end{tabular}
\end{table}

\subsection{Blackbody photon radiance and solar étendue}
The spectral photon flux per unit area per steradian per unit energy is
\begin{equation}
\Phi_\mathrm{bb}(E,T)=\frac{2E^2}{h^3c^2}\frac{1}{\exp(E/k_BT)-1}
\end{equation}
The sun subtends solid angle $\Omega_s=\pi \sin^2\theta_s$ with solar half-angle $\theta_s\simeq 0.266^\circ$. 
Optical concentration is represented by multiplying the incident flux by a factor $C$:
\begin{equation}
\Phi_\odot(E;C)=C\,\Omega_s\,\Phi_\mathrm{bb}(E,T_s)
\end{equation}
Throughout, $T_s=\SI{5778}{K}$ and $T_c=\SI{300}{K}$ unless otherwise noted.

\subsection{Full concentration convention}
In the classical SQ convention, the maximum concentration corresponds to increasing the effective solar étendue until the incident illumination fills $\pi$ steradians, giving
\begin{equation}
C_\mathrm{max}=\frac{1}{\sin^2\theta_s}\approx 4.64\times 10^{4}
\end{equation}
This convention is consistent with the use of a hemispherical emission factor $\Omega_\mathrm{emit}=2\pi$ for one-sided emission in the dark current.\cite{ShockleyQueisser1961,Green2006}

\subsection{Numerical integration range and high-energy cutoff}
All integrals are evaluated on an energy grid spanning \SIrange{0.01}{10}{eV}. 
For $T_s=\SI{5778}{K}$, the excluded power above 10~eV is negligible, while excluding power above 4.5~eV would omit $\sim 2\%$ of the incident power. 
Using \SI{10}{eV}, therefore, improves quantitative agreement with canonical thermodynamic limits without materially affecting computational cost.

% =========================================================
\section{Single-junction detailed balance: derivations and validation}
\subsection{General absorptance form}
For a device with absorptance $A(E)$, the photogenerated current density is
\begin{equation}
J_\mathrm{sc}=q\int_0^\infty A(E)\,\Phi_\odot(E;C)\,dE
\end{equation}
By Kirchhoff's law and reciprocity, the radiative emission spectrum under bias $V$ is
\begin{equation}
\phi_\mathrm{em}(E,V)=A(E)\,\Phi_\mathrm{bb}(E,T_c)\,\Omega_\mathrm{emit}\left(e^{qV/k_BT_c}-1\right)
\end{equation}
Integrating this photon flux yields the radiative recombination current:
\begin{equation}
J_\mathrm{rad}(V)=q\int_0^\infty A(E)\,\Phi_\mathrm{bb}(E,T_c)\,\Omega_\mathrm{emit}\left(e^{qV/k_BT_c}-1\right)\,dE
=J_{0,\mathrm{rad}}\left(e^{qV/k_BT_c}-1\right)
\end{equation}
where
\begin{equation}
J_{0,\mathrm{rad}}=q\int_0^\infty A(E)\,\Phi_\mathrm{bb}(E,T_c)\,\Omega_\mathrm{emit}\,dE
\end{equation}

\subsection{External radiative efficiency (ERE) and total dark current}
We incorporate nonradiative recombination via
\begin{equation}
J_0 = \frac{J_{0,\mathrm{rad}}}{\mathrm{ERE}}
\end{equation}
leading to the ideal-diode equation
\begin{equation}
J(V)=J_\mathrm{sc}-J_0\left(e^{qV/k_BT_c}-1\right)
\end{equation}
This reproduces the standard SQ triangle voltage penalty $\Delta V=(k_BT_c/q)\ln(\mathrm{ERE})$ and matches the thermodynamic voltage-loss decomposition in Rau et al.\cite{Rau2014,Rau2017}

\subsection{Device-level derating factors beyond the ideal limit}
The ideal detailed-balance curves in the main text are upper bounds. For a fabricated transfer-printed stack, only absorption in the photovoltaic absorber that is followed by successful exciton/free-carrier separation and transport contributes to $J_{\mathrm{sc}}$. We therefore define a collected active absorptance
\begin{equation}
A^{\mathrm{col}}_i(E;t)=\eta_{\mathrm{diss},i}(E)\,\eta_{\mathrm{tr},i}(E,t)\,A^{\mathrm{act}}_i(E;t),
\end{equation}
where $A^{\mathrm{act}}_i$ excludes parasitic absorption in electrodes, mirrors, and interlayers, $\eta_{\mathrm{diss},i}$ is an exciton/free-carrier dissociation probability, and $\eta_{\mathrm{tr},i}$ is a transport/collection factor. In the numerical limit calculations, $A^{\mathrm{col}}_i=A_i$; practical devices should instead replace $A_i$ by $A^{\mathrm{col}}_i$ in both $J_{\mathrm{sc}}$ and the reciprocity-related radiative dark current.

Interface defects, transfer residues, and vdW heterojunction recombination are represented by $\mathrm{ERE}_i<1$. Contact and sheet resistance can be included by converting the internal diode voltage $V_i$ to a terminal voltage,
\begin{equation}
V_{\mathrm{term},i}=V_i-J_iR_{s,i},\qquad P_i=V_{\mathrm{term},i}J_i,
\end{equation}
with additional shunt terms included, if needed, in the standard diode equation. Thus, nonideal factors do not invalidate the thermodynamic ladder calculation; they set a downward derating from the limit and identify the fabrication priorities for a given stack.

\subsection{Analytic maximum power point (MPP, Lambert-$W$)}
Setting $\partial(VJ)/\partial V=0$ yields
\begin{equation}
(1+v)e^v = 1+\frac{J_\mathrm{sc}}{J_0},\qquad v\equiv \frac{qV}{k_BT_c}
\end{equation}
so
\begin{equation}
v_\mathrm{mpp} = W\!\left(e\left[1+\frac{J_\mathrm{sc}}{J_0}\right]\right)-1,\qquad V_\mathrm{mpp}=\frac{k_BT_c}{q}v_\mathrm{mpp}
\end{equation}
The corresponding $J_\mathrm{mpp}$ and $P_\mathrm{mpp}$ follow directly.

\subsection{Canonical validation checks (order-of-magnitude)}
Using the blackbody solar model and $A(E)=\Theta(E-E_g)$ with one-sided emission ($\Omega_\mathrm{emit}=2\pi$), our code reproduces canonical SQ-like trends:
(i) full concentration increases $V_\mathrm{oc}$ and $\eta$ relative to 1 sun, 
(ii) the optimal $E_g$ shifts downward with increasing concentration, and
(iii) $V_\mathrm{oc}$ and $\eta$ penalize logarithmically with reduced ERE.
Absolute values differ modestly from AM1.5G SQ limits, because of the blackbody spectrum and geometric conventions.

% =========================================================
\section{Unlimited-junction split-spectrum multijunction model}
\subsection{Window definition and step absorbers}
We define descending bandgaps $E_{g,1}>E_{g,2}>\cdots>E_{g,N}$ and adopt split-spectrum windows:
\begin{equation}
E\in [E_{g,i}, E_{g,i-1}),\qquad E_{g,0}\equiv E_\mathrm{max}
\end{equation}
For step absorbers, $A_i(E)=1$ within the window and 0 outside. 
Current generation and radiative recombination are restricted to the same window, which is consistent with ideal spectral splitting optics (filters) that enforce identical acceptance for absorption and emission (reciprocity).

\subsection{Total power and dynamic-programming optimization}
The total maximum power is
\begin{equation}
P^\ast_\mathrm{tot}=\sum_{i=1}^N P_i^\ast(E_{g,i};E_{g,i-1})
\end{equation}
where each term is obtained from the single-junction analytic MPP in Section~2.
Discretizing $E_g$ on a grid $\{E^{(k)}\}$, define $P^\ast(E^{(k)},E^{(j)})$ for $j>k$ and use
\begin{equation}
F(n,k)=\max_{j>k}\left[F(n-1,j)+P^\ast(E^{(k)},E^{(j)})\right]
\end{equation}
with $F(1,k)=P^\ast(E^{(k)},E_\mathrm{max})$.
Backtracking yields the optimal ladder.

\subsection{Practical notes: grid choice and computational complexity}
If the grid size is $K$, precomputing $P^\ast$ requires $\mathcal{O}(K^2)$ evaluations. The DP recurrence requires $\mathcal{O}(NK^2)$ operations. 
For the conservative TMD window (1.0--2.1~eV with \SI{0.01}{eV} spacing; $K\approx 111$) and $N\le 50$, this is computationally light (seconds on a laptop).
For the unconstrained case, we use a coarser grid (0.02~eV spacing) to balance accuracy and cost.

\subsection{Meaning of the multi-terminal limit}
The multi-terminal assumption is used to isolate spectral partitioning from current matching. It corresponds to the best-case situation in which each subcell is biased at its own maximum power point and the powers are summed. A practical implementation requires separate selective contacts, lateral or transparent current collection, insulation/passivation between subcells, and external MPPT or submodule power conversion. These elements can be represented by a system-level expression such as
\begin{equation}
P_{\mathrm{system}}=\eta_{\mathrm{MPPT}}\sum_i \left(V_i-J_iR_{s,i}\right)J_i-P_{\mathrm{aux}},
\end{equation}
where $\eta_{\mathrm{MPPT}}$ is the electrical conversion efficiency and $P_{\mathrm{aux}}$ includes any auxiliary overhead. The thermodynamic efficiency in the main text corresponds to $\eta_{\mathrm{MPPT}}=1$, $R_{s,i}=0$, and $P_{\mathrm{aux}}=0$. Two-terminal stacks can be treated with the same optical model by adding current-matching constraints, but that would answer a different engineering question.

% =========================================================
\section{Reciprocity, emission channels, and the role of mirrors}
\subsection{Reciprocity relation}
 Rau et al.\cite{Rau2014,Rau2017} derived that the emitted electroluminescence spectrum is directly related to the external quantum efficiency (EQE) under reciprocity.\cite{Rau2007}
In our notation, using absorptance $A(E)$ as the reciprocity-relevant optical response,
\begin{equation}
\phi_\mathrm{em}(E,V) \propto A(E)\,\Phi_\mathrm{bb}(E,T_c)\left(e^{qV/k_BT_c}-1\right)
\end{equation}
Thus, any thickness- or nanophotonic modification of $A(E)$ changes both the current generation and the radiative emission channels.

\subsection{Mirror reflectivity and external luminescence}
Miller and Yablonovitch emphasized that approaching SQ requires strong external luminescence and that back-mirror reflectivity can be crucial for photon recycling and extraction.\cite{MillerYablonovitch2012}
In our framework, a perfect back reflector changes $\Omega_\mathrm{emit}$ from $4\pi$ to $2\pi$ (one-sided emission), reducing $J_{0,\mathrm{rad}}$ by a factor of two and increasing $V$ by $(k_BT_c/q)\ln 2$.
In multijunction stacks, analogous emission-channel control is typically achieved by intermediate mirrors or angular filters between TMD subcells.

% =========================================================
\section{Luminescent coupling chain model}
\subsection{Coupled current equations}
For junction $i$, define upward and downward hemispherical radiative prefactors
\begin{equation}
J_{0,i}^{\uparrow/\downarrow}=q\int_{E_{g,i}}^{E_{g,i-1}} A_i(E)\,\Phi_\mathrm{bb}(E,T_c)\,(2\pi)\,dE
\end{equation}
Then,
\begin{equation}
J_{\mathrm{rad},i}^{\uparrow/\downarrow}(V_i)=J_{0,i}^{\uparrow/\downarrow}\left(e^{qV_i/k_BT_c}-1\right)
\end{equation}
Assuming full absorption of downward photons by junction $i+1$, the current becomes
\begin{equation}
J_i(V_i)=J_{\odot,i}+J_{\mathrm{rad},i-1}^{\downarrow}(V_{i-1})
-\frac{J_{0,i}^{\uparrow}+J_{0,i}^{\downarrow}}{\mathrm{ERE}_i}\left(e^{qV_i/k_BT_c}-1\right)
\end{equation}
with $J_{\mathrm{rad},0}^{\downarrow}\equiv 0$.
The system output is
\begin{equation}
P=\sum_{i=1}^{N} V_i J_i(V_i)
\end{equation}
which is maximized over $\{V_i\}$.

\subsection{MPP optimization method}
We maximize $P(\{V_i\})$ by coordinate ascent: iteratively sweep each $V_i$ over a grid $[0,E_{g,i}]$, while holding other voltages fixed. 
Because coupling is unidirectional in this model, each $V_i$ primarily affects $J_i$ and $J_{i+1}$, and convergence is typically rapid for $N\le 50$.
We also provide hooks to replace coordinate ascent with gradient-free optimizers, if desired.

\subsection{Upward luminescence power as an entropy-loss proxy}
We define the upward emitted luminescence power
\begin{equation}
P^\uparrow_\mathrm{lum}=\sum_i \int_{E_{g,i}}^{E_{g,i-1}} E\,A_i(E)\,\Phi_\mathrm{bb}(E,T_c)\,(2\pi)\left(e^{qV_i/k_BT_c}-1\right)dE
\end{equation}
and report $P^\uparrow_\mathrm{lum}/P_\mathrm{in}$.

\subsection{Coupling thermalization}
The downward luminescence power emitted by junction $i$ is
\begin{equation}
P^\downarrow_{\mathrm{lum},i}=\int_{E_{g,i}}^{E_{g,i-1}} E\,A_i(E)\,\Phi_\mathrm{bb}(E,T_c)\,(2\pi)\left(e^{qV_i/k_BT_c}-1\right)dE
\end{equation}
Assuming one electron generated per absorbed photon in junction $i+1$ and extracted at $V_{i+1}$, the electrical work extracted from the coupling current $J_{\mathrm{LC},i}=J^\downarrow_{\mathrm{rad},i}$ is $V_{i+1}J_{\mathrm{LC},i}$.
We define coupling thermalization proxy
\begin{equation}
P^{\mathrm{LC}}_{\mathrm{therm}}=\sum_{i=1}^{N-1}\left(P^\downarrow_{\mathrm{lum},i}-V_{i+1}J_{\mathrm{LC},i}\right)
\end{equation}
This can separate upward entropy loss from the (partially recoverable) energy transferred downward.

\subsection{Relation between the large-$N$ plateau and coupling heat loss}
The large-$N$ efficiency plateau reported in the main text is obtained for an ideal split-spectrum, one-sided-emission model without downward luminescent coupling. The coupling calculation is a separate reciprocal-stack derating that becomes relevant when intermediate mirrors or angular filters are absent. Increasing $N$ narrows some adjacent bandgap spacings, which increases the coupling ratio $\gamma=J_{\mathrm{LC}}/J_{\odot,2}$ and the thermalization proxy above. Therefore, the two results are complementary rather than contradictory: after the spectral-partitioning benefit has nearly saturated, adding more junctions introduces more interfaces and potentially stronger small-$\Delta E_g$ coupling losses, making $N\sim5$ a practical target for the conservative TMD window.

% =========================================================

% =========================================================
\section{TMD optical response, candidate material library, and excitonic absorptance model}

\subsection{Reciprocity-consistent thickness-dependent absorptance}
For each junction, we use a thickness-dependent absorptance written in Beer--Lambert form,
\begin{equation}
A(E;t)=1-\exp\!\left[-\alpha(E)\,F(E)\,t\right]
\label{eq:beer_lambert}
\end{equation}
where $t$ is the absorber thickness, $\alpha(E)$ is the intrinsic absorption coefficient, and $F(E)$ is a dimensionless optical path-length enhancement factor.
At an elementary level, $F(E)=L_{\mathrm{eff}}(E)/t$ is the ratio between the average optical path length, that is experienced by photons inside the absorber and the physical thickness:
$F(E)=1$ corresponds to a single-pass planar film, while $F(E)>1$ captures multi-pass propagation and light trapping (e.g., back reflectors, resonant cavities, and scattering/angle randomization).

\textbf{Reciprocity:} the same $A(E;t)$ must be used in both the current generation term ($J_{\mathrm{sc}}$) and the radiative dark current ($J_{0,\mathrm{rad}}$) to enforce reciprocity.\cite{Rau2007}
This reciprocity-consistent treatment is essential, when discussing voltage losses and luminescence thermodynamics. This is because any optical design that increases absorption in a given spectral window must also increase emission in that same window.

\textbf{Exciton separation and transport.} The excitonic absorption model in this work describes the optical generation of bound excitons. Device-level conversion of those excitons to collected carriers can be included by the derating factors $\eta_{\mathrm{diss}}$ and $\eta_{\mathrm{tr}}$ defined in Section~2.3. Physically, $\eta_{\mathrm{diss}}$ depends on built-in fields, type-II band offsets, contact selectivity, and interfacial disorder, while $\eta_{\mathrm{tr}}$ depends on exciton diffusion length, free-carrier mobility, absorber thickness, and recombination during transport. If exciton dissociation is incomplete, $A^{\mathrm{col}}(E;t)<A(E;t)$ and the efficiency falls below the thermodynamic limit. In addition, exciton-dominated nonradiative recombination lowers $\mathrm{ERE}$ and therefore directly lowers voltage. The revised model therefore does not neglect strong-exciton physics; it separates the optical exciton resonance from the collection and radiative-quality derating that must be measured for a specific device architecture.

\textbf{Broadband nanophotonic proxy used in this work.}
In realistic ultrathin TMD stacks, the achievable enhancement is thickness-limited: the geometric-optics Yablonovitch bound $F\le 4n^2$ can only be approached, when the absorber is optically thick enough to support many radiative and guided channels (see Section~\ref{sec:nanophotonic_bounds}).
To map these ideas into a compact detailed-balance model, we introduce a normalized broadband enhancement function $g(t)\in[0,1]$ and define a thickness-dependent effective enhancement factor:
\begin{equation}
F_{\mathrm{proxy}}(t)=1+\left(4n^2-1\right)g(t)
\label{eq:Fproxy_SI}
\end{equation}
In our calculations, we approximate $F(E)\approx F_{\mathrm{proxy}}(t)$ (i.e., we use a wavelength-independent enhancement as a broadband average).
We adopt the simple saturating form $g(t)=1-\exp(-t/t_0)$ with $t_0=200~\mathrm{nm}$ and use $n=4.5$ as a representative TMD refractive index for the $4n^2$ limit (Section~\ref{sec:nanophotonic_bounds}).
This ``proxy'' is not meant to replace full wave-optics modeling for a specific architecture; rather, it provides a physically motivated way to translate nanophotonic bounds into thickness requirements in detailed-balance calculations.

\subsection{Optical-constant conventions: $n(\lambda)$, $k(\lambda)$, and $\alpha(\lambda)$}
We describe optical response using the complex refractive index
\begin{equation}
\tilde{n}(E)=n(E)+i\,k(E)
\end{equation}
where $n$ is the refractive index and $k$ is the extinction coefficient. 
The absorption coefficient $\alpha$ (in m$^{-1}$) and $k$ are related by\cite{KirchartzRau2018}
\begin{equation}
\alpha(\lambda)=\frac{4\pi k(\lambda)}{\lambda}\quad \Leftrightarrow\quad
k(\lambda)=\frac{\alpha(\lambda)\lambda}{4\pi}
\end{equation}
Writing $\lambda=hc/E$ yields the equivalent energy-domain relation (used in our code when $E$ is expressed in joules):\cite{ReichPop2023}
\begin{equation}
\alpha(E)=\frac{4\pi k(E)E}{hc}
\end{equation}
These relations enable consistent interconversion between (i) empirical optical-constant datasets reported as $k(E)$ and (ii) absorption models parameterized directly in terms of $\alpha(E)$.

\textbf{Representative $n$ values.} 
In the detailed-balance calculations, the optical constants enter in two ways:
(i) via the absorptance $A(E;t)$ (through $\alpha$) and
(ii) via nanophotonic/path-length enhancement proxies that depend on $n$ (e.g., the geometric-optics factor $4n^2$).
Reich et al. report representative refractive indices $n$ at the bandgap energy for several bulk group-VI TMDs.\cite{ReichPop2023}
Table~\ref{tab:tmd_n} reproduces these representative values for convenient reference.

\begin{table}[h]
\centering
\caption{Representative refractive index $n$ values at the bandgap energy for selected bulk TMDs (from Ref.~\cite{ReichPop2023}).}
\label{tab:tmd_n}
\begin{tabular}{@{}lccc@{}}
\toprule
Material (bulk) & $E_g$ (eV) & $n(E_g)$ & Notes \\
\midrule
MoS$_2$ & 1.29 & 4.48 & indirect gap (bulk) \\
MoSe$_2$ & 1.09 & 4.79 & indirect gap (bulk) \\
WS$_2$ & 1.38 & 4.17 & indirect gap (bulk) \\
WSe$_2$ & 1.20 & 4.70 & indirect gap (bulk) \\
\bottomrule
\end{tabular}
\end{table}

\textbf{Working refractive index $n$.}
Unless otherwise stated, we use a representative constant index $n=4.5$ in the nanophotonic proxy calculations (Figure~8) and in the broadband path-length enhancement proxy $F(E)$, consistent with Table~\ref{tab:tmd_n}. 
The thickness-dependent detailed-balance results are primarily controlled by $\alpha(E)$ rather than weak dispersion of $n(E)$; including full dispersion is straightforward by replacing $n$ with tabulated $n(E)$ and using a transfer-matrix method (TMM) for multilayers.

\subsection{Candidate library for vdW/TMD absorbers within 1.0--2.1~eV}
Table~\ref{tab:tmd_candidates} compiles representative optical-gap ranges for a subset of vdW or TMD semiconductors relevant to the conservative 1.0--2.1~eV window used in the manuscript (Option~A). 
Values depend on thickness (from direct to indirect transitions), dielectric environment, strain, and alloying; the table is intended as a design map rather than a definitive database.\cite{JariwalaAtwater2017,ReichPop2023,TMDReview2024} We also include anisotropic layered optoelectronic candidates discussed in recent room-temperature photodetector reviews.\cite{DuMser2024}

{\small\setlength{\tabcolsep}{3pt}\renewcommand{\arraystretch}{1.05}
\begin{longtable}{@{}>{\raggedright\arraybackslash}p{0.17\linewidth} >{\raggedright\arraybackslash}p{0.26\linewidth} >{\raggedright\arraybackslash}p{0.13\linewidth} >{\raggedright\arraybackslash}p{0.36\linewidth}@{}}
\caption{Candidate vdW and TMD absorber library within the conservative 1.0--2.1~eV optical-gap window. ``1L'' denotes monolayer; ``few-L'' denotes few-layer. Reported bandgaps are representative optical-transition energies; actual values depend on thickness, dielectric screening, strain, and temperature.\cite{JariwalaAtwater2017,ReichPop2023,TMDReview2024}}
\label{tab:tmd_candidates}\\
\toprule
Material family & Variant / tuning knob & $E_{g,\mathrm{opt}}$ (eV) & Notes / design relevance \\
\midrule
\endfirsthead
\toprule
Material family & Variant / tuning knob & $E_{g,\mathrm{opt}}$ (eV) & Notes / design relevance \\
\midrule
\endhead
\midrule
\multicolumn{4}{r}{Continued on next page} \\
\endfoot
\bottomrule
\endlastfoot

Group-VI TMDs & WS$_2$ (1L) & $\sim$2.0--2.1 & Wide-gap visible absorber; strong A/B excitons; good candidate for $\sim$2.1~eV top rung. \\
Group-VI TMDs & MoS$_2$ (1L) & $\sim$1.8--1.95 & Strong visible absorption; canonical monolayer TMD; candidate for $\sim$1.8~eV rung. \\
Group-VI TMDs & WSe$_2$ (1L) & $\sim$1.6--1.7 & Red/near-IR edge for monolayer; thickness-tunable; candidate for mid rungs. \\
Group-VI TMDs & MoSe$_2$ (1L) & $\sim$1.5--1.6 & Mid-gap monolayer; strong excitons; candidate for $\sim$1.5~eV rung. \\
Group-VI TMDs & MoTe$_2$ (few-L/bulk) & $\sim$0.95--1.1 & Narrow-gap within conservative window; candidate for $\sim$1.0~eV bottom rung; requires phase control, inert encapsulation, and high-ERE contacts because oxidation/phase instability and indirect-gap behavior can strongly derate voltage. \\
Group-VI TMDs & WS$_2$ (2L--few-L) & $\sim$1.6--1.9 & Thickness reduces gap; candidate for $\sim$1.7--1.8~eV rung if monolayer gap is too high. \\
Group-VI TMDs & MoS$_2$ (bulk) & $\sim$1.2--1.3 & Indirect gap; useful for $\sim$1.24~eV rung with increased thickness to compensate absorptance. \\
Group-VI TMDs & WSe$_2$ (few-L/bulk) & $\sim$1.2--1.4 & Thickness-tunable; candidate for $\sim$1.24~eV rung; indirectness may reduce radiative efficiency unless photon recycling is strong. \\
Group-VI TMDs & MoSe$_2$ (bulk) & $\sim$1.1--1.2 & Candidate near $\sim$1.1~eV; could serve as alternate bottom rung if MoTe$_2$ is unavailable. \\

Alloyed TMDs & MoS$_{2(1-x)}$Se$_{2x}$ & $\sim$1.55--1.9 & Alloying provides quasi-continuous tuning between MoS$_2$ and MoSe$_2$; can ``fill'' ladder gaps. \\
Alloyed TMDs & WS$_{2(1-x)}$Se$_{2x}$ & $\sim$1.6--2.1 & Tunable wide-gap branch; candidates for top/mid rungs while retaining strong excitons. \\

Re-based TMDs & ReS$_2$ (few-L) & $\sim$1.45--1.6 & In-plane anisotropic; often retains a direct-like optical transition across thickness; candidate near $\sim$1.5~eV rung. \\
Re-based TMDs & ReSe$_2$ (few-L) & $\sim$1.25--1.45 & Candidate for $\sim$1.24~eV rung; anisotropic optics may be leveraged for polarization-selective trapping. \\

Other layered chalcogenides & SnSe$_2$ (few-L) & $\sim$1.0--1.2 & Layered semiconductor; potential bottom-rung alternative; optical constants and contacts need evaluation. \\
Other layered chalcogenides & InSe (few-L) & $\sim$1.2--1.6 & High mobility vdW semiconductor; candidate for $\sim$1.24~eV rung and above; often used in heterostructures. \\
Other layered chalcogenides & GaSe (few-L) & $\sim$1.8--2.1 & Wide-gap layered chalcogenide; possible alternative for top rung if TMD options are limited. \\

Black phosphorus & BP (thickness-tuned, encapsulated) & $\sim$0.3--2.0 & Broadly tunable by thickness; environmental instability requires encapsulation; could ``bridge'' gaps but is not a TMD. It is best treated as a contingency route if a TMD-only 1.0~eV bottom cell is not stable enough. \\

\end{longtable}
} % end scope for Table~S (tmd_candidates)

\subsection{Prioritized mapping to the representative $N=5$ ladder}
The DP-optimized conservative-window ladder for $N=5$ is (from top to bottom)
\begin{equation}
(2.10,\ 1.78,\ 1.50,\ 1.24,\ and\ 1.00)\ \mathrm{eV}
\end{equation}
Table~\ref{tab:rung_mapping} summarizes one pragmatic rung-to-material mapping consistent with the conservative window and the transfer-printing motivation of the manuscript. We emphasize that the ``best'' mapping is application-specific and should consider ERE, achievable thickness, contact selectivity, and stability (encapsulation). The broader candidate library in Table~\ref{tab:tmd_candidates} can be used to generate alternative stacks.

\begin{table}[h]
\centering
\caption{Prioritized mapping of realistic vdW/TMD candidates to the representative $N=5$ target bandgap ladder (Option~A). ``Primary'' emphasizes group-VI TMDs; ``alternates'' illustrate thickness/alloy/other-vdW routes.}
\label{tab:rung_mapping}
\small
\setlength{\tabcolsep}{4pt}
\renewcommand{\arraystretch}{1.15}
\begin{tabularx}{\linewidth}{@{}c p{2.6cm} p{3.4cm} Y@{}}
\toprule
Target $E_g$ (eV) & Primary & Alternates & Practical notes \\
\midrule
2.10 & WS$_2$ (1L) & GaSe (few-L); WS$_{2(1-x)}$Se$_{2x}$ (alloy) & Top cell benefits strongly from high ERE because its voltage loss propagates through the stack power sum. \\
1.78 & MoS$_2$ (1L) & WS$_2$ (2L--few-L); MoS$_{2(1-x)}$Se$_{2x}$ & Contact resistance and optical parasitics at the top interface dominate because currents are largest in upper cells. \\
1.50 & MoSe$_2$ (1L) & ReS$_2$; WSe$_2$ (1L, strain) & Mid-gap rung is often easiest to realize with monolayers; exciton linewidth and outcoupling set ERE. \\
1.24 & WSe$_2$ (few-L/bulk) & MoS$_2$ (bulk); InSe (few-L); ReSe$_2$ & For indirect multilayers, thickness and photon recycling (mirrors) become central to maintain voltage. \\
1.00 & MoTe$_2$ (few-L/bulk) & SnSe$_2$ (few-L); MoSe$_2$ (bulk, $\sim$1.1); BP (encapsulated) & Bottom rung is the primary limiter in the conservative window; stability and selective contacts are critical. If 1.0~eV cannot be implemented with high ERE, the ladder should be re-optimized with $E_{g,\min}=1.1$--1.2~eV rather than forcing an unstable absorber. \\
\bottomrule
\end{tabularx}
\end{table}

\subsection{Minimal excitonic absorption coefficient model}
To reproduce a TMD-like spectral shape (strong A/B excitons superimposed on a continuum), we use
\begin{equation}
\alpha(E)=\alpha_U(E)+\alpha_\mathrm{cont}(E)+\alpha_A(E)+\alpha_B(E)
\label{eq:alpha_model}
\end{equation}
with
\begin{align}
\alpha_U(E) &= \alpha_{U,0}\exp\left(\frac{E-E_g}{E_U}\right)\Theta(E_g-E)\\
\alpha_\mathrm{cont}(E) &= \alpha_{c,0}\left[1-\exp\left(-\frac{E-E_g}{E_s}\right)\right]\Theta(E-E_g)\\
\alpha_{A/B}(E) &= A_{A/B}\frac{\Gamma_{A/B}^2}{(E-E_{A/B})^2+\Gamma_{A/B}^2}
\end{align}
where $E_A=E_g$ and $E_B=E_g+\Delta_B$.
Equation~\eqref{eq:alpha_model} is not intended to reproduce all excitonic fine structure; rather, it provides a compact parameterization that reproduces (i) a sharp excitonic onset, (ii) a broad continuum rise, and (iii) a second higher-energy exciton, consistent with typical 2D-TMD optical spectra.\cite{JariwalaAtwater2017,TMDReview2024}

From the perspective of thickness requirements, it is useful to note that the excitonic Lorentzians provide large $\alpha(E)$ only over a relatively narrow energy range near the band edge.
In contrast, the solar spectrum supplies a broad distribution of above-gap photons, and the approach to unity absorptance in an ultrathin film is often controlled by the continuum term $\alpha_{\mathrm{cont}}(E)$ and by the effective optical path length $F(E)t$.
As a result, incomplete broadband absorption can persist even when the band-edge exciton is strong, leading to substantial transmission losses at small $t$ (main text, Figure~7) and motivating broadband light trapping through $F_{\mathrm{proxy}}(t)$ (Eq.~\ref{eq:Fproxy_SI} and Section~\ref{sec:nanophotonic_bounds}).

\begin{table}[h]
\centering
\caption{Generic parameter values for the excitonic absorption coefficient model (used for the thickness-dependent examples in this work unless otherwise stated).}
\label{tab:exciton_generic_params}
\begin{tabular}{@{}lll@{}}
\toprule
Parameter & Symbol & Value \\
\midrule
Continuum prefactor & $\alpha_{c,0}$ & $1.5\times 10^7$ m$^{-1}$ \\
Continuum rise scale & $E_s$ & 0.25 eV \\
Continuum saturation & $\alpha_{\max}$ & $2.0\times 10^8$ m$^{-1}$ \\
Urbach prefactor & $\alpha_{U,0}$ & $1.0\times 10^5$ m$^{-1}$ \\
Urbach energy & $E_U$ & 0.03 eV \\
A-exciton amplitude & $A_A$ & $1.0\times 10^8$ m$^{-1}$ \\
A-exciton width & $\Gamma_A$ & 0.03 eV \\
B-exciton amplitude & $A_B$ & $0.6\times 10^8$ m$^{-1}$ \\
B-exciton width & $\Gamma_B$ & 0.04 eV \\
A--B splitting & $\Delta_B$ & 0.18 eV \\
\bottomrule
\end{tabular}
\end{table}

\subsection{Material-specific excitonic parameter suggestions}
Group-VI TMDs exhibit material-dependent A--B exciton splittings (driven by spin--orbit coupling) and linewidths that depend on temperature and disorder. 
Table~\ref{tab:exciton_material_params} provides representative values that can be substituted into Eq.~\eqref{eq:alpha_model} for sensitivity analysis. 
In the Figures presented in this work, unless explicitly indicated, we use the generic set in Table~\ref{tab:exciton_generic_params} and shift $E_g$ to each rung.

\begin{table}[h]
\centering
\caption{Illustrative material-specific excitonic parameters for sensitivity analysis (to be used with Eq.~\eqref{eq:alpha_model}). Values are representative and may vary with temperature, dielectric environment, and thickness.\cite{JariwalaAtwater2017,TMDReview2024}}
\label{tab:exciton_material_params}
\small
\setlength{\tabcolsep}{4pt}
\renewcommand{\arraystretch}{1.10}
\begin{tabularx}{\linewidth}{@{}l c c c c Y@{}}
\toprule
Material & $E_A$ (eV) & $\Delta_B$ (eV) & $\Gamma_A$ (eV) & $\Gamma_B$ (eV) & Notes \\
\midrule
MoS$_2$ & 1.88 & 0.15--0.20 & 0.03--0.05 & 0.04--0.06 & Mo-based: smaller spin--orbit splitting \\
WS$_2$ & 2.05 & 0.30--0.45 & 0.03--0.05 & 0.04--0.07 & W-based: larger splitting; strong visible excitons \\
MoSe$_2$ & 1.55 & 0.18--0.25 & 0.03--0.06 & 0.04--0.07 & Candidate for $\sim$1.5~eV rung \\
WSe$_2$ & 1.65 & 0.35--0.50 & 0.03--0.06 & 0.04--0.08 & Candidate for mid/near-IR rungs \\
MoTe$_2$ & 1.05 & 0.15--0.25 & 0.03--0.07 & 0.04--0.09 & Narrow-gap; encapsulation typically required \\
\bottomrule
\end{tabularx}
\end{table}

\subsection{Additional figures for thickness-resolved results and for the representative $N=5$ stack}
\begin{figure}[h]
\centering
\includegraphics[width=0.9\linewidth]{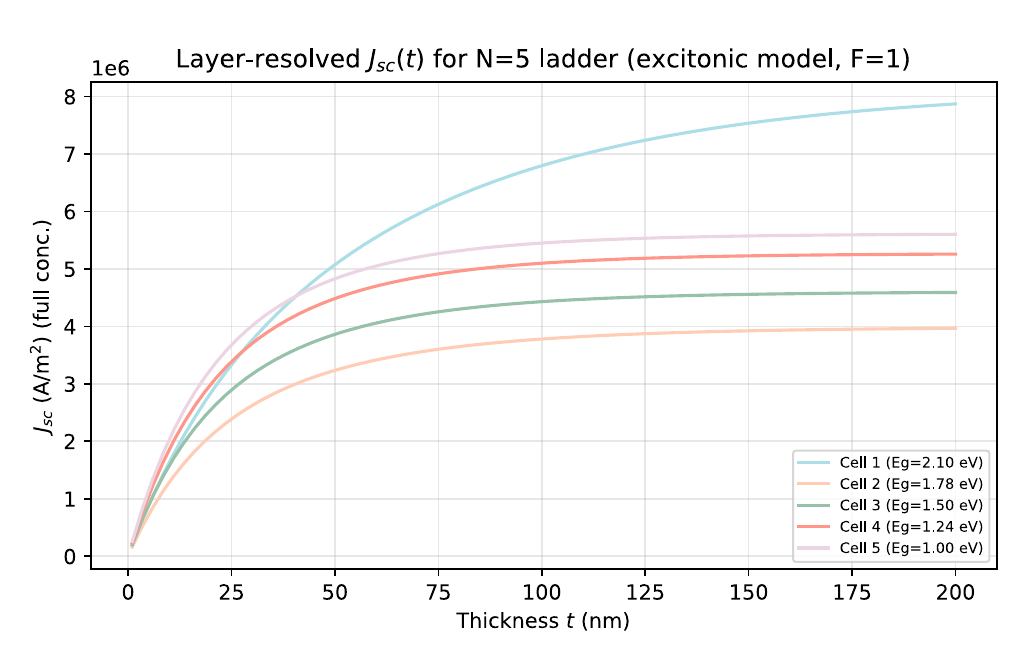}
\caption{Layer thickness-resolved $J_{sc}(t)$ for the representative $N=5$ ladder (excitonic model, single-pass).}
\label{fig:layer_resolved_jsc}
\end{figure}

\begin{figure}[h]
\centering
\includegraphics[width=0.9\linewidth]{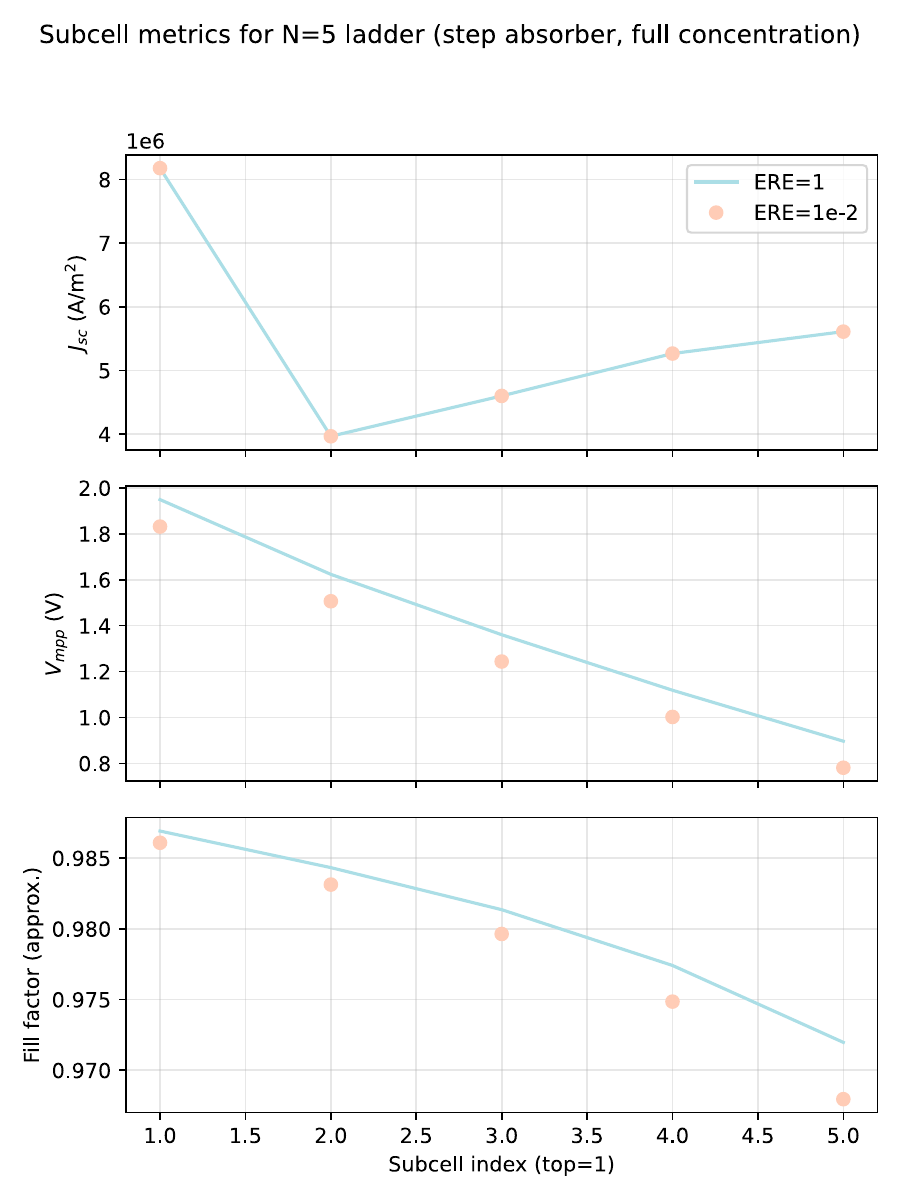}
\caption{\textbf{Subcell operating metrics for the representative $N=5$ ladder under full concentration (step-absorber model).}
Shown are $J_{sc}$, $V_\mathrm{mpp}$, and fill factor (approx.) for $\mathrm{ERE}=1$ and $\mathrm{ERE}=10^{-2}$. The two $J_{sc}$ traces overlap (and therefore visually appear as a single curve) because $\mathrm{ERE}$ affects the radiative recombination current that sets voltage, but does not change the optical-generation term in the detailed-balance model.}
\label{fig:appA1}
\end{figure}

For convenience, Figure~\ref{fig:layer_resolved_jsc} reproduces the layer thickness-resolved current trends, which are referenced in the main text. Figure~\ref{fig:appA1} reports layer-resolved operating metrics for the representative $N=5$ ladder under full concentration, highlighting the joint influence of finite absorptance and ERE on each subcell's $J_{sc}$, $V_{oc}$, and fill factor.

\FloatBarrier

\section{Nanophotonic bounds and Miller thickness constraints}
\label{sec:nanophotonic_bounds}

\subsection{Why introduce a thickness-dependent broadband proxy?}
The detailed-balance framework requires a model for how efficiently each subcell absorbs above-gap sunlight.
For an optically thin film, the Beer--Lambert form $A(E;t)=1-\exp[-\alpha(E)F(E)t]$ is a convenient way to connect microscopic material absorption ($\alpha$) to a macroscopic device response.
The dimensionless factor $F(E)$ summarizes optical design: it increases the effective photon dwell time (or average path length) inside the absorber relative to a single-pass film.
In the geometric-optics limit, where ray directions can be randomized many times before escape, the broadband average enhancement is bounded by the well-known Yablonovitch limit $F\le 4n^2$.
However, when the absorber thickness approaches the wavelength ($t\lesssim \lambda$), the number of radiative and guided modes that can participate is limited, and the enhancement that can be sustained over a broad spectrum is reduced.
This is the motivation for introducing a thickness-dependent broadband proxy (the normalized function $g(t)$ and the corresponding $F_{\mathrm{proxy}}(t)$ used in the main text and in the thickness sweeps of Figure~7).

\subsection{Nanophotonic bounds}
Yu, Raman, and Fan derived wave-optics bounds on absorption enhancement in nanophotonic structures as a function of thickness and in-plane periodicity, proving that the geometric-optics Lambertian $4n^2$ limit is not universally applicable in ultrathin films.\cite{YuRamanFan2010}
The key qualitative message is that wavelength-scale textures support a discrete set of optical channels; strong enhancement is possible. However, it is often resonant and, therefore, narrowband, while maintaining large enhancement across a broad bandwidth generally requires larger thickness (more modes and more channels).

\subsection{Thickness bounds}
Miller emphasized a related and more general viewpoint: many broadband optical functionalities intrinsically require finite thickness, because the number of independently controllable channels (degrees of freedom) scales with thickness.\cite{MillerThickness2023}
In the context of ultrathin multijunction absorbers, this reinforces that there is an unavoidable trade-off between extreme thickness reduction and broadband optical performance, even before considering practical issues such as contact losses, series resistance, or parasitic absorption.

\subsection{Thickness-dependent proxy used in this work}
To keep the detailed-balance model analytically transparent, we translate the above thickness-limited trend into a simple scalar proxy.
We define a normalized enhancement function $g(t)\in[0,1]$ and construct
\begin{equation}
F_{\mathrm{proxy}}(t)=1+\left(4n^2-1\right)g(t)
\end{equation}
in that $F_{\mathrm{proxy}}(t)=1$ for a single pass and $F_{\mathrm{proxy}}(t)\to 4n^2$ only in the thick limit.
In the calculations reported herein, we take $n=4.5$ as a representative TMD refractive index and use the saturating form $g(t)=1-\exp(-t/t_0)$ with $t_0=200~\mathrm{nm}$; the resulting $g(t)$ curve is plotted in Figure~8c of the main text.
This proxy is not intended to be a rigorous upper bound; rather, it provides a physically motivated way to explore how thickness constraints and broadband light trapping translate into multijunction performance metrics.
More architecture-specific modeling can be performed by replacing $g(t)$ (or $F(E)$) with a wavelength-dependent function computed from electromagnetic simulations or from published nanophotonic bounds.

% =========================================================
\section{Extended bandgap-ladder tables}
The following tables list DP-optimized bandgap ladders and efficiencies used throughout this paper (Tables~\ref{tab:A1}--\ref{tab:A4}). For convenience, we also cite them individually below as Tables~\ref{tab:A1}, \ref{tab:A2}, \ref{tab:A3}, and \ref{tab:A4}.

\subsection*{A1. Conservative TMD window (full concentration)}

{\scriptsize
\setlength{\tabcolsep}{4pt}
\renewcommand{\arraystretch}{1.1}
\begin{longtable}{@{}r r >{\raggedright\arraybackslash}p{0.72\linewidth}@{}}
\caption{Optimal TMD-window bandgap ladders (full concentration, ERE=1, split-spectrum). Bandgaps listed top-to-bottom.}\label{tab:A1}\\
\toprule
$N$ & $\eta$ (\%) & $E_g$ ladder (eV; top$\rightarrow$bottom) \\
\midrule
\endfirsthead
\toprule
$N$ & $\eta$ (\%) & $E_g$ ladder (eV; top$\rightarrow$bottom) \\
\midrule
\endhead
1  & 39.97 & 1.07 \\
2  & 53.25 & 1.85, 1.00 \\
3  & 58.47 & 2.10, 1.48, 1.00 \\
4  & 60.48 & 2.10, 1.68, 1.32, 1.00 \\
5  & 61.46 & 2.10, 1.78, 1.50, 1.24, 1.00 \\
6  & 62.02 & 2.10, 1.84, 1.61, 1.39, 1.19, 1.00 \\
7  & 62.38 & 2.10, 1.88, 1.68, 1.50, 1.33, 1.16, 1.00 \\
8  & 62.62 & 2.10, 1.91, 1.74, 1.58, 1.43, 1.28, 1.14, 1.00 \\
9  & 62.79 & 2.10, 1.94, 1.79, 1.64, 1.50, 1.37, 1.24, 1.12, 1.00 \\
10 & 62.91 & 2.10, 1.96, 1.82, 1.69, 1.57, 1.45, 1.33, 1.22, 1.11, 1.00 \\
\bottomrule
\end{longtable}
}

\subsection*{A2. Unconstrained bandgaps (full concentration)}

{\scriptsize
\setlength{\tabcolsep}{4pt}
\renewcommand{\arraystretch}{1.1}
\begin{longtable}{@{}r r >{\raggedright\arraybackslash}p{0.72\linewidth}@{}}
\caption{Optimal unconstrained bandgap ladders (full concentration, ERE=1, split-spectrum).}\label{tab:A2}\\
\toprule
$N$ & $\eta$ (\%) & $E_g$ ladder (eV; top$\rightarrow$bottom) \\
\midrule
\endfirsthead
\toprule
$N$ & $\eta$ (\%) & $E_g$ ladder (eV; top$\rightarrow$bottom) \\
\midrule
\endhead
1  & 39.97 & 1.07 \\
2  & 54.79 & 1.65, 0.75 \\
3  & 62.62 & 2.03, 1.21, 0.59 \\
4  & 67.46 & 2.33, 1.55, 0.99, 0.49 \\
5  & 70.74 & 2.55, 1.81, 1.29, 0.85, 0.43 \\
6  & 73.11 & 2.77, 2.05, 1.55, 1.13, 0.75, 0.37 \\
7  & 74.89 & 2.95, 2.23, 1.75, 1.35, 1.01, 0.69, 0.35 \\
8  & 76.27 & 3.09, 2.39, 1.91, 1.53, 1.21, 0.91, 0.61, 0.31 \\
9  & 77.38 & 3.23, 2.53, 2.07, 1.69, 1.37, 1.09, 0.83, 0.57, 0.29 \\
10 & 78.28 & 3.35, 2.67, 2.21, 1.85, 1.55, 1.27, 1.01, 0.77, 0.53, 0.27 \\
\bottomrule
\end{longtable}
}

\subsection*{A3. Conservative TMD window (1 sun)}

{\scriptsize
\setlength{\tabcolsep}{4pt}
\renewcommand{\arraystretch}{1.1}
\begin{longtable}{@{}r r >{\raggedright\arraybackslash}p{0.72\linewidth}@{}}
\caption{Optimal TMD-window bandgap ladders (1 sun, ERE=1, split-spectrum).}\label{tab:A3}\\
\toprule
$N$ & $\eta$ (\%) & $E_g$ ladder (eV; top$\rightarrow$bottom) \\
\midrule
\endfirsthead
\toprule
$N$ & $\eta$ (\%) & $E_g$ ladder (eV; top$\rightarrow$bottom) \\
\midrule
\endhead
1  & 29.92 & 1.28 \\
2  & 41.43 & 1.85, 1.00 \\
3  & 46.61 & 2.10, 1.49, 1.00 \\
4  & 48.61 & 2.10, 1.68, 1.32, 1.00 \\
5  & 49.58 & 2.10, 1.78, 1.50, 1.24, 1.00 \\
6  & 50.15 & 2.10, 1.84, 1.61, 1.39, 1.19, 1.00 \\
7  & 50.50 & 2.10, 1.88, 1.68, 1.50, 1.33, 1.16, 1.00 \\
8  & 50.74 & 2.10, 1.91, 1.74, 1.58, 1.43, 1.28, 1.14, 1.00 \\
9  & 50.91 & 2.10, 1.94, 1.79, 1.64, 1.50, 1.37, 1.24, 1.12, 1.00 \\
10 & 51.03 & 2.10, 1.96, 1.82, 1.69, 1.57, 1.45, 1.33, 1.22, 1.11, 1.00 \\
\bottomrule
\end{longtable}
}

\subsection*{A4. Unconstrained bandgaps (1 sun)}

{\scriptsize
\setlength{\tabcolsep}{4pt}
\renewcommand{\arraystretch}{1.1}
\begin{longtable}{@{}r r >{\raggedright\arraybackslash}p{0.72\linewidth}@{}}
\caption{Optimal unconstrained bandgap ladders (1 sun, ERE=1, split-spectrum).}\label{tab:A4}\\
\toprule
$N$ & $\eta$ (\%) & $E_g$ ladder (eV; top$\rightarrow$bottom) \\
\midrule
\endfirsthead
\toprule
$N$ & $\eta$ (\%) & $E_g$ ladder (eV; top$\rightarrow$bottom) \\
\midrule
\endhead
1  & 29.92 & 1.27 \\
2  & 41.46 & 1.83, 0.97 \\
3  & 47.66 & 2.19, 1.39, 0.81 \\
4  & 51.53 & 2.49, 1.73, 1.19, 0.73 \\
5  & 54.18 & 2.71, 1.97, 1.47, 1.05, 0.65 \\
6  & 56.10 & 2.91, 2.19, 1.71, 1.31, 0.95, 0.61 \\
7  & 57.55 & 3.07, 2.37, 1.89, 1.51, 1.17, 0.87, 0.57 \\
8  & 58.68 & 3.23, 2.53, 2.07, 1.71, 1.39, 1.11, 0.83, 0.55 \\
9  & 59.59 & 3.35, 2.67, 2.21, 1.85, 1.55, 1.27, 1.01, 0.77, 0.53 \\
10 & 60.32 & 3.49, 2.81, 2.35, 1.99, 1.69, 1.43, 1.19, 0.97, 0.75, 0.51 \\
\bottomrule

\end{longtable}
}

\subsection*{A5. Bottom-cell bandgap sensitivity}
To test whether the conclusions rely uniquely on a 1.0~eV MoTe$_2$-like bottom absorber, we repeated the DP optimization after raising the minimum accessible bandgap. The top limit remains fixed at 2.1~eV and the same full-concentration, split-spectrum, $\mathrm{ERE}=1$ assumptions are used. The absolute efficiency limit decreases as near-IR photons are lost, but the near-plateau behavior around $N\sim5$ remains.

\begin{table}[h]
\centering
\caption{Bottom-cell bandgap sensitivity for the conservative vdW/TMD window under full concentration. The table reports re-optimized ladders for selected $E_{g,\min}$ values; $E_{g,\max}=2.1$~eV in all cases.}
\label{tab:bottom_sensitivity}
\small
\setlength{\tabcolsep}{4pt}
\renewcommand{\arraystretch}{1.10}
\begin{tabularx}{\linewidth}{@{}c c c Y@{}}
\toprule
Allowed window (eV) & $N$ & $\eta$ (\%) & Optimized ladder (top$\rightarrow$bottom, eV) \\
\midrule
1.0--2.1 & 5  & 61.45 & 2.10, 1.78, 1.50, 1.24, 1.00 \\
1.0--2.1 & 50 & 63.43 & Dense ladder from 2.10 to 1.00; see full list above \\
1.1--2.1 & 5  & 58.63 & 2.10, 1.81, 1.55, 1.32, 1.10 \\
1.1--2.1 & 50 & 60.14 & Dense ladder from 2.10 to 1.10 \\
1.2--2.1 & 5  & 55.54 & 2.10, 1.84, 1.61, 1.40, 1.20 \\
1.2--2.1 & 50 & 56.66 & Dense ladder from 2.10 to 1.20 \\
\bottomrule
\end{tabularx}
\end{table}

% =========================================================

\section{Nonreciprocal multijunction benchmark}
\subsection{Ideal nonreciprocal multijunction model}
We implement an idealized nonreciprocal chain by suppressing upward emission ($J_{0,i}^\uparrow=0$), while retaining downward emission and full absorption in the next junction.\cite{FanNonreciprocalMJ2022}
This suppresses upward luminescence power (entropy-loss proxy) and provides additional efficiency gain in multijunction architectures.

\subsection{No SQ-breaking in single junction}
Fan and Park showed that nonreciprocity does not break the SQ limit for single-junction solar cells under detailed balance.\cite{FanAPL2022}
Accordingly, in our comparisons, we treat the $N=1$ case as having no nonreciprocal gain, while $N>1$ can benefit.

% =========================================================
\section{Full ladder lists}
For reproducibility, we list the DP-optimized ladders (from top to bottom) for $N=1\ldots 50$.

\subsection*{Conservative TMD window (full concentration)}
\begin{small}
\begin{lstlisting}[breaklines=true]
N= 1: 1.07
N= 2: 1.85, 1.00
N= 3: 2.10, 1.48, 1.00
N= 4: 2.10, 1.68, 1.32, 1.00
N= 5: 2.10, 1.78, 1.50, 1.24, 1.00
N= 6: 2.10, 1.84, 1.61, 1.39, 1.19, 1.00
N= 7: 2.10, 1.88, 1.68, 1.50, 1.33, 1.16, 1.00
N= 8: 2.10, 1.91, 1.74, 1.58, 1.43, 1.28, 1.14, 1.00
N= 9: 2.10, 1.94, 1.79, 1.64, 1.50, 1.37, 1.24, 1.12, 1.00
N=10: 2.10, 1.96, 1.82, 1.69, 1.57, 1.45, 1.33, 1.22, 1.11, 1.00
N=11: 2.10, 1.97, 1.85, 1.73, 1.62, 1.51, 1.40, 1.30, 1.20, 1.10, 1.00
N=12: 2.10, 1.98, 1.87, 1.76, 1.66, 1.56, 1.46, 1.36, 1.27, 1.18, 1.09, 1.00
N=13: 2.10, 1.99, 1.89, 1.79, 1.69, 1.60, 1.51, 1.42, 1.33, 1.24, 1.16, 1.08, 1.00
N=14: 2.10, 2.00, 1.90, 1.81, 1.72, 1.63, 1.55, 1.47, 1.39, 1.31, 1.23, 1.15, 1.07, 1.00
N=15: 2.10, 2.01, 1.92, 1.83, 1.75, 1.67, 1.59, 1.51, 1.43, 1.35, 1.28, 1.21, 1.14, 1.07, 1.00
N=16: 2.10, 2.01, 1.93, 1.85, 1.77, 1.69, 1.62, 1.55, 1.48, 1.41, 1.34, 1.27, 1.20, 1.13, 1.06, 1.00
N=17: 2.10, 2.02, 1.94, 1.86, 1.79, 1.72, 1.65, 1.58, 1.51, 1.44, 1.37, 1.30, 1.24, 1.18, 1.12, 1.06, 1.00
N=18: 2.10, 2.02, 1.95, 1.88, 1.81, 1.74, 1.67, 1.60, 1.54, 1.48, 1.42, 1.36, 1.30, 1.24, 1.18, 1.12, 1.06, 1.00
N=19: 2.10, 2.03, 1.96, 1.89, 1.82, 1.76, 1.70, 1.64, 1.58, 1.52, 1.46, 1.40, 1.34, 1.28, 1.22, 1.16, 1.10, 1.05, 1.00
N=20: 2.10, 2.03, 1.96, 1.89, 1.83, 1.77, 1.71, 1.65, 1.59, 1.53, 1.47, 1.41, 1.35, 1.30, 1.25, 1.20, 1.15, 1.10, 1.05, 1.00
N=21: 2.10, 2.03, 1.97, 1.91, 1.85, 1.79, 1.73, 1.67, 1.61, 1.55, 1.50, 1.45, 1.40, 1.35, 1.30, 1.25, 1.20, 1.15, 1.10, 1.05, 1.00
N=22: 2.10, 2.04, 1.98, 1.92, 1.86, 1.80, 1.75, 1.70, 1.65, 1.60, 1.55, 1.50, 1.45, 1.40, 1.35, 1.30, 1.25, 1.20, 1.15, 1.10, 1.05, 1.00
N=23: 2.10, 2.04, 1.98, 1.92, 1.87, 1.82, 1.77, 1.72, 1.67, 1.62, 1.57, 1.52, 1.47, 1.42, 1.37, 1.32, 1.27, 1.22, 1.17, 1.12, 1.08, 1.04, 1.00
N=24: 2.10, 2.04, 1.98, 1.93, 1.88, 1.83, 1.78, 1.73, 1.68, 1.63, 1.58, 1.53, 1.48, 1.43, 1.38, 1.33, 1.28, 1.24, 1.20, 1.16, 1.12, 1.08, 1.04, 1.00
N=25: 2.10, 2.04, 1.99, 1.94, 1.89, 1.84, 1.79, 1.74, 1.69, 1.64, 1.59, 1.54, 1.49, 1.44, 1.40, 1.36, 1.32, 1.28, 1.24, 1.20, 1.16, 1.12, 1.08, 1.04, 1.00
N=26: 2.10, 2.05, 2.00, 1.95, 1.90, 1.85, 1.80, 1.75, 1.70, 1.65, 1.60, 1.56, 1.52, 1.48, 1.44, 1.40, 1.36, 1.32, 1.28, 1.24, 1.20, 1.16, 1.12, 1.08, 1.04, 1.00
N=27: 2.10, 2.05, 2.00, 1.95, 1.90, 1.85, 1.80, 1.76, 1.72, 1.68, 1.64, 1.60, 1.56, 1.52, 1.48, 1.44, 1.40, 1.36, 1.32, 1.28, 1.24, 1.20, 1.16, 1.12, 1.08, 1.04, 1.00
N=28: 2.10, 2.05, 2.00, 1.96, 1.92, 1.88, 1.84, 1.80, 1.76, 1.72, 1.68, 1.64, 1.60, 1.56, 1.52, 1.48, 1.44, 1.40, 1.36, 1.32, 1.28, 1.24, 1.20, 1.16, 1.12, 1.08, 1.04, 1.00
N=29: 2.10, 2.05, 2.01, 1.97, 1.93, 1.89, 1.85, 1.81, 1.77, 1.73, 1.69, 1.65, 1.61, 1.57, 1.53, 1.49, 1.45, 1.41, 1.37, 1.33, 1.29, 1.25, 1.21, 1.17, 1.13, 1.09, 1.06, 1.03, 1.00
N=30: 2.10, 2.06, 2.02, 1.98, 1.94, 1.90, 1.86, 1.82, 1.78, 1.74, 1.70, 1.66, 1.62, 1.58, 1.54, 1.50, 1.46, 1.42, 1.38, 1.34, 1.30, 1.26, 1.22, 1.18, 1.15, 1.12, 1.09, 1.06, 1.03, 1.00
N=31: 2.10, 2.06, 2.02, 1.98, 1.94, 1.90, 1.86, 1.82, 1.78, 1.74, 1.70, 1.66, 1.62, 1.58, 1.54, 1.50, 1.46, 1.42, 1.38, 1.34, 1.30, 1.27, 1.24, 1.21, 1.18, 1.15, 1.12, 1.09, 1.06, 1.03, 1.00
N=32: 2.10, 2.06, 2.02, 1.98, 1.94, 1.90, 1.86, 1.82, 1.78, 1.74, 1.70, 1.66, 1.62, 1.58, 1.54, 1.50, 1.46, 1.42, 1.39, 1.36, 1.33, 1.30, 1.27, 1.24, 1.21, 1.18, 1.15, 1.12, 1.09, 1.06, 1.03, 1.00
N=33: 2.10, 2.06, 2.02, 1.98, 1.94, 1.90, 1.86, 1.82, 1.78, 1.74, 1.70, 1.66, 1.62, 1.58, 1.54, 1.51, 1.48, 1.45, 1.42, 1.39, 1.36, 1.33, 1.30, 1.27, 1.24, 1.21, 1.18, 1.15, 1.12, 1.09, 1.06, 1.03, 1.00
N=34: 2.10, 2.06, 2.02, 1.98, 1.94, 1.90, 1.86, 1.82, 1.78, 1.74, 1.70, 1.66, 1.63, 1.60, 1.57, 1.54, 1.51, 1.48, 1.45, 1.42, 1.39, 1.36, 1.33, 1.30, 1.27, 1.24, 1.21, 1.18, 1.15, 1.12, 1.09, 1.06, 1.03, 1.00
N=35: 2.10, 2.06, 2.02, 1.98, 1.94, 1.90, 1.86, 1.82, 1.78, 1.75, 1.72, 1.69, 1.66, 1.63, 1.60, 1.57, 1.54, 1.51, 1.48, 1.45, 1.42, 1.39, 1.36, 1.33, 1.30, 1.27, 1.24, 1.21, 1.18, 1.15, 1.12, 1.09, 1.06, 1.03, 1.00
N=36: 2.10, 2.06, 2.02, 1.98, 1.94, 1.90, 1.87, 1.84, 1.81, 1.78, 1.75, 1.72, 1.69, 1.66, 1.63, 1.60, 1.57, 1.54, 1.51, 1.48, 1.45, 1.42, 1.39, 1.36, 1.33, 1.30, 1.27, 1.24, 1.21, 1.18, 1.15, 1.12, 1.09, 1.06, 1.03, 1.00
N=37: 2.10, 2.06, 2.02, 1.99, 1.96, 1.93, 1.90, 1.87, 1.84, 1.81, 1.78, 1.75, 1.72, 1.69, 1.66, 1.63, 1.60, 1.57, 1.54, 1.51, 1.48, 1.45, 1.42, 1.39, 1.36, 1.33, 1.30, 1.27, 1.24, 1.21, 1.18, 1.15, 1.12, 1.09, 1.06, 1.03, 1.00
N=38: 2.10, 2.07, 2.04, 2.01, 1.98, 1.95, 1.92, 1.89, 1.86, 1.83, 1.80, 1.77, 1.74, 1.71, 1.68, 1.65, 1.62, 1.59, 1.56, 1.53, 1.50, 1.47, 1.44, 1.41, 1.38, 1.35, 1.32, 1.29, 1.26, 1.23, 1.20, 1.17, 1.14, 1.11, 1.08, 1.05, 1.02, 1.00
N=39: 2.10, 2.07, 2.04, 2.01, 1.98, 1.95, 1.92, 1.89, 1.86, 1.83, 1.80, 1.77, 1.74, 1.71, 1.68, 1.65, 1.62, 1.59, 1.56, 1.53, 1.50, 1.47, 1.44, 1.41, 1.38, 1.35, 1.32, 1.29, 1.26, 1.23, 1.20, 1.17, 1.14, 1.11, 1.08, 1.06, 1.04, 1.02, 1.00
N=40: 2.10, 2.07, 2.04, 2.01, 1.98, 1.95, 1.92, 1.89, 1.86, 1.83, 1.80, 1.77, 1.74, 1.71, 1.68, 1.65, 1.62, 1.59, 1.56, 1.53, 1.50, 1.47, 1.44, 1.41, 1.38, 1.35, 1.32, 1.29, 1.26, 1.23, 1.20, 1.17, 1.14, 1.12, 1.10, 1.08, 1.06, 1.04, 1.02, 1.00
N=41: 2.10, 2.07, 2.04, 2.01, 1.98, 1.95, 1.92, 1.89, 1.86, 1.83, 1.80, 1.77, 1.74, 1.71, 1.68, 1.65, 1.62, 1.59, 1.56, 1.53, 1.50, 1.47, 1.44, 1.41, 1.38, 1.35, 1.32, 1.29, 1.26, 1.23, 1.20, 1.18, 1.16, 1.14, 1.12, 1.10, 1.08, 1.06, 1.04, 1.02, 1.00
N=42: 2.10, 2.07, 2.04, 2.01, 1.98, 1.95, 1.92, 1.89, 1.86, 1.83, 1.80, 1.77, 1.74, 1.71, 1.68, 1.65, 1.62, 1.59, 1.56, 1.53, 1.50, 1.47, 1.44, 1.41, 1.38, 1.35, 1.32, 1.29, 1.26, 1.24, 1.22, 1.20, 1.18, 1.16, 1.14, 1.12, 1.10, 1.08, 1.06, 1.04, 1.02, 1.00
N=43: 2.10, 2.07, 2.04, 2.01, 1.98, 1.95, 1.92, 1.89, 1.86, 1.83, 1.80, 1.77, 1.74, 1.71, 1.68, 1.65, 1.62, 1.59, 1.56, 1.53, 1.50, 1.47, 1.44, 1.41, 1.38, 1.35, 1.32, 1.30, 1.28, 1.26, 1.24, 1.22, 1.20, 1.18, 1.16, 1.14, 1.12, 1.10, 1.08, 1.06, 1.04, 1.02, 1.00
N=44: 2.10, 2.07, 2.04, 2.01, 1.98, 1.95, 1.92, 1.89, 1.86, 1.83, 1.80, 1.77, 1.74, 1.71, 1.68, 1.65, 1.62, 1.59, 1.56, 1.53, 1.50, 1.47, 1.44, 1.41, 1.38, 1.36, 1.34, 1.32, 1.30, 1.28, 1.26, 1.24, 1.22, 1.20, 1.18, 1.16, 1.14, 1.12, 1.10, 1.08, 1.06, 1.04, 1.02, 1.00
N=45: 2.10, 2.07, 2.04, 2.01, 1.98, 1.95, 1.92, 1.89, 1.86, 1.83, 1.80, 1.77, 1.74, 1.71, 1.68, 1.65, 1.62, 1.59, 1.56, 1.53, 1.50, 1.47, 1.44, 1.42, 1.40, 1.38, 1.36, 1.34, 1.32, 1.30, 1.28, 1.26, 1.24, 1.22, 1.20, 1.18, 1.16, 1.14, 1.12, 1.10, 1.08, 1.06, 1.04, 1.02, 1.00
N=46: 2.10, 2.07, 2.04, 2.01, 1.98, 1.95, 1.92, 1.89, 1.86, 1.83, 1.80, 1.77, 1.74, 1.71, 1.68, 1.65, 1.62, 1.59, 1.56, 1.53, 1.50, 1.48, 1.46, 1.44, 1.42, 1.40, 1.38, 1.36, 1.34, 1.32, 1.30, 1.28, 1.26, 1.24, 1.22, 1.20, 1.18, 1.16, 1.14, 1.12, 1.10, 1.08, 1.06, 1.04, 1.02, 1.00
N=47: 2.10, 2.07, 2.04, 2.01, 1.98, 1.95, 1.92, 1.89, 1.86, 1.83, 1.80, 1.77, 1.74, 1.71, 1.68, 1.65, 1.62, 1.59, 1.56, 1.54, 1.52, 1.50, 1.48, 1.46, 1.44, 1.42, 1.40, 1.38, 1.36, 1.34, 1.32, 1.30, 1.28, 1.26, 1.24, 1.22, 1.20, 1.18, 1.16, 1.14, 1.12, 1.10, 1.08, 1.06, 1.04, 1.02, 1.00
N=48: 2.10, 2.07, 2.04, 2.01, 1.98, 1.95, 1.92, 1.89, 1.86, 1.83, 1.80, 1.77, 1.74, 1.71, 1.68, 1.65, 1.62, 1.60, 1.58, 1.56, 1.54, 1.52, 1.50, 1.48, 1.46, 1.44, 1.42, 1.40, 1.38, 1.36, 1.34, 1.32, 1.30, 1.28, 1.26, 1.24, 1.22, 1.20, 1.18, 1.16, 1.14, 1.12, 1.10, 1.08, 1.06, 1.04, 1.02, 1.00
N=49: 2.10, 2.07, 2.04, 2.01, 1.98, 1.95, 1.92, 1.89, 1.86, 1.83, 1.80, 1.77, 1.74, 1.71, 1.68, 1.66, 1.64, 1.62, 1.60, 1.58, 1.56, 1.54, 1.52, 1.50, 1.48, 1.46, 1.44, 1.42, 1.40, 1.38, 1.36, 1.34, 1.32, 1.30, 1.28, 1.26, 1.24, 1.22, 1.20, 1.18, 1.16, 1.14, 1.12, 1.10, 1.08, 1.06, 1.04, 1.02, 1.00
N=50: 2.10, 2.07, 2.04, 2.01, 1.98, 1.95, 1.92, 1.89, 1.86, 1.83, 1.80, 1.77, 1.74, 1.72, 1.70, 1.68, 1.66, 1.64, 1.62, 1.60, 1.58, 1.56, 1.54, 1.52, 1.50, 1.48, 1.46, 1.44, 1.42, 1.40, 1.38, 1.36, 1.34, 1.32, 1.30, 1.28, 1.26, 1.24, 1.22, 1.20, 1.18, 1.16, 1.14, 1.12, 1.10, 1.08, 1.06, 1.04, 1.02, 1.00
\end{lstlisting}
\end{small}

\subsection*{Unconstrained bandgaps (full concentration)}
\begin{small}
\begin{lstlisting}[breaklines=true]
N= 1: 1.07
N= 2: 1.65, 0.75
N= 3: 2.03, 1.21, 0.59
N= 4: 2.33, 1.55, 0.99, 0.49
N= 5: 2.55, 1.81, 1.29, 0.85, 0.43
N= 6: 2.77, 2.05, 1.55, 1.13, 0.75, 0.37
N= 7: 2.95, 2.23, 1.75, 1.35, 1.01, 0.69, 0.35
N= 8: 3.09, 2.39, 1.91, 1.53, 1.21, 0.91, 0.61, 0.31
N= 9: 3.23, 2.53, 2.07, 1.69, 1.37, 1.09, 0.83, 0.57, 0.29
N=10: 3.35, 2.67, 2.21, 1.85, 1.55, 1.27, 1.01, 0.77, 0.53, 0.27
N=11: 3.49, 2.81, 2.35, 1.99, 1.69, 1.43, 1.19, 0.95, 0.73, 0.51, 0.25
N=12: 3.59, 2.91, 2.47, 2.13, 1.83, 1.57, 1.33, 1.11, 0.89, 0.69, 0.47, 0.23
N=13: 3.67, 3.01, 2.57, 2.23, 1.95, 1.69, 1.45, 1.23, 1.03, 0.83, 0.63, 0.43, 0.21
N=14: 3.77, 3.11, 2.67, 2.33, 2.05, 1.79, 1.57, 1.37, 1.17, 0.99, 0.81, 0.63, 0.43, 0.21
N=15: 3.87, 3.21, 2.77, 2.43, 2.15, 1.91, 1.69, 1.49, 1.29, 1.11, 0.93, 0.75, 0.57, 0.39, 0.19
N=16: 3.95, 3.29, 2.85, 2.51, 2.23, 1.99, 1.77, 1.57, 1.39, 1.21, 1.05, 0.89, 0.73, 0.57, 0.39, 0.19
N=17: 4.07, 3.41, 2.97, 2.63, 2.35, 2.11, 1.89, 1.69, 1.51, 1.33, 1.17, 1.01, 0.85, 0.69, 0.53, 0.37, 0.17
N=18: 4.11, 3.47, 3.05, 2.71, 2.43, 2.19, 1.97, 1.77, 1.59, 1.43, 1.27, 1.11, 0.95, 0.81, 0.67, 0.51, 0.35, 0.17
N=19: 4.15, 3.51, 3.09, 2.77, 2.49, 2.25, 2.05, 1.85, 1.67, 1.51, 1.35, 1.19, 1.05, 0.91, 0.77, 0.63, 0.49, 0.33, 0.15
N=20: 4.23, 3.59, 3.17, 2.85, 2.57, 2.33, 2.13, 1.95, 1.77, 1.61, 1.45, 1.31, 1.17, 1.03, 0.89, 0.75, 0.61, 0.47, 0.33, 0.15
N=21: 4.29, 3.65, 3.23, 2.91, 2.65, 2.41, 2.21, 2.03, 1.85, 1.69, 1.53, 1.39, 1.25, 1.11, 0.97, 0.85, 0.73, 0.59, 0.45, 0.31, 0.13
N=22: 4.35, 3.71, 3.29, 2.97, 2.71, 2.47, 2.27, 2.09, 1.91, 1.75, 1.61, 1.47, 1.33, 1.19, 1.07, 0.95, 0.83, 0.71, 0.59, 0.45, 0.31, 0.13
N=23: 4.39, 3.75, 3.33, 3.01, 2.75, 2.53, 2.33, 2.15, 1.99, 1.83, 1.69, 1.55, 1.41, 1.27, 1.15, 1.03, 0.91, 0.79, 0.67, 0.55, 0.43, 0.29, 0.13
N=24: 4.49, 3.85, 3.43, 3.11, 2.85, 2.63, 2.43, 2.25, 2.09, 1.93, 1.79, 1.65, 1.51, 1.39, 1.27, 1.15, 1.03, 0.91, 0.79, 0.67, 0.55, 0.43, 0.29, 0.13
N=25: 4.57, 3.93, 3.51, 3.19, 2.93, 2.71, 2.51, 2.33, 2.17, 2.01, 1.87, 1.73, 1.59, 1.47, 1.35, 1.23, 1.11, 0.99, 0.87, 0.75, 0.63, 0.51, 0.39, 0.27, 0.11
N=26: 4.59, 3.95, 3.53, 3.21, 2.95, 2.73, 2.53, 2.35, 2.19, 2.03, 1.89, 1.75, 1.63, 1.51, 1.39, 1.27, 1.15, 1.03, 0.93, 0.83, 0.73, 0.63, 0.51, 0.39, 0.27, 0.11
N=27: 4.65, 4.01, 3.59, 3.27, 3.01, 2.79, 2.59, 2.41, 2.25, 2.09, 1.95, 1.81, 1.69, 1.57, 1.45, 1.33, 1.21, 1.11, 1.01, 0.91, 0.81, 0.71, 0.61, 0.51, 0.39, 0.27, 0.11
N=28: 4.69, 4.07, 3.65, 3.33, 3.07, 2.85, 2.65, 2.47, 2.31, 2.15, 2.01, 1.87, 1.75, 1.63, 1.51, 1.39, 1.29, 1.19, 1.09, 0.99, 0.89, 0.79, 0.69, 0.59, 0.49, 0.37, 0.25, 0.11
N=29: 4.75, 4.13, 3.71, 3.39, 3.13, 2.91, 2.71, 2.53, 2.37, 2.23, 2.09, 1.95, 1.83, 1.71, 1.59, 1.47, 1.37, 1.27, 1.17, 1.07, 0.97, 0.87, 0.77, 0.67, 0.57, 0.47, 0.37, 0.25, 0.11
N=30: 4.81, 4.19, 3.77, 3.45, 3.19, 2.97, 2.77, 2.59, 2.43, 2.29, 2.15, 2.01, 1.89, 1.77, 1.65, 1.55, 1.45, 1.35, 1.25, 1.15, 1.05, 0.95, 0.85, 0.75, 0.65, 0.55, 0.45, 0.35, 0.23, 0.09
N=31: 4.87, 4.25, 3.85, 3.53, 3.27, 3.05, 2.85, 2.67, 2.51, 2.37, 2.23, 2.11, 1.99, 1.87, 1.75, 1.65, 1.55, 1.45, 1.35, 1.25, 1.15, 1.05, 0.95, 0.85, 0.75, 0.65, 0.55, 0.45, 0.35, 0.23, 0.09
N=32: 4.89, 4.27, 3.87, 3.55, 3.29, 3.07, 2.87, 2.69, 2.53, 2.39, 2.25, 2.13, 2.01, 1.89, 1.77, 1.67, 1.57, 1.47, 1.37, 1.27, 1.17, 1.07, 0.97, 0.89, 0.81, 0.73, 0.63, 0.53, 0.43, 0.33, 0.23, 0.09
N=33: 4.93, 4.31, 3.91, 3.59, 3.33, 3.11, 2.91, 2.73, 2.57, 2.43, 2.29, 2.17, 2.05, 1.93, 1.81, 1.71, 1.61, 1.51, 1.41, 1.31, 1.21, 1.11, 1.03, 0.95, 0.87, 0.79, 0.71, 0.63, 0.53, 0.43, 0.33, 0.23, 0.09
N=34: 4.97, 4.35, 3.95, 3.63, 3.37, 3.15, 2.95, 2.77, 2.61, 2.47, 2.33, 2.21, 2.09, 1.97, 1.85, 1.75, 1.65, 1.55, 1.45, 1.35, 1.25, 1.17, 1.09, 1.01, 0.93, 0.85, 0.77, 0.69, 0.61, 0.53, 0.43, 0.33, 0.23, 0.09
N=35: 5.01, 4.39, 3.99, 3.67, 3.41, 3.19, 2.99, 2.81, 2.65, 2.51, 2.37, 2.25, 2.13, 2.01, 1.91, 1.81, 1.71, 1.61, 1.51, 1.41, 1.31, 1.23, 1.15, 1.07, 0.99, 0.91, 0.83, 0.75, 0.67, 0.59, 0.51, 0.41, 0.31, 0.21, 0.07
N=36: 5.03, 4.41, 4.01, 3.71, 3.45, 3.23, 3.03, 2.85, 2.69, 2.55, 2.41, 2.29, 2.17, 2.05, 1.95, 1.85, 1.75, 1.65, 1.55, 1.45, 1.37, 1.29, 1.21, 1.13, 1.05, 0.97, 0.89, 0.81, 0.73, 0.65, 0.57, 0.49, 0.41, 0.31, 0.21, 0.07
N=37: 5.11, 4.49, 4.09, 3.79, 3.53, 3.31, 3.11, 2.93, 2.77, 2.63, 2.49, 2.37, 2.25, 2.13, 2.03, 1.93, 1.83, 1.73, 1.63, 1.53, 1.45, 1.37, 1.29, 1.21, 1.13, 1.05, 0.97, 0.89, 0.81, 0.73, 0.65, 0.57, 0.49, 0.41, 0.31, 0.21, 0.07
N=38: 5.15, 4.53, 4.13, 3.83, 3.57, 3.35, 3.15, 2.97, 2.81, 2.67, 2.53, 2.41, 2.29, 2.17, 2.07, 1.97, 1.87, 1.77, 1.67, 1.59, 1.51, 1.43, 1.35, 1.27, 1.19, 1.11, 1.03, 0.95, 0.87, 0.79, 0.71, 0.63, 0.55, 0.47, 0.39, 0.31, 0.21, 0.07
N=39: 5.19, 4.57, 4.17, 3.87, 3.61, 3.39, 3.21, 3.03, 2.87, 2.73, 2.59, 2.47, 2.35, 2.23, 2.13, 2.03, 1.93, 1.83, 1.73, 1.65, 1.57, 1.49, 1.41, 1.33, 1.25, 1.17, 1.09, 1.01, 0.93, 0.85, 0.77, 0.69, 0.61, 0.53, 0.45, 0.37, 0.29, 0.19, 0.07
N=40: 5.25, 4.63, 4.23, 3.93, 3.67, 3.45, 3.27, 3.11, 2.95, 2.81, 2.67, 2.55, 2.43, 2.31, 2.21, 2.11, 2.01, 1.91, 1.81, 1.73, 1.65, 1.57, 1.49, 1.41, 1.33, 1.25, 1.17, 1.09, 1.01, 0.93, 0.85, 0.77, 0.69, 0.61, 0.53, 0.45, 0.37, 0.29, 0.19, 0.07
N=41: 5.29, 4.67, 4.27, 3.97, 3.71, 3.49, 3.31, 3.15, 2.99, 2.85, 2.71, 2.59, 2.47, 2.35, 2.25, 2.15, 2.05, 1.95, 1.87, 1.79, 1.71, 1.63, 1.55, 1.47, 1.39, 1.31, 1.23, 1.15, 1.07, 0.99, 0.91, 0.83, 0.75, 0.67, 0.59, 0.51, 0.43, 0.35, 0.27, 0.17, 0.05
N=42: 5.33, 4.71, 4.31, 4.01, 3.75, 3.53, 3.35, 3.19, 3.03, 2.89, 2.75, 2.63, 2.51, 2.39, 2.29, 2.19, 2.09, 1.99, 1.91, 1.83, 1.75, 1.67, 1.59, 1.51, 1.43, 1.35, 1.27, 1.19, 1.11, 1.03, 0.95, 0.87, 0.81, 0.75, 0.67, 0.59, 0.51, 0.43, 0.35, 0.27, 0.17, 0.05
N=43: 5.33, 4.71, 4.31, 4.01, 3.75, 3.53, 3.35, 3.19, 3.03, 2.89, 2.75, 2.63, 2.51, 2.39, 2.29, 2.19, 2.09, 1.99, 1.91, 1.83, 1.75, 1.67, 1.59, 1.51, 1.43, 1.35, 1.27, 1.19, 1.11, 1.03, 0.97, 0.91, 0.85, 0.79, 0.73, 0.67, 0.59, 0.51, 0.43, 0.35, 0.27, 0.17, 0.05
N=44: 5.33, 4.71, 4.31, 4.01, 3.77, 3.55, 3.37, 3.21, 3.05, 2.91, 2.77, 2.65, 2.53, 2.43, 2.33, 2.23, 2.13, 2.03, 1.95, 1.87, 1.79, 1.71, 1.63, 1.55, 1.47, 1.39, 1.31, 1.23, 1.15, 1.07, 1.01, 0.95, 0.89, 0.83, 0.77, 0.71, 0.65, 0.59, 0.51, 0.43, 0.35, 0.27, 0.17, 0.05
N=45: 5.37, 4.75, 4.35, 4.05, 3.81, 3.59, 3.41, 3.25, 3.09, 2.95, 2.81, 2.69, 2.57, 2.47, 2.37, 2.27, 2.17, 2.07, 1.99, 1.91, 1.83, 1.75, 1.67, 1.59, 1.51, 1.43, 1.35, 1.27, 1.19, 1.13, 1.07, 1.01, 0.95, 0.89, 0.83, 0.77, 0.71, 0.65, 0.59, 0.51, 0.43, 0.35, 0.27, 0.17, 0.05
N=46: 5.39, 4.77, 4.37, 4.07, 3.83, 3.61, 3.43, 3.27, 3.11, 2.97, 2.83, 2.71, 2.59, 2.49, 2.39, 2.29, 2.19, 2.09, 2.01, 1.93, 1.85, 1.77, 1.69, 1.61, 1.53, 1.45, 1.37, 1.29, 1.23, 1.17, 1.11, 1.05, 0.99, 0.93, 0.87, 0.81, 0.75, 0.69, 0.63, 0.57, 0.51, 0.43, 0.35, 0.27, 0.17, 0.05
N=47: 5.43, 4.81, 4.41, 4.11, 3.87, 3.65, 3.47, 3.31, 3.15, 3.01, 2.87, 2.75, 2.63, 2.53, 2.43, 2.33, 2.23, 2.13, 2.05, 1.97, 1.89, 1.81, 1.73, 1.65, 1.57, 1.49, 1.41, 1.33, 1.27, 1.21, 1.15, 1.09, 1.03, 0.97, 0.91, 0.85, 0.79, 0.73, 0.67, 0.61, 0.55, 0.49, 0.41, 0.33, 0.25, 0.17, 0.05
N=48: 5.45, 4.83, 4.43, 4.13, 3.89, 3.67, 3.49, 3.33, 3.17, 3.03, 2.89, 2.77, 2.65, 2.55, 2.45, 2.35, 2.25, 2.15, 2.07, 1.99, 1.91, 1.83, 1.75, 1.67, 1.59, 1.51, 1.43, 1.37, 1.31, 1.25, 1.19, 1.13, 1.07, 1.01, 0.95, 0.89, 0.83, 0.77, 0.71, 0.65, 0.59, 0.53, 0.47, 0.41, 0.33, 0.25, 0.17, 0.05
N=49: 5.49, 4.87, 4.47, 4.17, 3.93, 3.71, 3.53, 3.37, 3.21, 3.07, 2.95, 2.83, 2.71, 2.61, 2.51, 2.41, 2.31, 2.21, 2.13, 2.05, 1.97, 1.89, 1.81, 1.73, 1.65, 1.57, 1.49, 1.43, 1.37, 1.31, 1.25, 1.19, 1.13, 1.07, 1.01, 0.95, 0.89, 0.83, 0.77, 0.71, 0.65, 0.59, 0.53, 0.47, 0.41, 0.33, 0.25, 0.17, 0.05
N=50: 5.51, 4.89, 4.49, 4.19, 3.95, 3.73, 3.55, 3.39, 3.23, 3.09, 2.97, 2.85, 2.73, 2.63, 2.53, 2.43, 2.33, 2.23, 2.15, 2.07, 1.99, 1.91, 1.83, 1.75, 1.67, 1.59, 1.53, 1.47, 1.41, 1.35, 1.29, 1.23, 1.17, 1.11, 1.05, 0.99, 0.93, 0.87, 0.81, 0.75, 0.69, 0.63, 0.57, 0.51, 0.45, 0.39, 0.33, 0.25, 0.17, 0.05
\end{lstlisting}
\end{small}

\subsection*{Conservative TMD window (1 sun)}
\begin{small}
\begin{lstlisting}[breaklines=true]
N= 1: 1.28
N= 2: 1.85, 1.00
N= 3: 2.10, 1.49, 1.00
N= 4: 2.10, 1.68, 1.32, 1.00
N= 5: 2.10, 1.78, 1.50, 1.24, 1.00
N= 6: 2.10, 1.84, 1.61, 1.39, 1.19, 1.00
N= 7: 2.10, 1.88, 1.68, 1.50, 1.33, 1.16, 1.00
N= 8: 2.10, 1.91, 1.74, 1.58, 1.43, 1.28, 1.14, 1.00
N= 9: 2.10, 1.94, 1.79, 1.64, 1.50, 1.37, 1.24, 1.12, 1.00
N=10: 2.10, 1.96, 1.82, 1.69, 1.57, 1.45, 1.33, 1.22, 1.11, 1.00
N=11: 2.10, 1.97, 1.85, 1.73, 1.62, 1.51, 1.40, 1.30, 1.20, 1.10, 1.00
N=12: 2.10, 1.98, 1.87, 1.76, 1.66, 1.56, 1.46, 1.36, 1.27, 1.18, 1.09, 1.00
N=13: 2.10, 1.99, 1.89, 1.79, 1.69, 1.60, 1.51, 1.42, 1.33, 1.24, 1.16, 1.08, 1.00
N=14: 2.10, 2.00, 1.90, 1.81, 1.72, 1.63, 1.55, 1.47, 1.39, 1.31, 1.23, 1.15, 1.07, 1.00
N=15: 2.10, 2.01, 1.92, 1.83, 1.75, 1.67, 1.59, 1.51, 1.43, 1.35, 1.28, 1.21, 1.14, 1.07, 1.00
N=16: 2.10, 2.01, 1.93, 1.85, 1.77, 1.69, 1.62, 1.55, 1.48, 1.41, 1.34, 1.27, 1.20, 1.13, 1.06, 1.00
N=17: 2.10, 2.02, 1.94, 1.86, 1.79, 1.72, 1.65, 1.58, 1.51, 1.44, 1.37, 1.30, 1.24, 1.18, 1.12, 1.06, 1.00
N=18: 2.10, 2.02, 1.95, 1.88, 1.81, 1.74, 1.67, 1.60, 1.54, 1.48, 1.42, 1.36, 1.30, 1.24, 1.18, 1.12, 1.06, 1.00
N=19: 2.10, 2.03, 1.96, 1.89, 1.82, 1.76, 1.70, 1.64, 1.58, 1.52, 1.46, 1.40, 1.34, 1.28, 1.22, 1.16, 1.10, 1.05, 1.00
N=20: 2.10, 2.03, 1.96, 1.89, 1.83, 1.77, 1.71, 1.65, 1.59, 1.53, 1.47, 1.41, 1.35, 1.30, 1.25, 1.20, 1.15, 1.10, 1.05, 1.00
N=21: 2.10, 2.03, 1.97, 1.91, 1.85, 1.79, 1.73, 1.67, 1.61, 1.55, 1.50, 1.45, 1.40, 1.35, 1.30, 1.25, 1.20, 1.15, 1.10, 1.05, 1.00
N=22: 2.10, 2.04, 1.98, 1.92, 1.86, 1.80, 1.75, 1.70, 1.65, 1.60, 1.55, 1.50, 1.45, 1.40, 1.35, 1.30, 1.25, 1.20, 1.15, 1.10, 1.05, 1.00
N=23: 2.10, 2.04, 1.98, 1.92, 1.87, 1.82, 1.77, 1.72, 1.67, 1.62, 1.57, 1.52, 1.47, 1.42, 1.37, 1.32, 1.27, 1.22, 1.17, 1.12, 1.08, 1.04, 1.00
N=24: 2.10, 2.04, 1.98, 1.93, 1.88, 1.83, 1.78, 1.73, 1.68, 1.63, 1.58, 1.53, 1.48, 1.43, 1.38, 1.33, 1.28, 1.24, 1.20, 1.16, 1.12, 1.08, 1.04, 1.00
N=25: 2.10, 2.05, 2.00, 1.95, 1.90, 1.85, 1.80, 1.75, 1.70, 1.65, 1.60, 1.55, 1.50, 1.45, 1.40, 1.36, 1.32, 1.28, 1.24, 1.20, 1.16, 1.12, 1.08, 1.04, 1.00
N=26: 2.10, 2.05, 2.00, 1.95, 1.90, 1.85, 1.80, 1.75, 1.70, 1.65, 1.60, 1.56, 1.52, 1.48, 1.44, 1.40, 1.36, 1.32, 1.28, 1.24, 1.20, 1.16, 1.12, 1.08, 1.04, 1.00
N=27: 2.10, 2.05, 2.00, 1.95, 1.90, 1.85, 1.80, 1.76, 1.72, 1.68, 1.64, 1.60, 1.56, 1.52, 1.48, 1.44, 1.40, 1.36, 1.32, 1.28, 1.24, 1.20, 1.16, 1.12, 1.08, 1.04, 1.00
N=28: 2.10, 2.05, 2.00, 1.96, 1.92, 1.88, 1.84, 1.80, 1.76, 1.72, 1.68, 1.64, 1.60, 1.56, 1.52, 1.48, 1.44, 1.40, 1.36, 1.32, 1.28, 1.24, 1.20, 1.16, 1.12, 1.08, 1.04, 1.00
N=29: 2.10, 2.05, 2.01, 1.97, 1.93, 1.89, 1.85, 1.81, 1.77, 1.73, 1.69, 1.65, 1.61, 1.57, 1.53, 1.49, 1.45, 1.41, 1.37, 1.33, 1.29, 1.25, 1.21, 1.17, 1.13, 1.09, 1.06, 1.03, 1.00
N=30: 2.10, 2.06, 2.02, 1.98, 1.94, 1.90, 1.86, 1.82, 1.78, 1.74, 1.70, 1.66, 1.62, 1.58, 1.54, 1.50, 1.46, 1.42, 1.38, 1.34, 1.30, 1.26, 1.22, 1.18, 1.15, 1.12, 1.09, 1.06, 1.03, 1.00
N=31: 2.10, 2.06, 2.02, 1.98, 1.94, 1.90, 1.86, 1.82, 1.78, 1.74, 1.70, 1.66, 1.62, 1.58, 1.54, 1.50, 1.46, 1.42, 1.38, 1.34, 1.30, 1.27, 1.24, 1.21, 1.18, 1.15, 1.12, 1.09, 1.06, 1.03, 1.00
N=32: 2.10, 2.06, 2.02, 1.98, 1.94, 1.90, 1.86, 1.82, 1.78, 1.74, 1.70, 1.66, 1.62, 1.58, 1.54, 1.50, 1.46, 1.42, 1.39, 1.36, 1.33, 1.30, 1.27, 1.24, 1.21, 1.18, 1.15, 1.12, 1.09, 1.06, 1.03, 1.00
N=33: 2.10, 2.06, 2.02, 1.98, 1.94, 1.90, 1.86, 1.82, 1.78, 1.74, 1.70, 1.66, 1.62, 1.58, 1.54, 1.51, 1.48, 1.45, 1.42, 1.39, 1.36, 1.33, 1.30, 1.27, 1.24, 1.21, 1.18, 1.15, 1.12, 1.09, 1.06, 1.03, 1.00
N=34: 2.10, 2.06, 2.02, 1.98, 1.94, 1.90, 1.86, 1.82, 1.78, 1.74, 1.70, 1.66, 1.63, 1.60, 1.57, 1.54, 1.51, 1.48, 1.45, 1.42, 1.39, 1.36, 1.33, 1.30, 1.27, 1.24, 1.21, 1.18, 1.15, 1.12, 1.09, 1.06, 1.03, 1.00
N=35: 2.10, 2.06, 2.02, 1.98, 1.94, 1.90, 1.86, 1.82, 1.78, 1.75, 1.72, 1.69, 1.66, 1.63, 1.60, 1.57, 1.54, 1.51, 1.48, 1.45, 1.42, 1.39, 1.36, 1.33, 1.30, 1.27, 1.24, 1.21, 1.18, 1.15, 1.12, 1.09, 1.06, 1.03, 1.00
N=36: 2.10, 2.06, 2.02, 1.98, 1.94, 1.90, 1.87, 1.84, 1.81, 1.78, 1.75, 1.72, 1.69, 1.66, 1.63, 1.60, 1.57, 1.54, 1.51, 1.48, 1.45, 1.42, 1.39, 1.36, 1.33, 1.30, 1.27, 1.24, 1.21, 1.18, 1.15, 1.12, 1.09, 1.06, 1.03, 1.00
N=37: 2.10, 2.06, 2.02, 1.99, 1.96, 1.93, 1.90, 1.87, 1.84, 1.81, 1.78, 1.75, 1.72, 1.69, 1.66, 1.63, 1.60, 1.57, 1.54, 1.51, 1.48, 1.45, 1.42, 1.39, 1.36, 1.33, 1.30, 1.27, 1.24, 1.21, 1.18, 1.15, 1.12, 1.09, 1.06, 1.03, 1.00
N=38: 2.10, 2.07, 2.04, 2.01, 1.98, 1.95, 1.92, 1.89, 1.86, 1.83, 1.80, 1.77, 1.74, 1.71, 1.68, 1.65, 1.62, 1.59, 1.56, 1.53, 1.50, 1.47, 1.44, 1.41, 1.38, 1.35, 1.32, 1.29, 1.26, 1.23, 1.20, 1.17, 1.14, 1.11, 1.08, 1.05, 1.02, 1.00
N=39: 2.10, 2.07, 2.04, 2.01, 1.98, 1.95, 1.92, 1.89, 1.86, 1.83, 1.80, 1.77, 1.74, 1.71, 1.68, 1.65, 1.62, 1.59, 1.56, 1.53, 1.50, 1.47, 1.44, 1.41, 1.38, 1.35, 1.32, 1.29, 1.26, 1.23, 1.20, 1.17, 1.14, 1.11, 1.08, 1.06, 1.04, 1.02, 1.00
N=40: 2.10, 2.07, 2.04, 2.01, 1.98, 1.95, 1.92, 1.89, 1.86, 1.83, 1.80, 1.77, 1.74, 1.71, 1.68, 1.65, 1.62, 1.59, 1.56, 1.53, 1.50, 1.47, 1.44, 1.41, 1.38, 1.35, 1.32, 1.29, 1.26, 1.23, 1.20, 1.17, 1.14, 1.12, 1.10, 1.08, 1.06, 1.04, 1.02, 1.00
N=41: 2.10, 2.07, 2.04, 2.01, 1.98, 1.95, 1.92, 1.89, 1.86, 1.83, 1.80, 1.77, 1.74, 1.71, 1.68, 1.65, 1.62, 1.59, 1.56, 1.53, 1.50, 1.47, 1.44, 1.41, 1.38, 1.35, 1.32, 1.29, 1.26, 1.23, 1.20, 1.18, 1.16, 1.14, 1.12, 1.10, 1.08, 1.06, 1.04, 1.02, 1.00
N=42: 2.10, 2.07, 2.04, 2.01, 1.98, 1.95, 1.92, 1.89, 1.86, 1.83, 1.80, 1.77, 1.74, 1.71, 1.68, 1.65, 1.62, 1.59, 1.56, 1.53, 1.50, 1.47, 1.44, 1.41, 1.38, 1.35, 1.32, 1.29, 1.26, 1.24, 1.22, 1.20, 1.18, 1.16, 1.14, 1.12, 1.10, 1.08, 1.06, 1.04, 1.02, 1.00
N=43: 2.10, 2.07, 2.04, 2.01, 1.98, 1.95, 1.92, 1.89, 1.86, 1.83, 1.80, 1.77, 1.74, 1.71, 1.68, 1.65, 1.62, 1.59, 1.56, 1.53, 1.50, 1.47, 1.44, 1.41, 1.38, 1.35, 1.32, 1.30, 1.28, 1.26, 1.24, 1.22, 1.20, 1.18, 1.16, 1.14, 1.12, 1.10, 1.08, 1.06, 1.04, 1.02, 1.00
N=44: 2.10, 2.07, 2.04, 2.01, 1.98, 1.95, 1.92, 1.89, 1.86, 1.83, 1.80, 1.77, 1.74, 1.71, 1.68, 1.65, 1.62, 1.59, 1.56, 1.53, 1.50, 1.47, 1.44, 1.41, 1.38, 1.36, 1.34, 1.32, 1.30, 1.28, 1.26, 1.24, 1.22, 1.20, 1.18, 1.16, 1.14, 1.12, 1.10, 1.08, 1.06, 1.04, 1.02, 1.00
N=45: 2.10, 2.07, 2.04, 2.01, 1.98, 1.95, 1.92, 1.89, 1.86, 1.83, 1.80, 1.77, 1.74, 1.71, 1.68, 1.65, 1.62, 1.59, 1.56, 1.53, 1.50, 1.47, 1.44, 1.42, 1.40, 1.38, 1.36, 1.34, 1.32, 1.30, 1.28, 1.26, 1.24, 1.22, 1.20, 1.18, 1.16, 1.14, 1.12, 1.10, 1.08, 1.06, 1.04, 1.02, 1.00
N=46: 2.10, 2.07, 2.04, 2.01, 1.98, 1.95, 1.92, 1.89, 1.86, 1.83, 1.80, 1.77, 1.74, 1.71, 1.68, 1.65, 1.62, 1.59, 1.56, 1.53, 1.50, 1.48, 1.46, 1.44, 1.42, 1.40, 1.38, 1.36, 1.34, 1.32, 1.30, 1.28, 1.26, 1.24, 1.22, 1.20, 1.18, 1.16, 1.14, 1.12, 1.10, 1.08, 1.06, 1.04, 1.02, 1.00
N=47: 2.10, 2.07, 2.04, 2.01, 1.98, 1.95, 1.92, 1.89, 1.86, 1.83, 1.80, 1.77, 1.74, 1.71, 1.68, 1.65, 1.62, 1.59, 1.56, 1.54, 1.52, 1.50, 1.48, 1.46, 1.44, 1.42, 1.40, 1.38, 1.36, 1.34, 1.32, 1.30, 1.28, 1.26, 1.24, 1.22, 1.20, 1.18, 1.16, 1.14, 1.12, 1.10, 1.08, 1.06, 1.04, 1.02, 1.00
N=48: 2.10, 2.07, 2.04, 2.01, 1.98, 1.95, 1.92, 1.89, 1.86, 1.83, 1.80, 1.77, 1.74, 1.71, 1.68, 1.65, 1.62, 1.60, 1.58, 1.56, 1.54, 1.52, 1.50, 1.48, 1.46, 1.44, 1.42, 1.40, 1.38, 1.36, 1.34, 1.32, 1.30, 1.28, 1.26, 1.24, 1.22, 1.20, 1.18, 1.16, 1.14, 1.12, 1.10, 1.08, 1.06, 1.04, 1.02, 1.00
N=49: 2.10, 2.07, 2.04, 2.01, 1.98, 1.95, 1.92, 1.89, 1.86, 1.83, 1.80, 1.77, 1.74, 1.71, 1.68, 1.66, 1.64, 1.62, 1.60, 1.58, 1.56, 1.54, 1.52, 1.50, 1.48, 1.46, 1.44, 1.42, 1.40, 1.38, 1.36, 1.34, 1.32, 1.30, 1.28, 1.26, 1.24, 1.22, 1.20, 1.18, 1.16, 1.14, 1.12, 1.10, 1.08, 1.06, 1.04, 1.02, 1.00
N=50: 2.10, 2.07, 2.04, 2.01, 1.98, 1.95, 1.92, 1.89, 1.86, 1.83, 1.80, 1.77, 1.74, 1.72, 1.70, 1.68, 1.66, 1.64, 1.62, 1.60, 1.58, 1.56, 1.54, 1.52, 1.50, 1.48, 1.46, 1.44, 1.42, 1.40, 1.38, 1.36, 1.34, 1.32, 1.30, 1.28, 1.26, 1.24, 1.22, 1.20, 1.18, 1.16, 1.14, 1.12, 1.10, 1.08, 1.06, 1.04, 1.02, 1.00
\end{lstlisting}
\end{small}

\subsection*{Unconstrained bandgaps (1 sun)}
\begin{small}
\begin{lstlisting}[breaklines=true]
N= 1: 1.27
N= 2: 1.83, 0.97
N= 3: 2.19, 1.39, 0.81
N= 4: 2.49, 1.73, 1.19, 0.73
N= 5: 2.71, 1.97, 1.47, 1.05, 0.65
N= 6: 2.91, 2.19, 1.71, 1.31, 0.95, 0.61
N= 7: 3.07, 2.37, 1.89, 1.51, 1.17, 0.87, 0.57
N= 8: 3.23, 2.53, 2.07, 1.71, 1.39, 1.11, 0.83, 0.55
N= 9: 3.35, 2.67, 2.21, 1.85, 1.55, 1.27, 1.01, 0.77, 0.53
N=10: 3.49, 2.81, 2.35, 1.99, 1.69, 1.43, 1.19, 0.97, 0.75, 0.51
N=11: 3.61, 2.93, 2.49, 2.15, 1.85, 1.59, 1.35, 1.13, 0.91, 0.71, 0.49
N=12: 3.69, 3.03, 2.59, 2.25, 1.97, 1.71, 1.49, 1.27, 1.07, 0.87, 0.67, 0.47
N=13: 3.79, 3.13, 2.69, 2.35, 2.07, 1.83, 1.61, 1.41, 1.21, 1.03, 0.85, 0.67, 0.47
N=14: 3.91, 3.25, 2.81, 2.47, 2.19, 1.95, 1.73, 1.53, 1.35, 1.17, 0.99, 0.81, 0.63, 0.45
N=15: 4.01, 3.35, 2.91, 2.57, 2.29, 2.05, 1.83, 1.63, 1.45, 1.27, 1.11, 0.95, 0.79, 0.63, 0.45
N=16: 4.07, 3.43, 3.01, 2.67, 2.39, 2.15, 1.93, 1.73, 1.55, 1.39, 1.23, 1.07, 0.91, 0.75, 0.59, 0.43
N=17: 4.13, 3.49, 3.07, 2.75, 2.47, 2.23, 2.01, 1.81, 1.63, 1.47, 1.31, 1.15, 1.01, 0.87, 0.73, 0.59, 0.43
N=18: 4.19, 3.55, 3.13, 2.81, 2.53, 2.29, 2.09, 1.91, 1.73, 1.57, 1.41, 1.27, 1.13, 0.99, 0.85, 0.71, 0.57, 0.41
N=19: 4.29, 3.65, 3.23, 2.91, 2.65, 2.41, 2.21, 2.03, 1.85, 1.69, 1.53, 1.39, 1.25, 1.11, 0.97, 0.83, 0.69, 0.55, 0.41
N=20: 4.35, 3.71, 3.29, 2.97, 2.71, 2.47, 2.27, 2.09, 1.91, 1.75, 1.59, 1.45, 1.31, 1.17, 1.05, 0.93, 0.81, 0.69, 0.55, 0.41
N=21: 4.39, 3.75, 3.33, 3.01, 2.75, 2.53, 2.33, 2.15, 1.99, 1.83, 1.69, 1.55, 1.41, 1.27, 1.15, 1.03, 0.91, 0.79, 0.67, 0.55, 0.41
N=22: 4.47, 3.83, 3.41, 3.09, 2.83, 2.61, 2.41, 2.23, 2.07, 1.91, 1.77, 1.63, 1.49, 1.37, 1.25, 1.13, 1.01, 0.89, 0.77, 0.65, 0.53, 0.39
N=23: 4.57, 3.93, 3.51, 3.19, 2.93, 2.71, 2.51, 2.33, 2.17, 2.01, 1.87, 1.73, 1.59, 1.47, 1.35, 1.23, 1.11, 0.99, 0.87, 0.75, 0.63, 0.51, 0.39
N=24: 4.61, 3.97, 3.55, 3.23, 2.97, 2.75, 2.55, 2.37, 2.21, 2.05, 1.91, 1.77, 1.65, 1.53, 1.41, 1.29, 1.17, 1.05, 0.95, 0.85, 0.75, 0.63, 0.51, 0.39
N=25: 4.67, 4.03, 3.61, 3.29, 3.03, 2.81, 2.61, 2.43, 2.27, 2.11, 1.97, 1.83, 1.71, 1.59, 1.47, 1.35, 1.23, 1.13, 1.03, 0.93, 0.83, 0.73, 0.63, 0.51, 0.39
N=26: 4.69, 4.07, 3.65, 3.33, 3.07, 2.85, 2.65, 2.47, 2.31, 2.15, 2.01, 1.87, 1.75, 1.63, 1.51, 1.39, 1.29, 1.19, 1.09, 0.99, 0.89, 0.79, 0.69, 0.59, 0.49, 0.37
N=27: 4.77, 4.15, 3.73, 3.41, 3.15, 2.93, 2.73, 2.55, 2.39, 2.25, 2.11, 1.97, 1.85, 1.73, 1.61, 1.49, 1.39, 1.29, 1.19, 1.09, 0.99, 0.89, 0.79, 0.69, 0.59, 0.49, 0.37
N=28: 4.83, 4.21, 3.79, 3.47, 3.21, 2.99, 2.79, 2.61, 2.45, 2.31, 2.17, 2.03, 1.91, 1.79, 1.67, 1.57, 1.47, 1.37, 1.27, 1.17, 1.07, 0.97, 0.87, 0.77, 0.67, 0.57, 0.47, 0.37
N=29: 4.89, 4.27, 3.87, 3.55, 3.29, 3.07, 2.87, 2.69, 2.53, 2.39, 2.25, 2.13, 2.01, 1.89, 1.77, 1.67, 1.57, 1.47, 1.37, 1.27, 1.17, 1.07, 0.97, 0.87, 0.77, 0.67, 0.57, 0.47, 0.37
N=30: 4.93, 4.31, 3.91, 3.59, 3.33, 3.11, 2.91, 2.73, 2.57, 2.43, 2.29, 2.17, 2.05, 1.93, 1.81, 1.71, 1.61, 1.51, 1.41, 1.31, 1.21, 1.11, 1.01, 0.93, 0.85, 0.77, 0.67, 0.57, 0.47, 0.37
N=31: 4.97, 4.35, 3.95, 3.63, 3.37, 3.15, 2.95, 2.77, 2.61, 2.47, 2.33, 2.21, 2.09, 1.97, 1.85, 1.75, 1.65, 1.55, 1.45, 1.35, 1.25, 1.15, 1.07, 0.99, 0.91, 0.83, 0.75, 0.67, 0.57, 0.47, 0.37
N=32: 4.99, 4.37, 3.97, 3.65, 3.39, 3.17, 2.97, 2.79, 2.63, 2.49, 2.35, 2.23, 2.11, 1.99, 1.87, 1.77, 1.67, 1.57, 1.47, 1.37, 1.27, 1.19, 1.11, 1.03, 0.95, 0.87, 0.79, 0.71, 0.63, 0.55, 0.45, 0.35
N=33: 5.03, 4.41, 4.01, 3.69, 3.43, 3.21, 3.01, 2.83, 2.67, 2.53, 2.39, 2.27, 2.15, 2.03, 1.93, 1.83, 1.73, 1.63, 1.53, 1.43, 1.35, 1.27, 1.19, 1.11, 1.03, 0.95, 0.87, 0.79, 0.71, 0.63, 0.55, 0.45, 0.35
N=34: 5.07, 4.45, 4.05, 3.75, 3.49, 3.27, 3.07, 2.89, 2.73, 2.59, 2.45, 2.33, 2.21, 2.09, 1.99, 1.89, 1.79, 1.69, 1.59, 1.49, 1.41, 1.33, 1.25, 1.17, 1.09, 1.01, 0.93, 0.85, 0.77, 0.69, 0.61, 0.53, 0.45, 0.35
N=35: 5.13, 4.51, 4.11, 3.81, 3.55, 3.33, 3.13, 2.95, 2.79, 2.65, 2.51, 2.39, 2.27, 2.15, 2.05, 1.95, 1.85, 1.75, 1.65, 1.57, 1.49, 1.41, 1.33, 1.25, 1.17, 1.09, 1.01, 0.93, 0.85, 0.77, 0.69, 0.61, 0.53, 0.45, 0.35
N=36: 5.19, 4.57, 4.17, 3.87, 3.61, 3.39, 3.21, 3.03, 2.87, 2.73, 2.59, 2.47, 2.35, 2.23, 2.13, 2.03, 1.93, 1.83, 1.73, 1.65, 1.57, 1.49, 1.41, 1.33, 1.25, 1.17, 1.09, 1.01, 0.93, 0.85, 0.77, 0.69, 0.61, 0.53, 0.45, 0.35
N=37: 5.23, 4.61, 4.21, 3.91, 3.65, 3.43, 3.25, 3.09, 2.93, 2.79, 2.65, 2.53, 2.41, 2.29, 2.19, 2.09, 1.99, 1.89, 1.79, 1.71, 1.63, 1.55, 1.47, 1.39, 1.31, 1.23, 1.15, 1.07, 0.99, 0.91, 0.83, 0.75, 0.67, 0.59, 0.51, 0.43, 0.35
N=38: 5.29, 4.67, 4.27, 3.97, 3.71, 3.49, 3.31, 3.15, 2.99, 2.85, 2.71, 2.59, 2.47, 2.35, 2.25, 2.15, 2.05, 1.95, 1.87, 1.79, 1.71, 1.63, 1.55, 1.47, 1.39, 1.31, 1.23, 1.15, 1.07, 0.99, 0.91, 0.83, 0.75, 0.67, 0.59, 0.51, 0.43, 0.35
N=39: 5.33, 4.71, 4.31, 4.01, 3.75, 3.53, 3.35, 3.19, 3.03, 2.89, 2.75, 2.63, 2.51, 2.39, 2.29, 2.19, 2.09, 1.99, 1.91, 1.83, 1.75, 1.67, 1.59, 1.51, 1.43, 1.35, 1.27, 1.19, 1.11, 1.03, 0.95, 0.87, 0.81, 0.73, 0.65, 0.57, 0.49, 0.41, 0.33
N=40: 5.33, 4.71, 4.31, 4.01, 3.75, 3.53, 3.35, 3.19, 3.03, 2.89, 2.75, 2.63, 2.51, 2.41, 2.31, 2.21, 2.11, 2.01, 1.93, 1.85, 1.77, 1.69, 1.61, 1.53, 1.45, 1.37, 1.29, 1.21, 1.13, 1.05, 0.97, 0.91, 0.85, 0.79, 0.73, 0.65, 0.57, 0.49, 0.41, 0.33
N=41: 5.33, 4.71, 4.31, 4.01, 3.77, 3.55, 3.37, 3.21, 3.05, 2.91, 2.77, 2.65, 2.53, 2.43, 2.33, 2.23, 2.13, 2.03, 1.95, 1.87, 1.79, 1.71, 1.63, 1.55, 1.47, 1.39, 1.31, 1.23, 1.15, 1.07, 1.01, 0.95, 0.89, 0.83, 0.77, 0.71, 0.65, 0.57, 0.49, 0.41, 0.33
N=42: 5.37, 4.75, 4.35, 4.05, 3.81, 3.59, 3.41, 3.25, 3.09, 2.95, 2.81, 2.69, 2.57, 2.47, 2.37, 2.27, 2.17, 2.07, 1.99, 1.91, 1.83, 1.75, 1.67, 1.59, 1.51, 1.43, 1.35, 1.27, 1.19, 1.13, 1.07, 1.01, 0.95, 0.89, 0.83, 0.77, 0.71, 0.65, 0.57, 0.49, 0.41, 0.33
N=43: 5.39, 4.77, 4.37, 4.07, 3.83, 3.61, 3.43, 3.27, 3.11, 2.97, 2.83, 2.71, 2.59, 2.49, 2.39, 2.29, 2.19, 2.09, 2.01, 1.93, 1.85, 1.77, 1.69, 1.61, 1.53, 1.45, 1.37, 1.29, 1.23, 1.17, 1.11, 1.05, 0.99, 0.93, 0.87, 0.81, 0.75, 0.69, 0.63, 0.57, 0.49, 0.41, 0.33
N=44: 5.43, 4.81, 4.41, 4.11, 3.87, 3.65, 3.47, 3.31, 3.15, 3.01, 2.87, 2.75, 2.63, 2.53, 2.43, 2.33, 2.23, 2.13, 2.05, 1.97, 1.89, 1.81, 1.73, 1.65, 1.57, 1.49, 1.41, 1.33, 1.27, 1.21, 1.15, 1.09, 1.03, 0.97, 0.91, 0.85, 0.79, 0.73, 0.67, 0.61, 0.55, 0.49, 0.41, 0.33
N=45: 5.47, 4.85, 4.45, 4.15, 3.91, 3.69, 3.51, 3.35, 3.19, 3.05, 2.91, 2.79, 2.67, 2.57, 2.47, 2.37, 2.27, 2.17, 2.09, 2.01, 1.93, 1.85, 1.77, 1.69, 1.61, 1.53, 1.45, 1.39, 1.33, 1.27, 1.21, 1.15, 1.09, 1.03, 0.97, 0.91, 0.85, 0.79, 0.73, 0.67, 0.61, 0.55, 0.49, 0.41, 0.33
N=46: 5.51, 4.89, 4.49, 4.19, 3.95, 3.73, 3.55, 3.39, 3.23, 3.09, 2.97, 2.85, 2.73, 2.63, 2.53, 2.43, 2.33, 2.23, 2.15, 2.07, 1.99, 1.91, 1.83, 1.75, 1.67, 1.59, 1.51, 1.45, 1.39, 1.33, 1.27, 1.21, 1.15, 1.09, 1.03, 0.97, 0.91, 0.85, 0.79, 0.73, 0.67, 0.61, 0.55, 0.49, 0.41, 0.33
N=47: 5.53, 4.91, 4.51, 4.21, 3.97, 3.75, 3.57, 3.41, 3.25, 3.11, 2.99, 2.87, 2.75, 2.65, 2.55, 2.45, 2.35, 2.25, 2.17, 2.09, 2.01, 1.93, 1.85, 1.77, 1.69, 1.61, 1.55, 1.49, 1.43, 1.37, 1.31, 1.25, 1.19, 1.13, 1.07, 1.01, 0.95, 0.89, 0.83, 0.77, 0.71, 0.65, 0.59, 0.53, 0.47, 0.41, 0.33
N=48: 5.57, 4.95, 4.55, 4.25, 4.01, 3.79, 3.61, 3.45, 3.29, 3.15, 3.03, 2.91, 2.79, 2.69, 2.59, 2.49, 2.39, 2.31, 2.23, 2.15, 2.07, 1.99, 1.91, 1.83, 1.75, 1.67, 1.61, 1.55, 1.49, 1.43, 1.37, 1.31, 1.25, 1.19, 1.13, 1.07, 1.01, 0.95, 0.89, 0.83, 0.77, 0.71, 0.65, 0.59, 0.53, 0.47, 0.41, 0.33
N=49: 5.59, 4.99, 4.59, 4.29, 4.05, 3.83, 3.65, 3.49, 3.33, 3.19, 3.07, 2.95, 2.83, 2.73, 2.63, 2.53, 2.43, 2.35, 2.27, 2.19, 2.11, 2.03, 1.95, 1.87, 1.79, 1.73, 1.67, 1.61, 1.55, 1.49, 1.43, 1.37, 1.31, 1.25, 1.19, 1.13, 1.07, 1.01, 0.95, 0.89, 0.83, 0.77, 0.71, 0.65, 0.59, 0.53, 0.47, 0.41, 0.33
N=50: 5.63, 5.03, 4.63, 4.33, 4.09, 3.87, 3.69, 3.53, 3.37, 3.23, 3.11, 2.99, 2.87, 2.77, 2.67, 2.57, 2.47, 2.39, 2.31, 2.23, 2.15, 2.07, 1.99, 1.91, 1.83, 1.77, 1.71, 1.65, 1.59, 1.53, 1.47, 1.41, 1.35, 1.29, 1.23, 1.17, 1.11, 1.05, 0.99, 0.93, 0.87, 0.81, 0.75, 0.69, 0.63, 0.57, 0.51, 0.45, 0.39, 0.31
\end{lstlisting}
\end{small}

% =========================================================

\FloatBarrier